\documentclass[useAMS,usenatbib]{mn2e}
\usepackage[dvips]{graphicx}
\usepackage{float}
\newcommand{\bz}{$\langle B_z \rangle$}
\newcommand{\nz}{$\langle N_z \rangle$}
\newcommand{\vsini}{$v \sin i$}
\newcommand{\kms}{km\,s$^{-1}$}
\newcommand{\halp}{H$\alpha$}

\newcommand{\mdot}{$\dot{M}$}
\newcommand{\vinf}{$v_\infty$}
\newcommand{\rsun}{$R_\odot$}
\newcommand{\msun}{$M_\odot$}

\newcommand{\teff}{$T_{\rm eff}$}
\newcommand{\lgg}{$\log{g}$}
\newcommand{\ra}{$R_{\rm A}$}
\newcommand{\rk}{$R_{\rm K}$}
\newcommand{\rark}{$\log_{10}{R_{\rm A}/R_{\rm K}}$}
\newcommand{\rs}{$R_*$}
\newcommand{\bd}{$B_{\rm d}$}

\def\gtrsim{\mathrel{\hbox{\rlap{\hbox{\lower4pt\hbox{$\sim$}}}\hbox{$>$}}}}
\def\ltsim{\mathrel{\hbox{\rlap{\hbox{\lower4pt\hbox{$\sim$}}}\hbox{$<$}}}}

\title[Magnetic field and spectral variability of HR 2949]{The magnetic field and spectral variability of the He-weak star HR 2949}
\author[M. Shultz et al.]
{M. Shultz$^{1,2}$\thanks{E-mail: mshultz@eso.org, matt.shultz@gmail.com},
Th. Rivinius$^2$,
C. P. Folsom$^{3,4}$,
G. A. Wade$^5$,
R. H. D. Townsend$^6$,
\newauthor{J. Sikora$^{1,5}$, J. Grunhut$^{7}$, O. Stahl$^{8}$, and the MiMeS Collaboration}\\
$^1$Department of Physics, Engineering Physics \& Astronomy, Queen's University, Canada \\
$^2$European Southern Observatory, Santiago de Chile \\
$^3$CNRS, Institut de Recherche en Astrophysique et Planetologie, 14, avenue Edouard Belin, F-31400 Toulouse, France \\
$^45$Universit\'e de Toulouse, UPS-OMP, IRAP, Toulouse, France \\
$^5$Department of Physics, Royal Military College of Canada, Ontario, Canada\\
$^6$Department of Astronomy, University of Wisconsin-Madison, USA\\
$^7$European Southern Observatory, Garching, Germany \\
$^8$Zentrum f\"ur Astronomie Heidelberg, Landessternwarte K\"onigstuhl, Heidelberg, Germany\\
}

\begin{document}


\pagerange{\pageref{firstpage}--\pageref{lastpage}} \pubyear{2002}

\maketitle

\label{firstpage}

\begin{abstract}
We analyze a high resolution spectropolarimetric dataset collected for the He-weak B3p IV star HR 2949. The Zeeman effect is visible in the circularly polarized component of numerous spectral lines. The longitudinal magnetic field varies between approximately $-650$ and $+150$ G. The polar strength of the surface magnetic dipole is calculated to be 2.4$^{+0.3}_{-0.2}$ kG. The star has strong overabundances of Fe-peak elements, along with extremely strong overabundances of rare-earth elements; however, He, Al, and S are underabundant. This implies that HR 2949 is a chemically peculiar star. Variability is seen in all photospheric lines, likely due to abundance patches as seen in many Ap/Bp stars. Longitudinal magnetic field variations measured from different spectral lines yield different results, likely a consequence of uneven sampling of the photospheric magnetic field by the abundance patches. Analysis of photometric and spectroscopic data for both HR 2949 and its companion star, HR 2948, suggests a  revision of HR 2949's fundamental parameters: in particular, it is somewhat larger, hotter, and more luminous than previously believed. There is no evidence of optical or ultraviolet emission originating in HR 2949's magnetosphere, despite its moderately strong magnetic field and relatively rapid rotation; however, when calculated using theoretical and empirical boundaries on the initial rotational velocity, the spindown age is compatible with the stellar age. With the extensive phase coverage presented here, HR 2949 will make an excellent subject for Zeeman Doppler Imaging. 
 \end{abstract}

\begin{keywords}
Stars : individual : HR 2949 -- Stars: magnetic fields -- Stars: binaries: visual -- Stars: chemically peculiar -- Stars: abundances -- Stars: early-type.
\end{keywords}

\section{Introduction}

Magnetic, massive stars have recently emerged as a major new sub-class of early-type stars. The Magnetism in Massive Stars (MiMeS) Survey Component has established that approximately 10\% of OBA-type stars host large-scale, organized magnetic fields \citep{grun2012c}. These fields range in strength from hundreds of G to several kG at the magnetic poles, and are capable of confining their radiative, ionized winds within stellar magnetospheres \citep{ud2002}. This both reduces a star's mass-loss rate, and spins it down due to angular momentum loss via the corotating magnetosphere \citep{mestel1968, ud2009}, a process that has been measured in the case of the He-strong B2p stars HD 37776 \citep{miku2008} and $\sigma$ Ori E \citep{town2010}. This rapid spindown has consequences for stellar evolution, in particular strongly inhibiting mixing \citep{meynet2011}. 

Stars detected as magnetic by the MiMeS Survey Component were moved into the Targeted Component, which aimed to collect large datasets of high-resolution spectropolarimetry for all stars of interest. One such star is HR 2949. For many years this mid-B type star was considered to be a standard star, and no variability was suspected in either its integrated light or spectral features. However, \cite{rivi2003} pointed out that the star is actually variable both photometrically and spectroscopically, suggesting it to be a He-variable and/or a He-weak star.

The He-variable stars, both He-weak and He-strong, are almost invariably magnetic \citep{preston1974}, and indeed the discovery of a magnetic field in HR 2949 was announced by \cite{rivi2011}, who noted the apparently non-sinusoidal variation of the longitudinal magnetic field curve, along with correlated He line profile variations, implying chemical abundance patches as commonly seen in the magnetic Ap/Bp stars. 

Many magnetic, massive stars display variable line emission in optical and/or ultraviolet resonance lines. This is an observational signature of their magnetospheres \citep{petit2013}. Such emission is not universal amongst the magnetic stars, and HR 2949 is amongst the stars that do not display it. Determining the conditions under which magnetic stars do and do not show emission requires precise determination of the stellar, rotational, and magnetic properties of the individual stars. 

In this paper we analyze the photometric, spectropolarimetric, and spectral data available for HR 2949 in detail. As a control we perform a simultaneous analysis of HR 2948, a visual binary companion of similar spectral type but with no detectable magnetic field. The stars are of similar magnitudes ($V = 4.53$ and 4.78, respectively), and are separated by 7.3".

In Section 2, we describe the observations. In Section 3, we define a rotational ephemeris for HR 2949. In Section 4, we derive stellar parameters for HR 2949 and its binary companion, HR 2948, along with the inclination angle of HR 2949's rotational axis. HR 2949's spectral variability is examined in Section 5. Mean surface abundances for both stars are presented in Section 6. The measurement and analysis of the magnetic data is presented in Section 7, where we analyze HR 2949's magnetic field using both individual spectral lines as well as single-element least-squares deconvolution profiles, and then present a model for the dipolar component of the photospheric magnetic field. In the discussion in Section 8 we consider the implications of an apparent discrepancy in the ages of the two stars; interpret HR 2949's spectral and photometric variability in the context of the star's magnetic field; and explore the theoretical characteristics of HR 2949's magnetosphere, including the extent of magnetic confinement, degree of rotational support, and estimated spindown time. Finally, our conclusions are summarized in Section 9. 

\section{Observations}

\begin{table}
\centering
\caption[Log of ESPaDOnS Observations]{Log of ESPaDOnS observations. When binned spectra have been used in the analysis, HJD and phase are given as a range from the first to the last spectrum. Exposure times correspond to the total exposure time of all polarized sub-exposures. Peak signal-to-noise ratios (SNRs) are per spectral pixel in the binned spectra.}
\begin{tabular}{lrrrr}
\hline
\hline
Obs. & HJD & Phase & $t_{\rm exp}$ & Peak\\
Date & -- 2450000 & &  (s) & SNR\\
\hline
\\
\multicolumn{5}{c}{HR 2949} \\
\\
2010-01-01 & 5198.0726 & 0.020 &  600 &  826\\
2010-01-03 & 5200.0880 & 0.076 &  600 & 1245\\
2010-01-25 & 5222.0194 & 0.566 &  600 & 1397\\
2010-01-26 & 5222.8473 & 1.000 &  600 & 1372\\
2010-01-26 & 5223.0030 & 0.082 &  600 & 1279\\
2010-01-27 & 5223.9870 & 0.597 &  600 & 1306\\
2010-01-28 & 5224.9357 & 0.094 &  600 & 1284\\
2010-01-29 & 5225.9513 & 0.626 &  600 & 1387\\
2010-01-30 & 5227.0052 & 0.178 &  600 & 1103\\
2010-01-31 & 5227.9687 & 0.683 &  600 & 1040\\
2010-02-01 & 5228.8705 & 0.156 &  600 & 1121\\
2010-02-02 & 5230.0433 & 0.770 &  600 & 1032\\
2010-02-26 & 5253.9359 & 0.288 &  600 & 1375\\
2010-02-27 & 5254.9081 & 0.797 &  600 & 1225\\
2010-02-28 & 5255.9602 & 0.348 &  600 & 1281\\
2010-03-03 & 5258.9242 & 0.901 &  600 &  758\\
2012-09-28 & 6199.1307 & 0.489 &  600 & 1510\\
2012-11-24 & 6256.1651 & 0.370 &  600 & 1298\\
2012-11-26 & 6258.1399 & 0.404 & 2400 & 1105\\
 & -- 0.1670 & -- 0.419 & & \\
2012-11-30 & 6262.1494 & 0.505 & 1800 &  794\\
 & -- 0.1683 & -- 0.515 & &\\
2014-02-15 & 6703.9418 & 0.966 & 2240 & 1000\\
 & -- 0.9746 & -- 0.983  & &\\
2014-02-16 & 6704.8340 & 0.434 & 2880 & 2256\\
 & -- 0.8779 & -- 0.457 & &\\
\hline
\\
\multicolumn{5}{c}{HR 2948} \\
\\
2010-10-19 & 5489.1147 & -- & 1800 & 1540\\
& -- 0.1329 & -- & &\\
\hline
\hline
\end{tabular}
\label{obs_log}
\end{table}

\begin{table}
\centering
\caption[Log of FEROS Observations]{Log of FEROS observations of HR 2949. Spectra discarded from the analysis are marked with an asterisk next to the calendar date.}
\begin{tabular}{lrrr}
\hline
\hline
Obs.  & HJD & Phase & Exp.\\
Date & --2450000 & & Time (s)  \\
\hline
1998-04-12 & 1151.7704 & 0.150 & 240 \\
1999-01-18 & 1196.7644 & 0.723 & 300 \\ 
2003-02-16$^*$ & 2686.5113 & 0.206 & 300 \\ 
2003-02-17$^*$ & 2687.5075 & 0.728 & 300 \\ 
2004-05-03$^*$ & 3129.4594 & 0.268 & 360 \\
2004-05-03 & 3129.4646 & 0.270 & 240 \\ 
2004-05-03 & 3129.4680 & 0.272 & 240 \\ 
2006-02-08 & 3774.6919 & 0.307 & 180 \\ 
2006-02-08 & 3774.6946 & 0.308 & 180 \\ 
\hline
\hline
\end{tabular}
\label{obs_log_feros}
\end{table}

Spectropolarimetric observations of HR 2948 and HR 2949 have been obtained with ESPaDOnS at the 3.6m Canada-France-Hawaii Telescope (CFHT) in the 2009/2010, 2012/2013, and the 2013/2014 fall/winter seasons. For each observation, the instrument delivered four Stokes $I$ (intensity) spectra and one Stokes $V$ (circular polarization) spectrum from 370 to 1040 nm, with a resolving power of $\Delta\lambda/\lambda\sim65,000$ at 500 nm. Most of the observations (25) of HR 2949 were acquired in the context of the Magnetism in Massive Stars (MiMeS) Large Program; the remainder (15) were acquired in the context of the BRITE-Constellation Spectropolarimetric Survey (PIs C. Neiner and G. A. Wade; Neiner et al., in prep.) In total 40 observations of HR 2949 and 3 of HR 2948 are available. In order to increase the SNR, observations acquired on the same night were binned, yielding 22 measurements of HR 2949 and 1 of HR 2948 (the maximum time spanned by a given sequence of binned spectra is approximately 1 hour). The binned spectra are summarized in the log of ESPaDOnS observations in Table~\ref{obs_log}.

A total of nine spectra of HR 2949 have been obtained with the FEROS spectrograph at La Silla providing a resolving power of $\sim$48 000 over a spectral range of 375 to 890 nm. Four spectra, obtained with FEROS mounted at the ESO 1.52m and MPG 2.2m telescopes, were already described by \cite{rivi2003}. Subsequent spectra were obtained at three more epochs with FEROS mounted at the MPG 2.2 m telescope. These data are available from the ESO public archive. Three spectra (those acquired in 2003, and the first acquired in 2004) were overexposed and excluded from the analysis. The log of FEROS observations is given in Table~\ref{obs_log_feros}.

\section{Ephemeris}

\begin{figure}
\centering
\includegraphics[width=9cm]{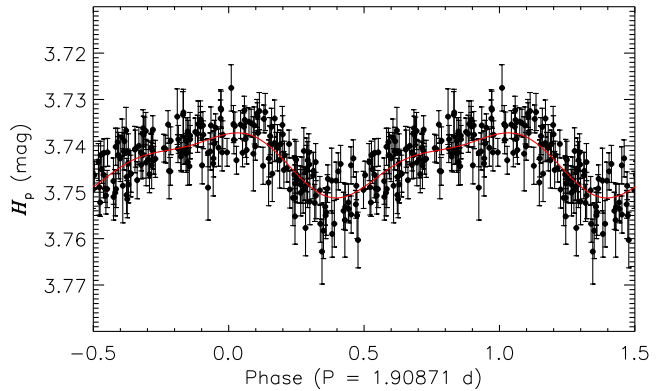} 
\caption{Hipparcos photometry (black circles) phased with the ephemeris in equation \ref{ephemeris}. The solid red line indicates the 2$^{nd}$-order sinusoid used to $\sigma$-clip the data.}
\label{hip}
\end{figure}

To constrain HR 2949's rotational period two independent data sets were used. First, there is the Hipparcos photometry, for which a period of 1.9093 d was published by \cite{koeneyer2002}. The Hipparcos photometry is shown in Fig. \ref{hip}, phased with the spectroscopic ephemeris found below. The mean $H_p$ magnitude, 3.75 mag, is much brighter than the individual $V$ magnitudes of the two stars (4.53 and 4.78 for HR 2949 and 2948, respectively), and is consistent with both stars having been observed by Hipparcos simultaneously. Our ESPaDOnS observations of HR 2948 span only 26 min, so variability of this star cannoted be ruled out. However, as explored in detail in the remainder of this paper, the spectroscopic and magnetic variations of HR 2949 follow the same period as the photometric variation. Therefore we suspect that the photometric variability is entirely due to HR 2949.

The data in Fig. \ref{hip} have been cleaned by first discarding all measurements with $H_{\rm p} > 3.77$, as in such cases one of the two stars was not fully in the aperature. Re-analysis of the cleaned data, based upon Scargle's statistics as implemented by \cite{k1996a}, gives $P = 1.9083 \pm 0.00025$ d. For presentation purposes, the data were then sigma-clipped by removing observations with uncertainties greater than twice the mean, and further cleaned by removing points lying more than 2$\sigma$ from a 2$^{\rm nd}$-order sinusoidal fit to the remaining points. 

In the spectra themselves, the He lines show the strongest variability. The ESPaDOnS and FEROS spectra were combined into a single set, and the equivalent width (EW) of several lines measured, namely He~{\sc i}~400.9~nm and 402.6~nm together as combined EW, as well as He~{\sc i}~438.8~nm and 447.1~nm individually. As the spectra are relatively sparse across a long time baseline (1999 -- 2009), severe aliasing makes it impossible to pick out a single peak in the periodogram based on these data alone. However, the aliases are relatively narrow, and there is only one peak, at $P = 1.90871 \pm 0.00007$ d, that is in reasonable agreement with the Hipparcos period. We adopt this period, being the most precise.

As epoch, we choose the ESPaDOnS data showing the strongest positive longitudinal magnetic field, so that the ephemeris is 

\begin{equation}\label{ephemeris}
T_{\rm max(B)} ({\rm HJD}) = 2455222.84726 + 1.90871(7) \times E.
\end{equation}

\section{Stellar Parameters}

\begin{table}
\centering
\caption[Stellar Parameters]{Stellar parameters. The second column gives the reference. Numerical references point to the section in this work in which the parameter is derived. Alphabetical references point to the literature according to the following reference key: a) \cite{lindroos1983}, b) \cite{nieva2013}, c) \cite{vanleeuwen2007}, d) \cite{hog2000}. }
\begin{tabular}{llrr}
\hline
\hline
& Ref. & HR 2949 & HR 2948 \\[2pt]
\hline
$V$ (mag) & a & 4.53 & 4.78 \\[2pt]
$E(B-V)$ & 4.1 & \multicolumn{2}{c}{0.04} \\[2pt]
BC (mag) & b & -1.80$\pm$0.06 & -1.05$\pm$0.21 \\[2pt]
$\pi$ (mas) & c & \multicolumn{2}{c}{7.18$\pm$1.06} \\[2pt]
$d$ (pc) & c & \multicolumn{2}{c}{139$^{+24}_{-18}$} \\[2pt]
PM (RA) (mas yr$^{-1}$) & d & -19.4$\pm$1.2 & -10.6$\pm$1.4 \\[2pt]
PM (Dec) (mas yr$^{-1}$) & d & 24.1$\pm$1.2 & 16.3$\pm$1.4\\[2pt]
$M_V$ (mag) & 4.2 &  -1.19$^{+0.26}_{-0.34}$ & -0.94$^{+0.26}_{-0.34}$ \\[2pt]
$M_{\rm bol}$ (mag) & 4.2 & -2.97$^{+0.32}_{-0.41}$ & -1.99$^{+0.46}_{-0.56}$ \\[2pt]
\hline
\teff~(kK) & 4.1 & 18.5$\pm$0.5 & 13.6$\pm$1.2 \\[2pt]
$\log{(L_*/L_\odot)}$ & 4.2 & 3.08$^{+0.16}_{-0.13}$ & 2.69$^{+0.22}_{-0.18}$ \\[2pt]
\lgg~(cgs) & 4.2 & 4.10 $\pm$ 0.15 & 4.24$\pm$0.15 \\[2pt]
Age (Myr) & 4.2 & 11$^{+15}_{-6}$ & 112$^{+20}_{-17}$ \\[2pt]
$R_* (R_\odot)$ & 4.2 & $3.0^{+0.7}_{-0.3}$ & $3.7^{+1.2}_{-0.6}$ \\[2pt]
$M_*$ ($M_\odot$) & 4.2 & 6.0$\pm 0.3$ & 4.3$\pm$0.3 \\[2pt]
\vsini~(\kms) & 4.3 & 61 $\pm$ 5 & 114 $\pm$ 5\\[2pt]
$v_{\rm eq}$ (\kms) & 4.3 & 90 $^{+39}_{-30}$ & -- \\[2pt]
\hline
Magnetic Parameters & & \\[2pt]
$i~(^\circ)$ & 4.3 & 43$^{+23}_{-12}$ & -- \\[2pt]
$\beta~(^\circ)$ & 7.2 & 50$\pm$16 & -- \\[2pt]
$B_p$ (kG) & 7.2 & 2.4$\pm$0.3 & -- \\[2pt]
\hline
\hline
\end{tabular}
\label{params}
\end{table}

HR 2949 is usually classified as a B5 or B6 dwarf to subgiant star \citep{houk1982}. However, magnetic, chemically peculiar Ap/Bp stars are not well-suited to the usual calibrations for spectral type, as the principle of spectral classification (to strictly classify a star by nearest resemblance of the spectrum to a standard star) yields misleading results when confronted with highly over- or under-abundant atomic species, especially when lines commonly used for spectral classification (such as He) are affected by chemical peculiarity. This necessitates a more careful analysis to constrain the physical parameters. 

We use the TLUSTY NLTE grid of model atmospheres for B-stars, BSTAR2006 \citep{lanzhubeny2007}, generating our own grids via linear interpolation between models of adjacent \teff~or \lgg, and convolving the model spectra with the ESPaDOnS instrumental broadening and the stellar rotational broadening. As the BSTAR2006 grid does not extend to effective temperatures cooler than 15 kK, we supplement this analysis with the LTE grid calculated by \cite{coelho2014}. 

\subsection{Effective temperature}

\begin{figure}
\centering
\includegraphics[width=8cm]{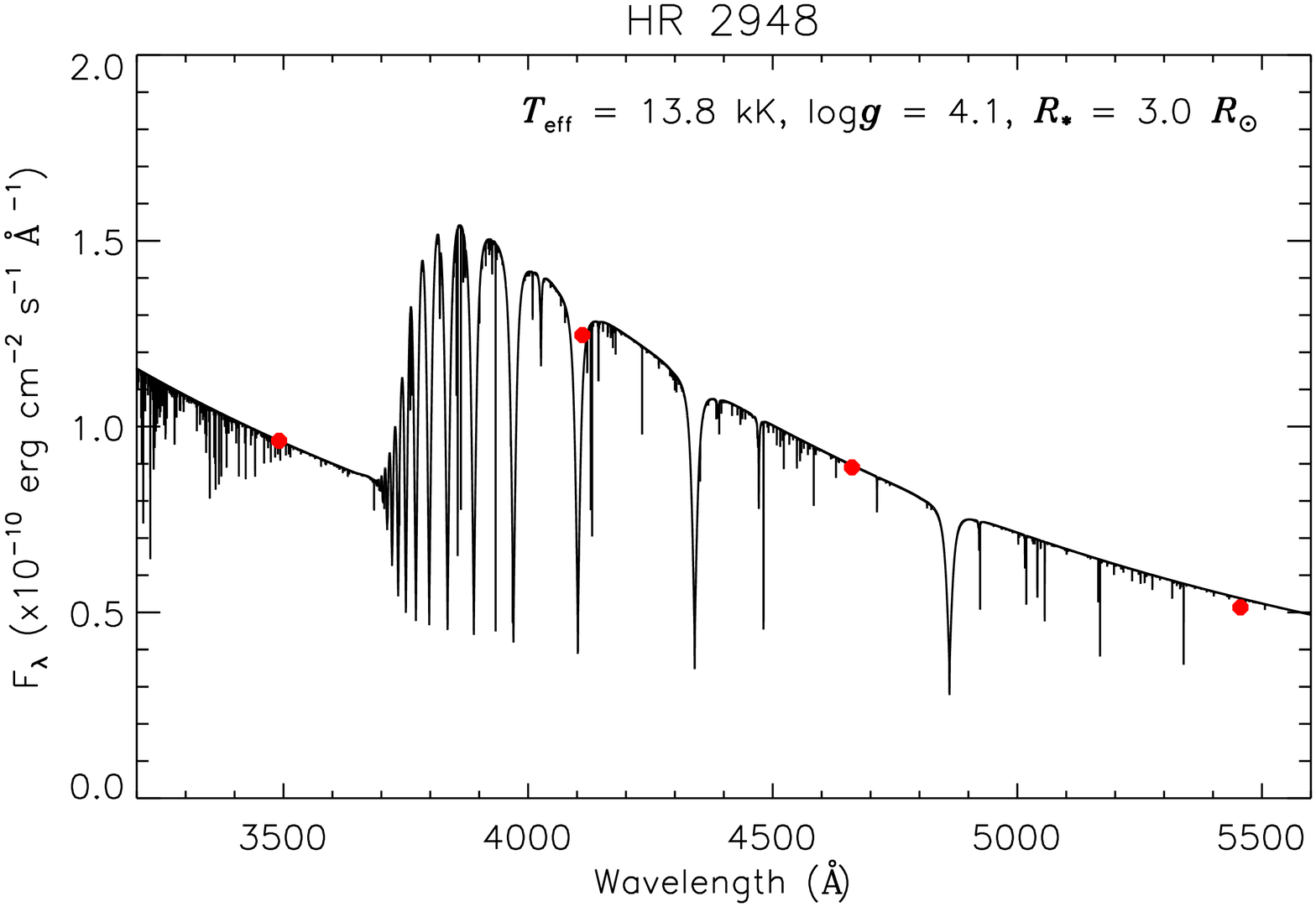}
\includegraphics[width=8cm]{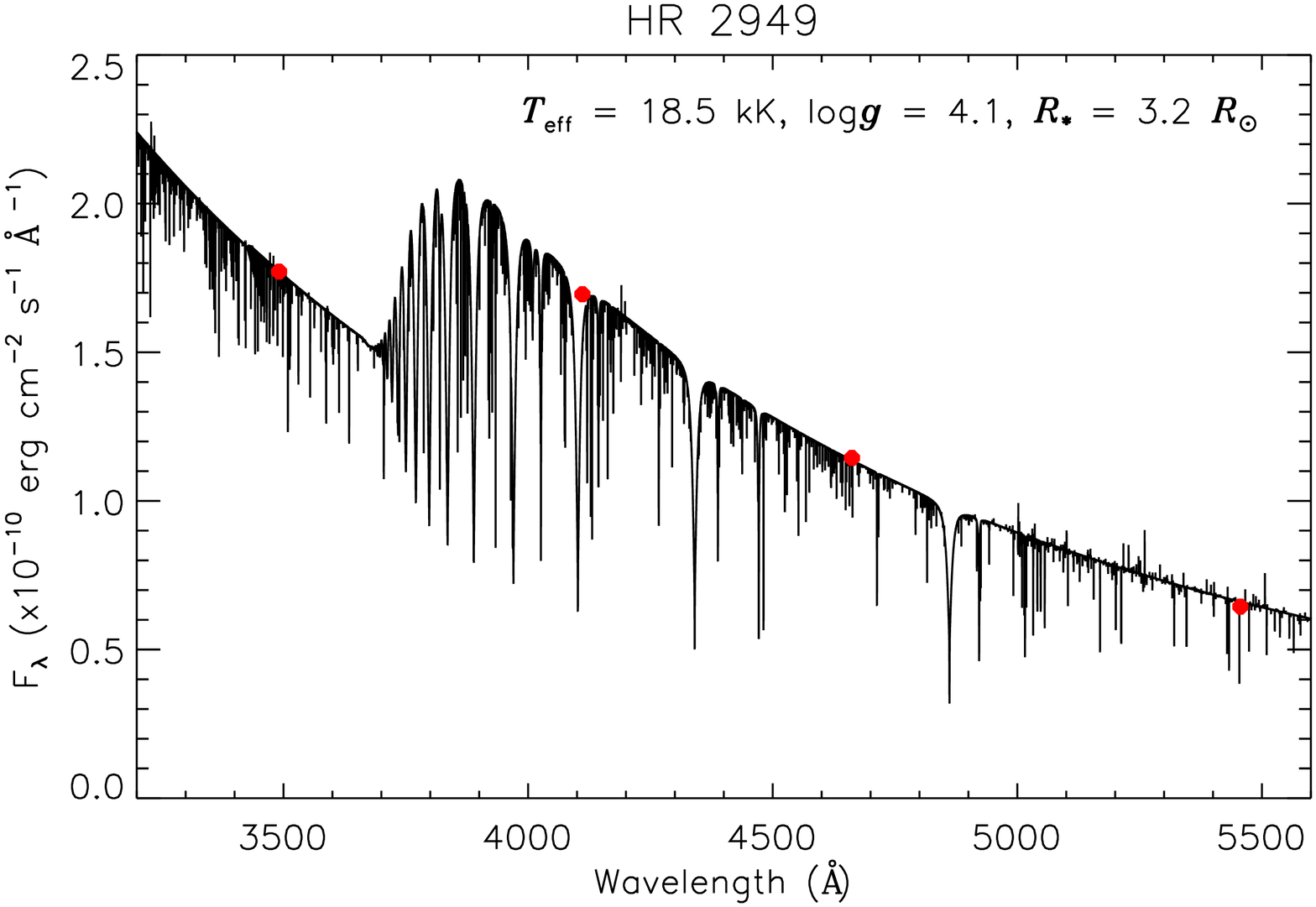} 
\caption{$uvby$ photometry for HR 2948 (top) and HR 2949 (bottom) compared to the best-fit model (black), corresponding to the parameters in Table \ref{params}. As discussed in Section 4.1, the original photometric data seems to have been mis-labeled.}
\label{uvby}
\end{figure}

One standard approach to determine \teff~is by comparing spectral energy distributions (SEDs) to photometric data. While most published flux measurements give only the combined magnitude and flux of the HR 2948+2949 system, there is one published {\em uvby} dataset containing individual measurements for each stellar component \citep{lindroos1983}. These data are reproduced in Fig. \ref{uvby}. Our analysis suggests these data were mislabelled in the original catalogue. The reasoning is explained below. 

The photometry was de-reddened using a standard \cite{cardelli1989} reddening curve, assuming a small reddening of $E(B-V) = 0.04$ given the stars' relative proximity to the Sun \citep{vanleeuwen2007}.

Chemically peculiar magnetic stars are UV-deficient compared to non-magnetic stars of similar spectral type. Since the measured values are to be compared to synthetic spectra calculated without taking the magnetic field and chemical peculiarities into account, correction factors must be applied \citep{kochukhov2005}. However, as the magnetic field strength is moderate (see Section 7), this correction is minimal for HR 2949 (smaller than the data points in Fig. \ref{uvby}). 

Using the designations given by \citeauthor{lindroos1983}, comparison of the photometry to model SEDs suggests a temperature of 13.5$\pm$0.5~kK for HR 2949, and 18.5$\pm$0.5~kK for HR 2948. The model spectral libraries provided by \cite{lanzhubeny2007} and \cite{coelho2014} give essentially identical results. 

\begin{figure}
\centering
\includegraphics[width=8cm]{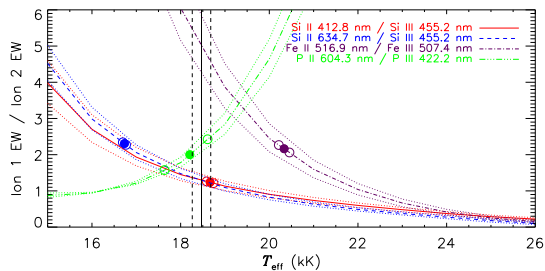}

\caption{Ionization balances as functions of \teff~for HR 2949. Curved lines indicate model EW ratios for \lgg~as given in Table \ref{params}; dotted lines indicate the EW ratios at the 1$\sigma$ uncertainties in \lgg. Filled circles indicate the weighted mean of the measured EW ratios, open circles the 1$\sigma$ uncertainties in the weighted means. The solid vertical line indicates the mean \teff, while the vertical dashed lines indicate the standard deviation from the different measurements.}
\label{ion_teff}
\end{figure}

An alternate means of determining \teff~is via ionization ratios. We performed this analysis using ESPaDOnS and FEROS observations for a selection of ions (Si~{\sc ii}/Si~{\sc iii}, Fe~{\sc ii}/Fe~{\sc iii}, and P~{\sc ii}/P~{\sc iii}), for each case taking the weighted mean of measurements from all observations. Equivalent Widths (EWs) of the same lines were then measured in the synthetic BSTAR2006 spectra, with uncertainties in the model EW ratios coming from the range in surface gravity (determined below). The observed EW ratios are compared to the model EW ratios in Fig. \ref{ion_teff}. While the results differ somewhat for different lines, they indicate \teff$=18.5\pm$1.5~kK for HR 2949, clearly inconsistent with the photometric results. 

In the course of the abundance analysis (see Section 6), a similar ionization balance analysis was performed for both stars using {\sc atlas9} and {\sc zeeman}. This analysis indicated \teff$=18.3 \pm 0.8$ kK for HR 2949, and \teff$=13.6 \pm 1.2$ kK for HR 2948. These results are, once again, inconsistent with the photometric results. 

There may be ionization imbalances in the Si {\sc II/III} lines.
\cite{2013AA...551A..30B} found abundances determined using Si {\sc II/III} lines to be
discordant by up to 1.5 dex in all normal B and Bp stars, and concluded that
vertical stratification is the likely reason. Whether this is the case for other
elements in late type Bp stars does not seem to have been investigated.
Furthermore, there are discrepancies in \teff~determined spectroscopically and
spectrophotometrically in some other cases. However, the effect is rather small:
(e.g., 1000 K in the case of HR 7355; \citealt{rivi2013}) and in some cases
entirely absent (e.g. HR 5907, for which spectral modeling and spectrophotometry
give identical results; \citealt{grun2012}). While we did not conduct a
systematic search for signs of stratification, in the course of our abundance
analysis, we did not detect any strong discrepancies between fits for different
ions of the same elements, suggesting that stratification, if present, is not
particularly pronounced.  Furthermore, our photometric and spectroscopic
analyses yield essentially identical results. Given these considerations, we
believe that while the assumption of ionization balance may well prove to be
incorrect, any errors thereby introduced into our determination of \teff~are
likely to be on the order of our uncertainties.

The discrepancy between the photometric and spectroscopic determinations of \teff~can be easily reconciled if the photometric data was mis-labeled. There are reasons to suspect this may be the case: first, the depths of the H Balmer lines are clearly inconsistent with the photometric \teff, but consistent with the ionization balance \teff; second, both stars are listed in the 2MASS catalogue \citep{cutri2003}, with HR 2949 the brighter of the two, whereas \cite{lindroos1983} lists it as the dimmer star; finally, \citeauthor{lindroos1983} does not provide coordinates for his targets, making it somewhat more likely for a typographical error to have entered in. 

We thus take HR 2949 to be the hotter, and brighter, of the two stars. In light of this, we reclassify HR 2949 as a B3p IV star, and HR 2948 as B6 IV. 

\subsection{Luminosity, radius, mass, and surface gravity}

To obtain the luminosity, we first calculate absolute magnitudes $M_V$ from the apparent $V$ magnitudes and the distance. From $M_V$ we calculate the bolometric magnitude $M_{\rm bol}$, using the bolometric correction $BC$ from the calibration of \cite{nieva2013}. Luminosities are then calculated from $M_{\rm bol}$. For HR 2949, we find $\log{L/L_\odot} = 3.08^{+0.16}_{-0.13}$; for HR 2948, $\log{L/L_\odot} = 2.69^{+0.22}_{-0.18}$.

\begin{figure}
\centering
\includegraphics[width=8cm]{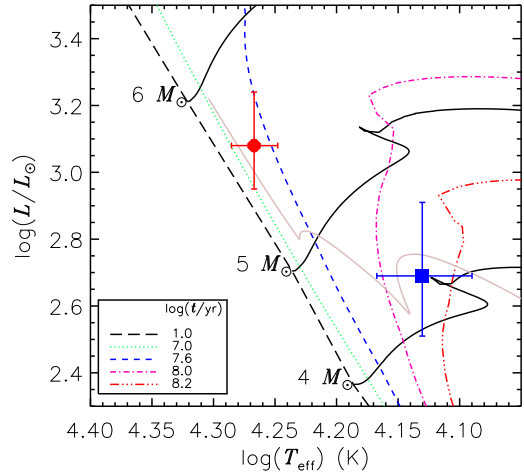}
\caption{Positions of HR 2949 (red circle) and HR 2948 (blue square) on the Hertzsprung-Russel diagram. Solid lines indicate the evolutionary tracks for 4, 5, and 6 \msun~models. Dashed and dotted lines indicate isochrones as per the legend. The grey line indicates a pre-main sequence isochrone.}
\label{hrd}
\end{figure}

HR 2949's position on the Hertzsprung-Russell Diagram (HRD) (Fig. \ref{hrd}) in comparison with the Geneva evolutionary tracks and isochrones \citep{ekstrom2012} implies a zero-age main-sequence mass of $6.0\pm 0.3$~\msun, a radius of $3.0^{+0.7}_{-0.3}$~\rsun, a surface gravity of \lgg~$=4.25^{+0.08}_{-0.17}$, and an age of $11^{+15}_{-6}$~Myr. HR 2948 should have $M_{\rm ZAMS} = 4.3\pm0.3$~\msun, $R_* = 3.7^{+1.2}_{-0.6}$~\rsun, \lgg$= 3.94^{+0.13}_{-0.22}$, and an age of 112$^{+20}_{-17}$~Myr. This large discrepancy in the apparent ages of the two stars is considered further in Section 8. 

Both stars are in the instability strip for slowly pulsating B stars. Only HR 2949 can be evaluated for line profile variability. We detect no sign of pulsational variability in this star: as detailed in section 5, all line profile variability can be explained as a consequence of chemical abundance spots.

\begin{figure}
\centering
\includegraphics[width=8cm]{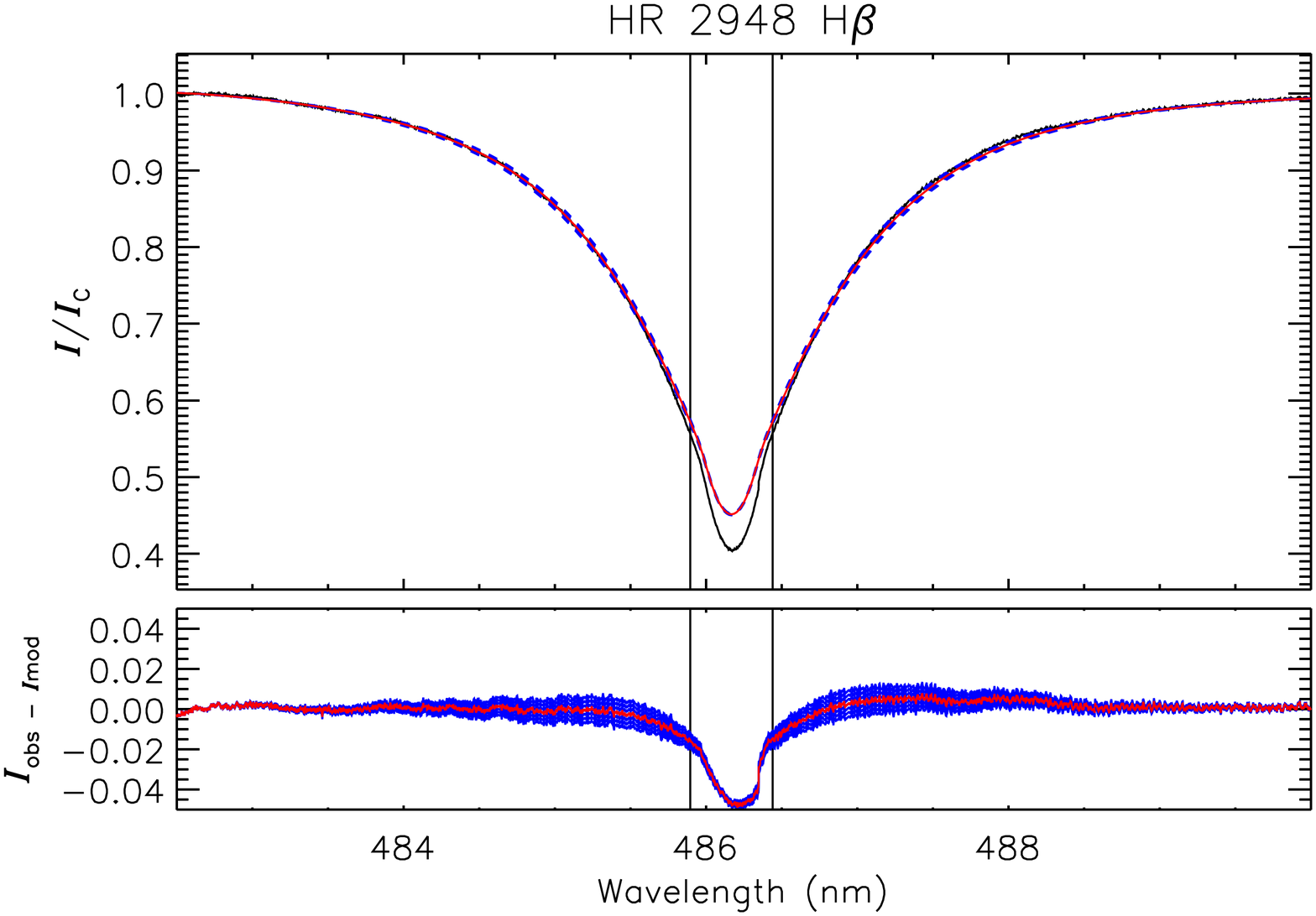}
\includegraphics[width=8cm]{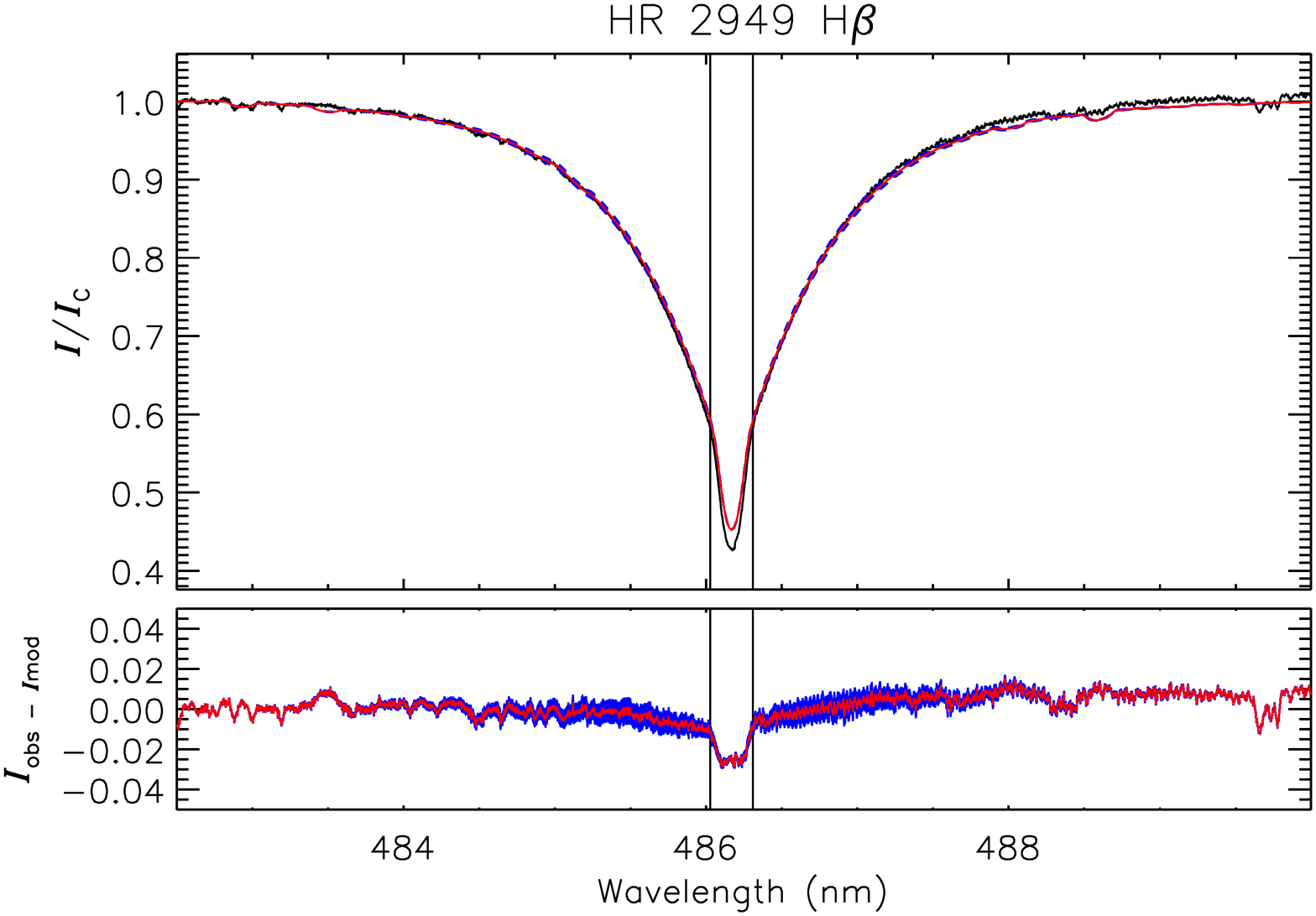}
\caption{H$\beta$ lines of HR 2948 (top) and HR 2949 (bottom) as compared to model fits. Residuals are shown below the 1-dimensional spectra. Vertical black lines demarcate the region excluded from the fit. The best-fit models are in red; blue indicate the $\pm$0.15 dex uncertainty.}
\label{hbeta_logg}
\end{figure}

HR 2949's surface gravity is consistent with model fits to the H$\beta$ line (Fig. \ref{hbeta_logg}, bottom), which yield the best fit for \lgg$=4.10\pm0.05$. Care is required in fitting models to H$\beta$ in ESPaDOnS spectra, as the red wing of H$\beta$ crosses two spectral orders. In determining \lgg, we first merged the un-normalized H$\beta$ lines across the two spectra, then normalized the line via a linear fit to nearby continuum regions: this avoided warping of the line profile during the order-by-order polynomial normalization procedure typically used for ESPaDOnS data. As the wings of H$\beta$ are weakly variable (see Section 5), we selected a phase found in the middle of the range of variation. In the course of this analysis, it was clear that the choice of continuum regions for renormalization of the line had an effect on \lgg~larger than the formal error bars. To account for this source of systematic error, we increase the uncertainty by 0.1, such that our final surface gravity for HR 2949 is \lgg$=4.10\pm0.15$. 

The same analysis for HR 2948 yields a higher surface gravity, \lgg$=4.24 \pm 0.15$, than is obtained from the HRD. However, the results are again compatible within error bars. 

There is some small blue-red asymmetry in the fits to the H$\beta$ shown in Fig. \ref{hbeta_logg}. In the case of HR 2949 this may be a consequence of the weak variability in the line wings, however, as the asymmetry affects both stars it is more likely that this is an artifact introduced by normalization.

\begin{figure}
\centering
\includegraphics[width=8cm]{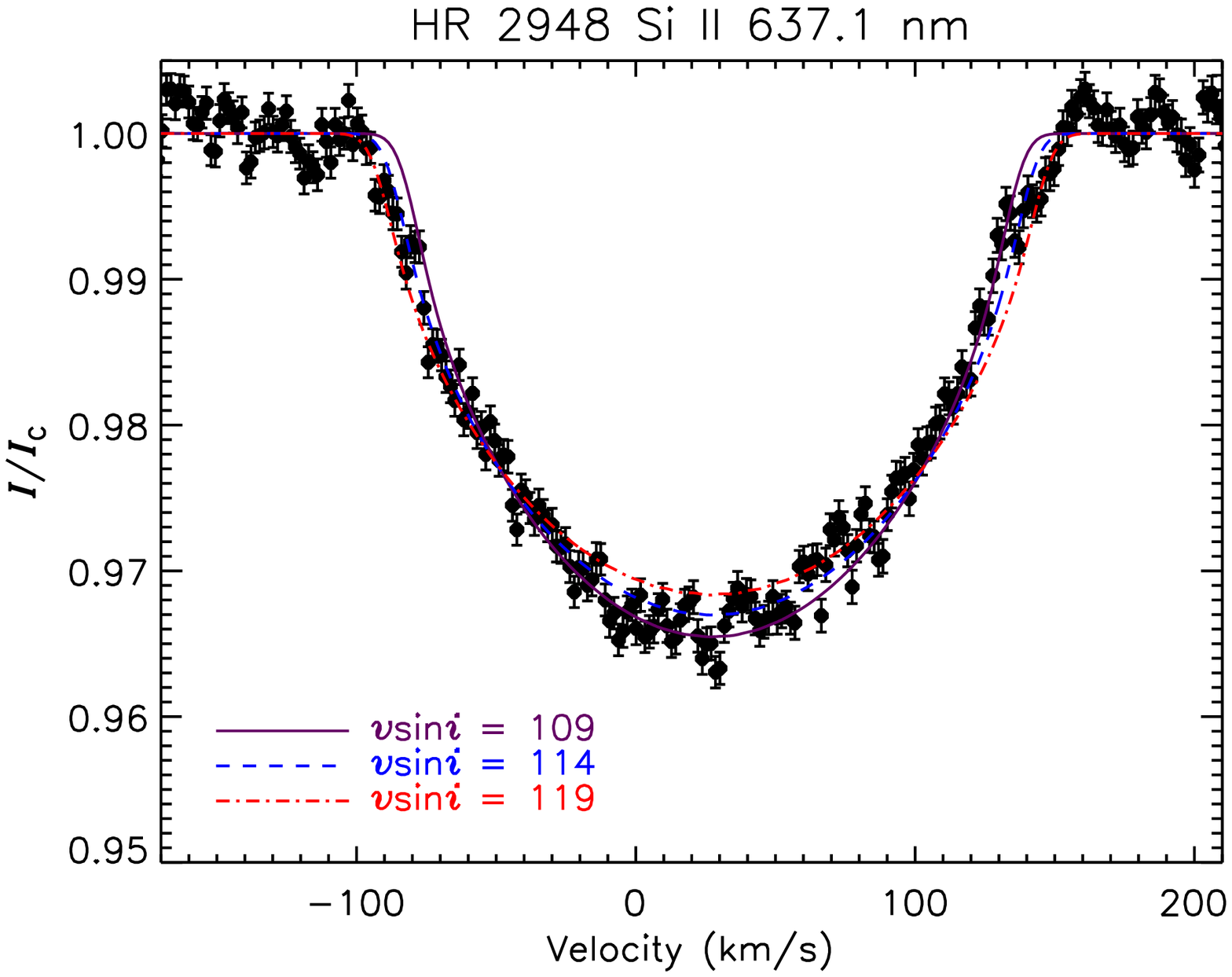}
\includegraphics[width=8cm]{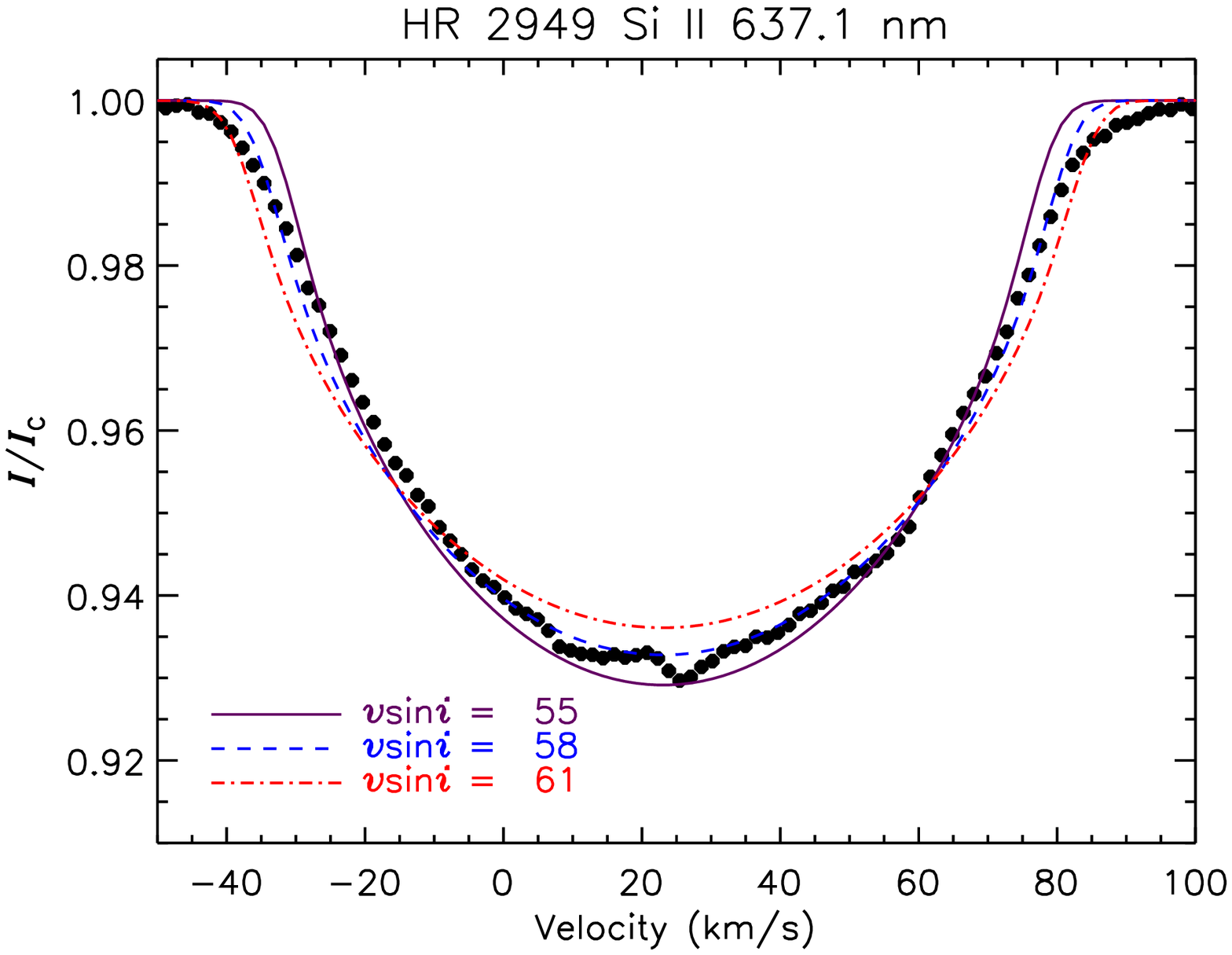}
\caption{Line profile models for the two stars.}
\label{vsini}
\end{figure}

\subsection{Rotation}

The high-resolution ESPaDOnS and FEROS data enable very good constraints on \vsini~via line profile modeling. This was performed for several species, including Mg~{\sc ii}, C~{\sc ii}, Si~{\sc ii}, and Si~{\sc iii}. Mean line profiles created from all observations were used in order to average out the distortions caused by abundance spots. For HR 2949, we find \vsini~$= 58 \pm 4$~\kms; for HR 2948, \vsini~$= 112 \pm 4$~\kms. An independant determination of \vsini~was performed in the course of the abundance analysis, obtaining slightly higher velocities, although identical within error bars: we thus expand our uncertainties to encompass systematic error, settling on \vsini~$= 61 \pm 5$~\kms~for HR 2949 and \vsini~$=114 \pm 5$~\kms~for HR 2948. Fig. \ref{vsini} shows model line profiles spanning these parameters, compared to the mean Si~{\sc ii}~637.1~nm line. 

In the case of HR 2949, there is some slight disagreement in the wings between the observations and the models. This is likely a consequence of saturated local line profiles associated with chemical spots in which abundances are in excess of the photospheric mean. As discussed below, this phenomenon affects essentially all metallic lines in the star's spectrum. 

In the course of obtaining \vsini, the systemic velocity was determined to be $v_{\rm sys} = 22 \pm 2$~\kms, and is identical for both stars. 

In the case of HR 2949, as both $P_{\rm rot}$ and \vsini~are known, the stellar radius $R_*$ is constrained by $i$. With $P_{\rm rot}~=~1.90871$~d, this yields $R_*\sin{i} = 2.18~R_\odot$. The radius found above via the HRD would then indicate an inclination $i~=~42^{+23\circ}_{-12}$, and consequently an equatorial rotational velocity $v_{\rm eq}~=~91^{+39}_{-30}$ \kms. 

\section{Spectral Variability}

\begin{figure}
\centering
\includegraphics[width=8cm]{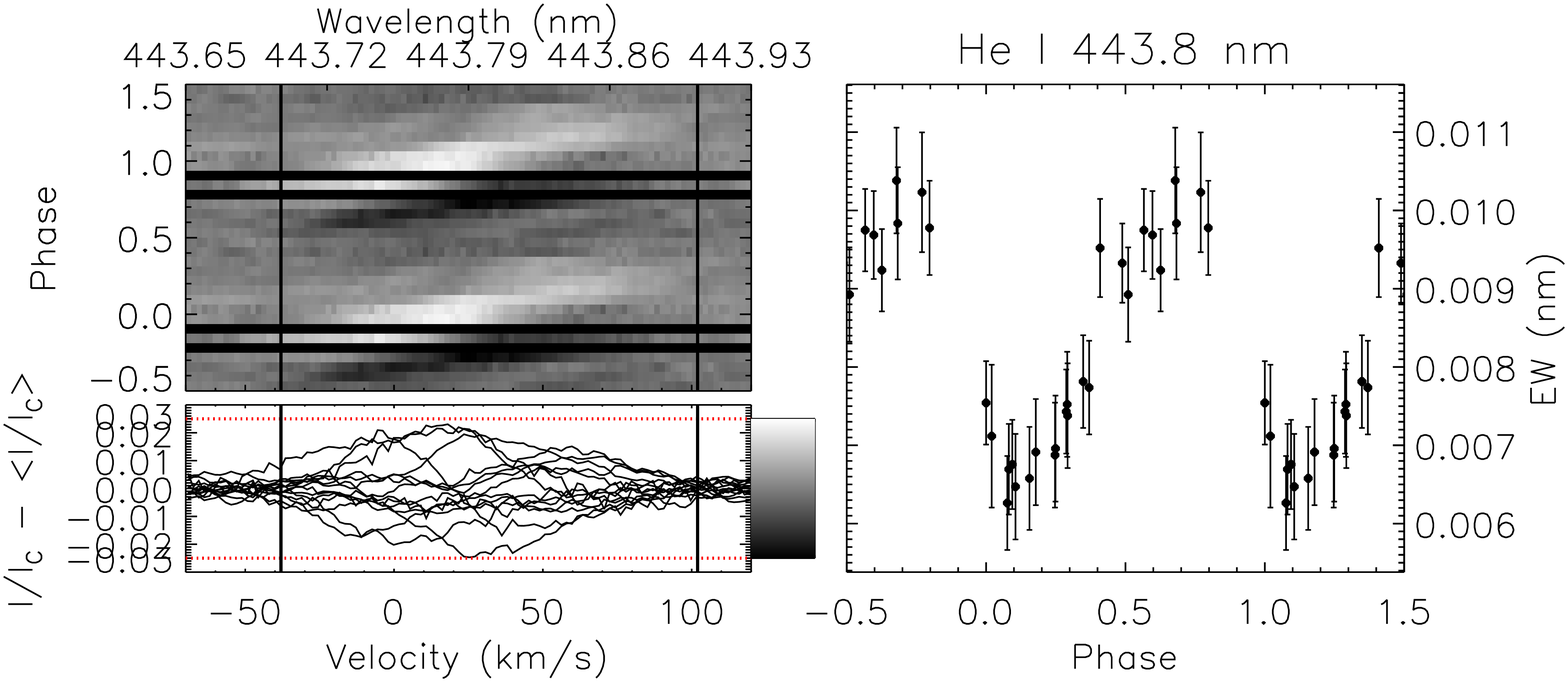} 
\includegraphics[width=8cm]{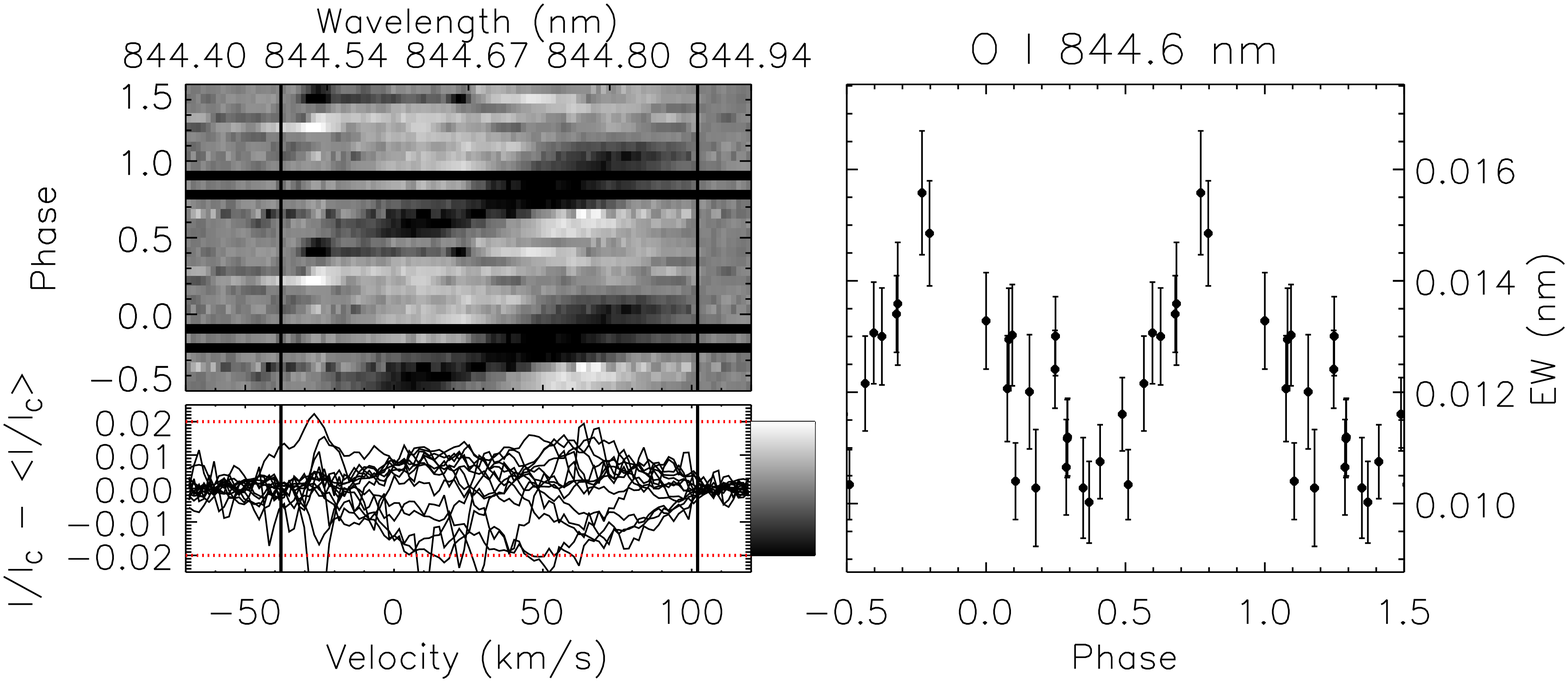} 
\includegraphics[width=8cm]{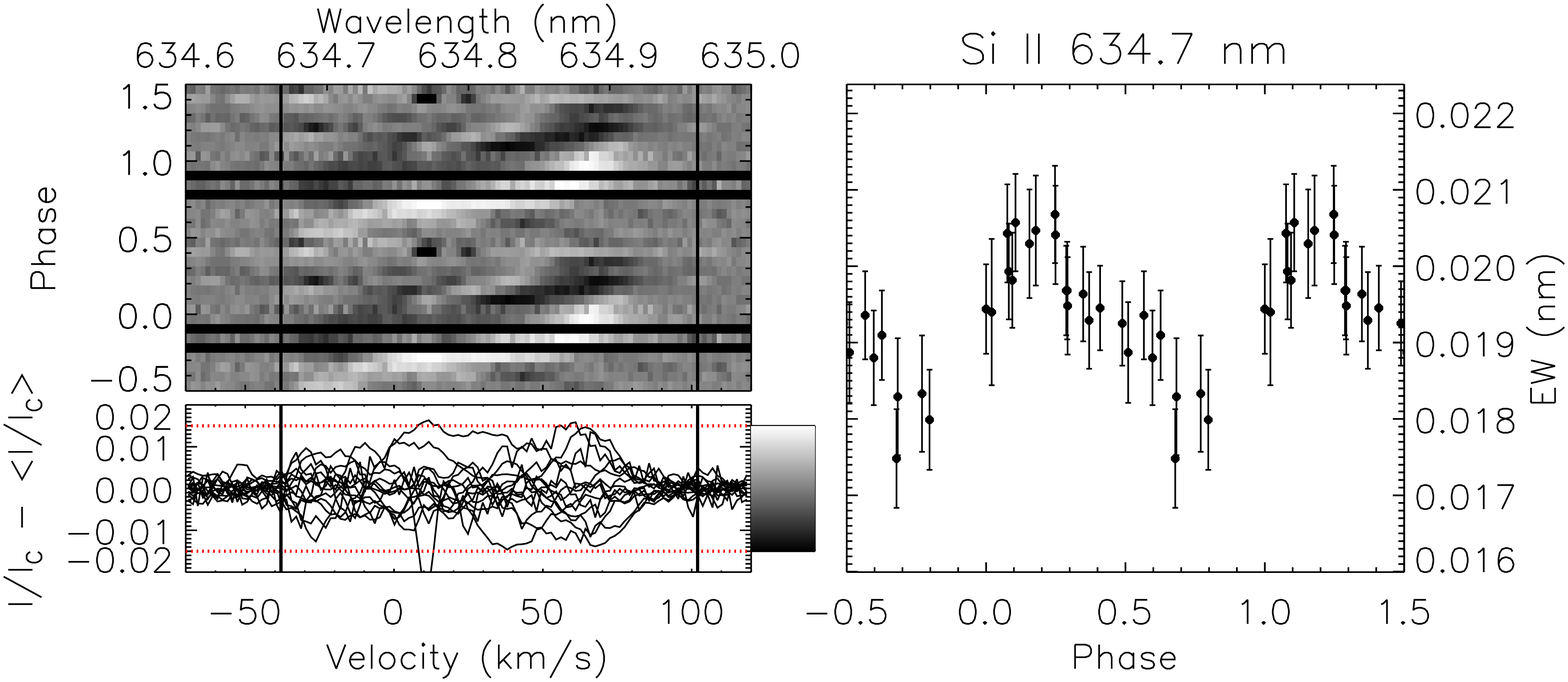} 
\includegraphics[width=8cm]{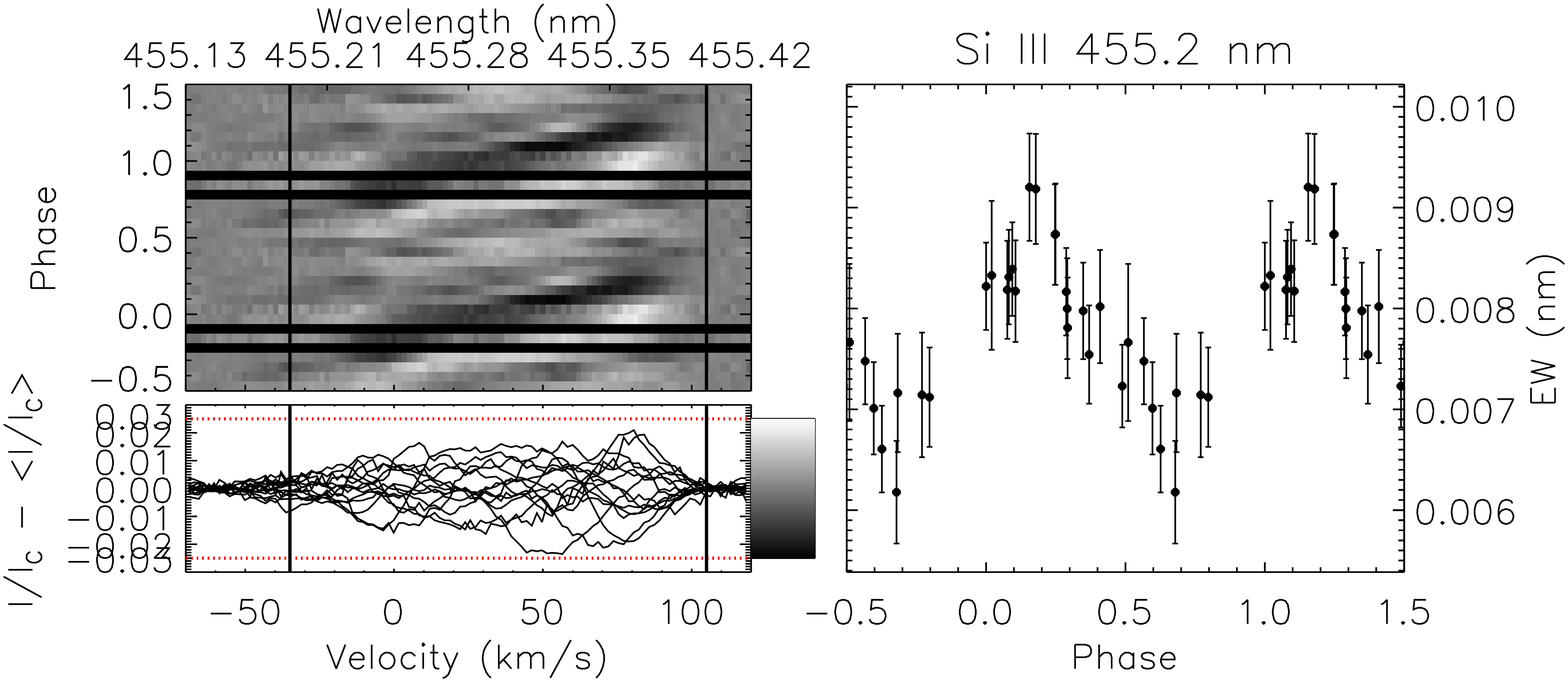} 
\includegraphics[width=8cm]{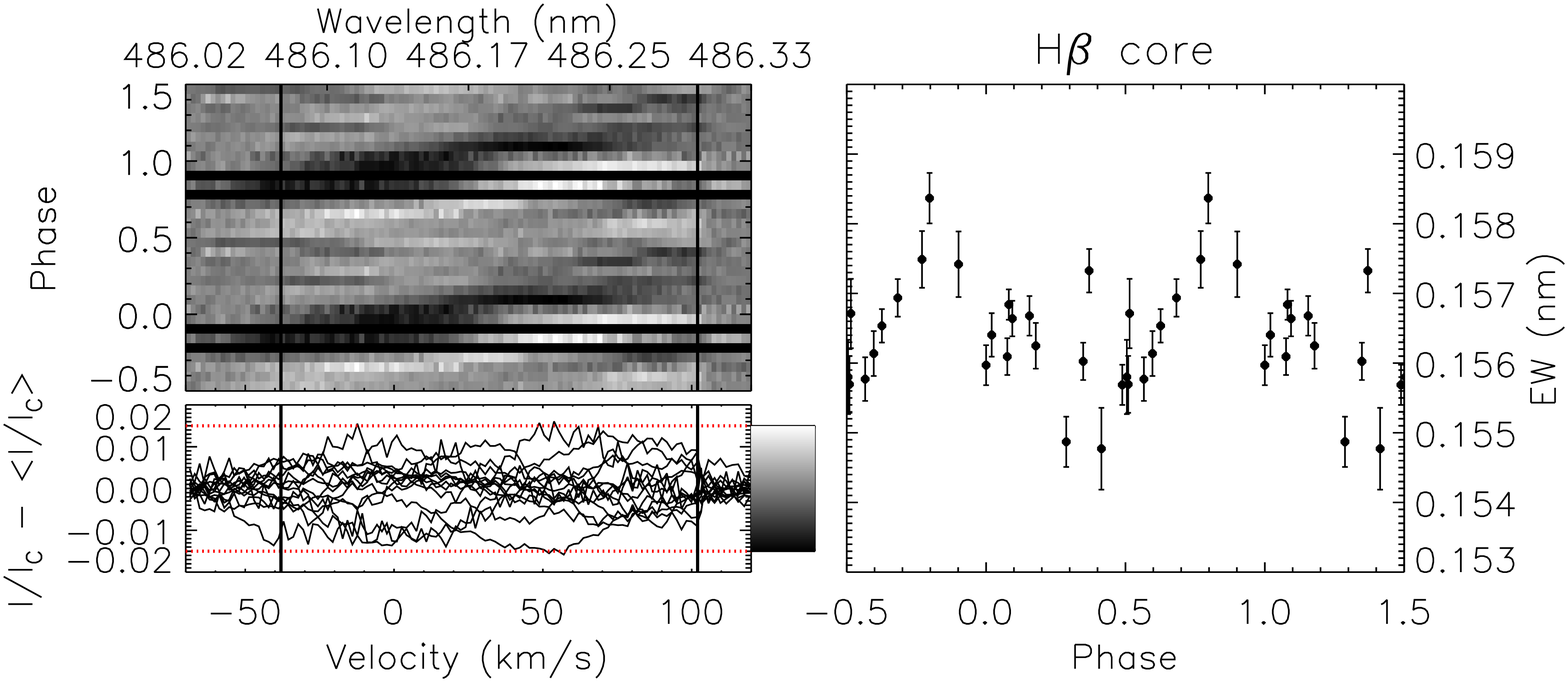} 
\includegraphics[width=8cm]{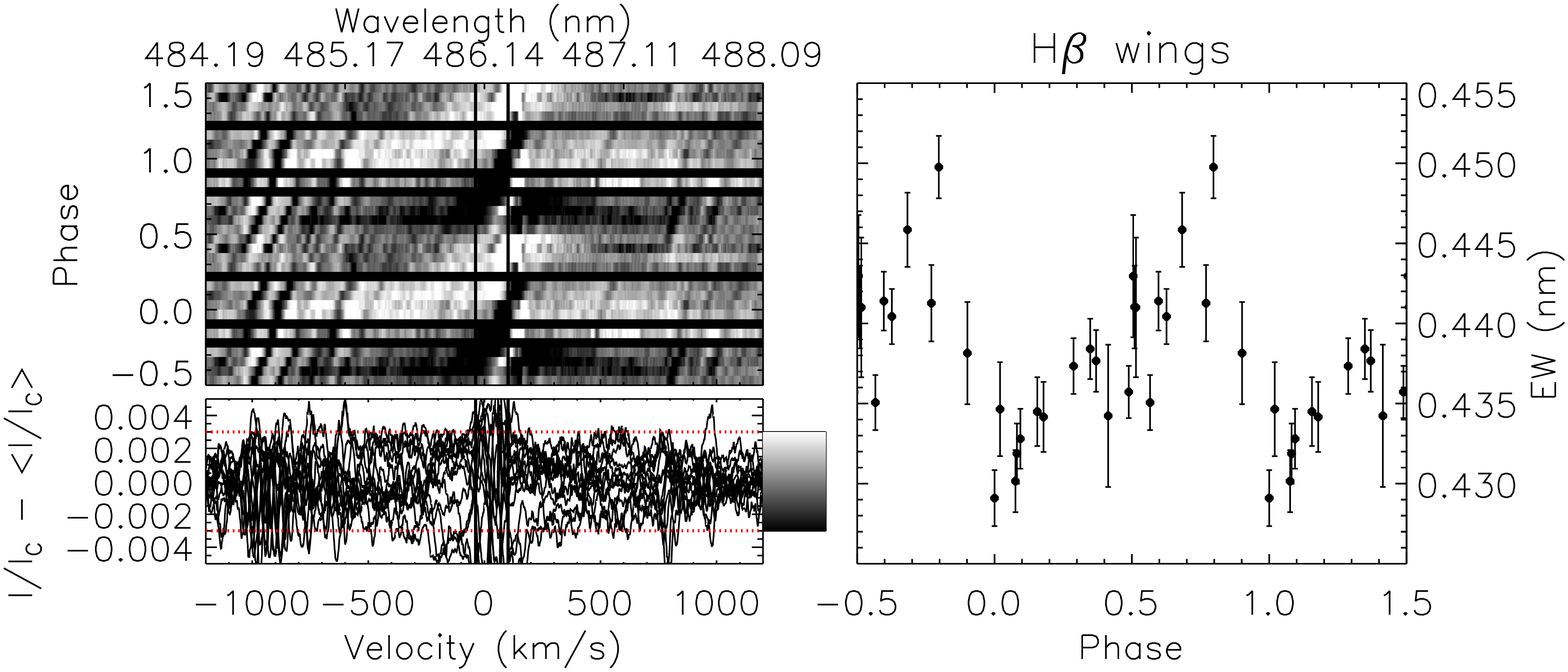} 
\caption{Dynamic spectra (left panels) and EWs (right panels) of He {\sc i}, O {\sc i}, Si {\sc ii} and {\sc iii}, and H$\beta$. The dynamic spectra are relative to mean line profiles. The colour-mapping scheme is shown relative to the 1D residual flux, with the limits of the colour map shown with dotted red horizontal lines. The dynamic spectra are plotted in order of rotational phase, with the residual flux binned into 16 phase bins. Black vertical lines indicate $\pm$\vsini.}
\label{dyn_he}
\end{figure}

HR 2949's spectrum displays a rich variety of line profile variability (LPV), with distinct patterns in different groups of chemical elements. Fig. \ref{dyn_he} shows dynamic spectra and equivalent width (EW) measurements for He, O, Si, and H. Dynamic spectra and EWs for C, N, Ne, Mg, P, S, Cl, Ca, Ti, Cr, and Fe are shown in the Appendix in Figs. \ref{dyn_2} and \ref{dyn_3}. 

The dynamic spectra are calculated as the residual flux of each line as compared to a mean line profile. Excesses in absorption at a given Doppler velocity appear as flux deficits, while deficits of absorption appear as pseudo-emission. This residual 1D flux is plotted in the bottom panels of Fig. \ref{dyn_he}. The residual flux is then mapped to a grayscale colour bar, shown next to the 1D flux, with dotted red horizontal lines indicating the mapping from the 1D flux to the colour bar, such that stronger absorption is darker while pseudo-emission is lighter. The color-mapped flux is then plotted as a function of rotational phase, with the spectra binned into 16 phase-bins, a number chosen so as to reach a compromise between higher SNR and phase resolution. Black vertical lines represent $\pm$\vsini.

The strongest, simplest variations are seen in He (first row of Fig. \ref{dyn_he}). An absorption excess crosses the stellar disk between phases 0.5 and 1.0. Comparison to magnetic data (see Section 7) indicates this is associated with the negative magnetic pole. O appears to be distributed in a similar fashion (Fig. \ref{dyn_he}, second row), as do C, Ne, and S  (see Fig. \ref{dyn_2} in the Appendix). A second, weaker absorption excess may be present around phases 0.2 to 0.5.

\begin{figure}
\centering
\includegraphics[width=8.5cm]{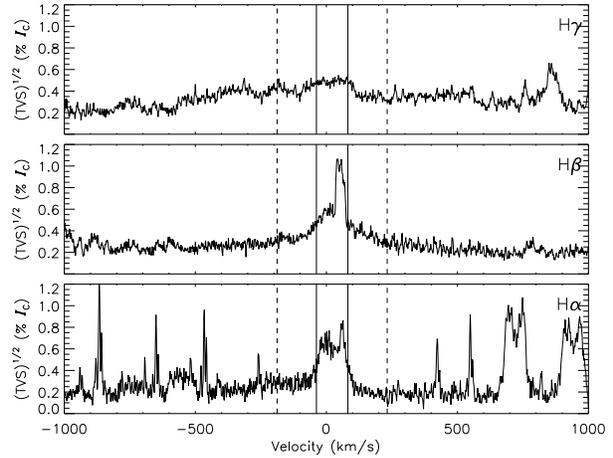}
\caption{Temporal variance spectra for (bottom--top) H$\alpha$, H$\beta$, and H$\gamma$. Solid black vertical lines indicate $\pm$\vsini. Dashed black vertical lines indicate the Kepler corotation radius $\pm$\rk~(as defined in Section 8.3.1).}
\label{tvs}
\end{figure}

Si {\sc ii} and Si {\sc iii} LPV anticorrelates with He (Fig. \ref{dyn_he}, third and fourth rows). Comparison of the EWs of He and Si lines makes the anticorrelation quite obvious: peaks and troughs of the EW variations are separated by $\sim$0.5 of a rotational cycle. Similarly, variation in the core of the H$\beta$ line also seems to approximately anticorrelate with He (Fig. \ref{dyn_he}, second panel from bottom).  

The most complex variations are seen in N, Mg, Ti, Ca, Cr, and Fe. These are shown in the Appendix in Fig. \ref{dyn_3}. These elements neither precisely correlate nor anti-correlate with He. Furthermore, their EWs do not exhibit smooth single-wave variations as in He or Si lines. 

The wings of H$\beta$ (Fig. \ref{dyn_he}, bottom panel) vary in a different way, with red and blue wings moving towards and away from the continuum approximately simultaneously. This variation is extremely weak. To evaluate the statistical significance of the Balmer line variability, we created Temporal Variance Spectra (TVS; \citealt{fullerton1996}) for H$\alpha$, H$\beta$, and H$\gamma$ (see Fig. \ref{tvs}). A TVS evaluates the deviations in individual pixels as compared to scatter in the continuum. In order to minimize spurious variability caused by polynomial normalization, in the case of H$\beta$ the same procedure was followed as described in Section 4 for measurement of \lgg. 

In all cases, the continuum varies at approximately the 0.2\% level. The strongest variability is seen in the line core. This is similar in character to TVS for metallic lines, as e.g. the two C {\sc ii} lines in the red wing of \halp. However, H$\beta$ and H$\gamma$ both clearly show weak variability in the wings. Evaluating the variability in H$\alpha$ is more difficult due to the large number of telluric lines in this region of the spectrum. In order to minimize their influence, only relatively un-contaminated spectra were used in calculating the \halp~TVS, with 9/22 spectra making the cut. There is no variability in the wings of this line exceeding that seen in H$\beta$ or H$\gamma$: indeed, there is no evidence of any variation at all in the wings. In order to check that discarding spectra was not hiding variability, the same observations were used to create test TVS for H$\beta$ and H$\gamma$: in both cases, the TVS showed variability at a similar level as seen in TVS created using the full dataset.

\begin{figure}
\centering
\includegraphics[width=8.5cm]{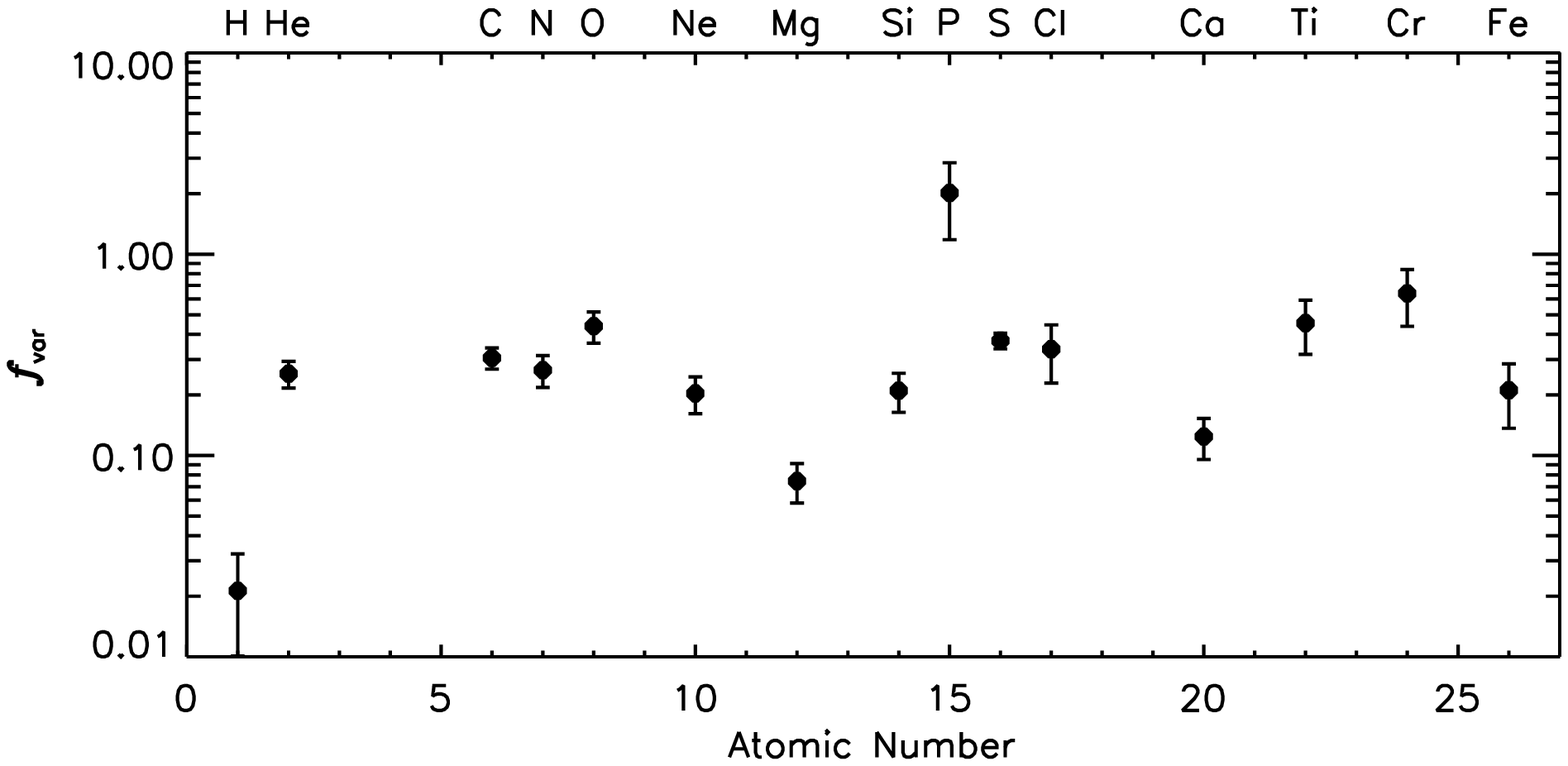} 
\includegraphics[width=8.5cm]{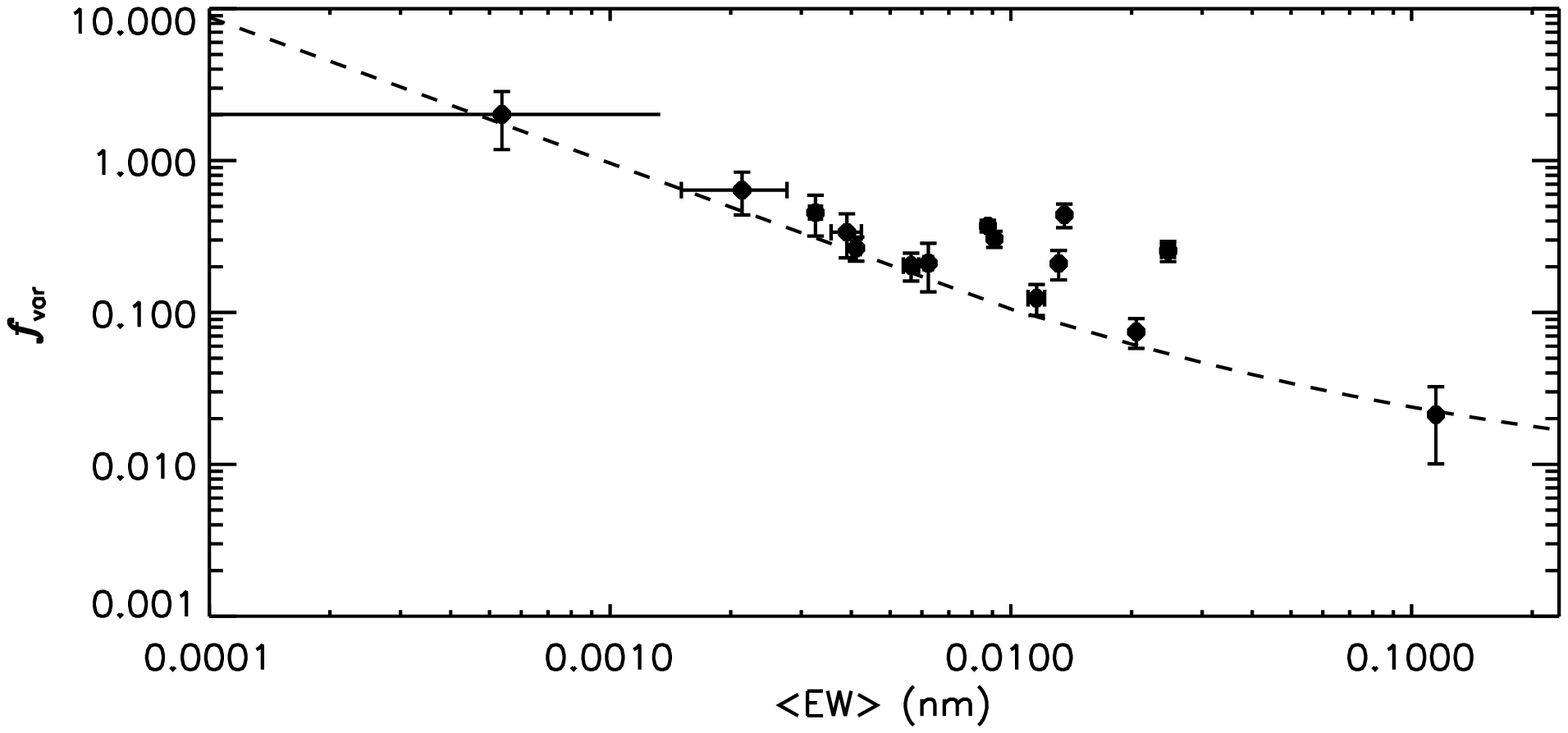} 
\includegraphics[width=8.5cm]{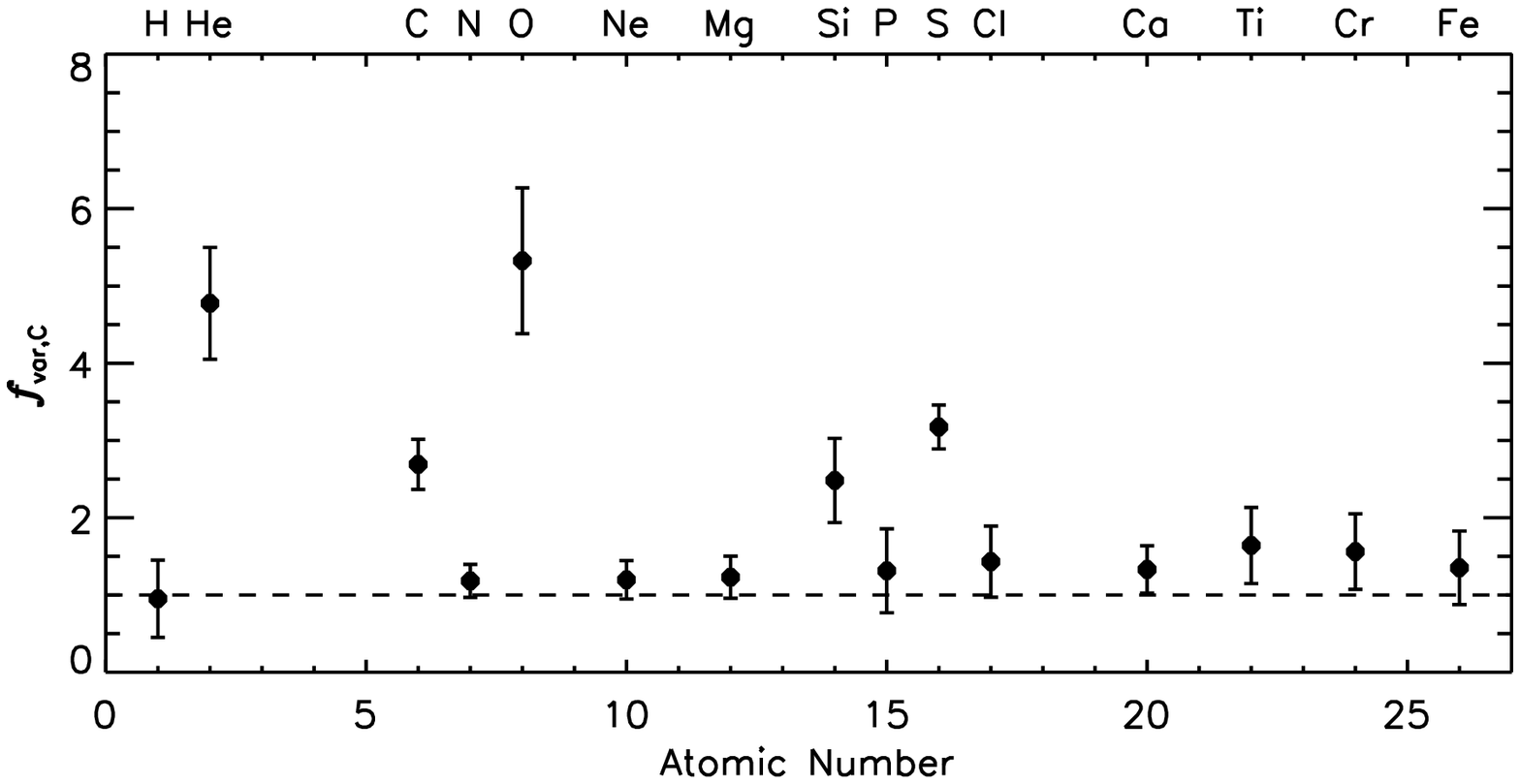}
\caption{The variability index $f_{\rm var}$, as a function of atomic number (top panel) and as compared to the mean EW of the line used to measure $f_{\rm var}$ (middle panel). The dashed line is the best-fit power law. The bottom panel shows $f_{\rm var, C}$, the variability index corrected by the hyperbolic fit.} 
\label{varind}
\end{figure}

In order to compare the variability in different elemental species, Fig. \ref{varind} shows the variability index 

\begin{equation}\label{fvar}
f_{\rm var} \equiv \frac{{\rm EW}^{\rm max} - {\rm EW}^{\rm min}}{\langle{\rm EW}\rangle}
\end{equation}

\noindent for all elemental species for which relatively strong, unblended lines are available. Measurement of the quantity $f_{\rm var}$ was based 3$^{rd}$-order sinusoids fit to the data (chosen as the minimum order necessary to achieve a good fit to the more complex EW curves), with uncertainties in the EW and the mean EW coming from the uncertainty in the fitting parameters. By thus making use of the full dataset, rather than simply the two most extreme measurements, uncertainties are minimized.


Weaker lines are expected to be more strongly affected by smaller relative changes than will strong lines, and indeed, when $f_{\rm var}$ is compared to the mean EW (middle panel of Fig. \ref{varind}), there is an obvious trend of decreasing $f_{\rm var}$ with increasing $\langle EW\rangle$. In order to correct for this trend, we fit a purely empirical two-part power law $C$ to the lower boundary of $f_{\rm var}$, with $C = 0.0008(EW)^{-1} + 0.01(EW)^{-0.2}$. This is indicated by the dashed line in Fig. \ref{varind}. It is plain that the differences in variability amongst the lines cannot be due to line strength alone. 

The bottom panel of Fig. \ref{varind} shows $f_{\rm var, C} = f_{\rm var} / C$. Elements with $f_{\rm var, C} = 1$  are of course still variable. For instance, P disappears almost entirely around phase 0.5 (see the dynamic spectra and EWs in Fig. \ref{dyn_3}), and its extremely high $f_{\rm var}$ reflects this; however, when corrected for line strength, it is no more variable than most other lines. $f_{\rm var, C}$ thus serves to highlight elements showing particularly large line strength variations. In particular, He, C, O, Si, and S all possess $f_{\rm var, C} > 1$.

\section{Abundances}

\begin{figure}
\centering
\includegraphics[width=6.cm, angle=270]{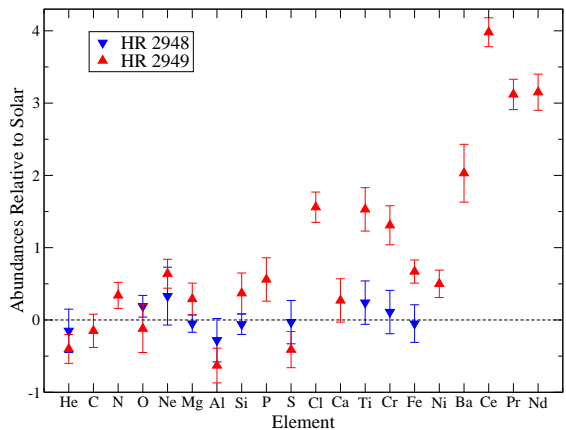} 
\caption{Abundances with respect to solar values.}
\label{abundances}
\end{figure}

\begin{table}
\centering
\caption[Abundances]{Absolute chemical abundances. Solar abundances [X/H]$_\odot$ are from \cite{asplund2009}. The number of spectral windows used in the analysis is given by $N$. Abundances relative to solar are shown in Fig. \ref{abundances}. For HR 2949, the variability quantity $f_{\rm var}$ is also provided (see Fig. \ref{varind}).}
\resizebox{8.5 cm}{!}{
\begin{tabular}{lrrrrrr}
\hline
\hline
& & \multicolumn{3}{c}{HR 2949} & \multicolumn{2}{c}{HR 2948}\\
El. & [X/H]$_\odot$ & [X/H] & $f_{\rm var}$ & $N$ & [X/H] &  $N$ \\
\hline
He & -1.07$\pm$0.01 & -1.47$\pm$0.20 & 0.25$\pm$0.04 & 3 & -1.22$\pm$0.30 & 1 \\
C & -3.57$\pm$0.05 & -3.72$\pm$0.23 & 0.31$\pm$0.04 & 8 & -- & -- \\
N & -4.17$\pm$0.05 &  -3.41$\pm$0.18 & 0.27$\pm$0.05 & 8 & -- & -- \\
O & -3.31$\pm$0.05 & -3.43$\pm$0.33 & 0.44$\pm$0.08 & 8 & -3.12$\pm$0.15 & 1\\
Ne & -4.07$\pm$0.05 &  -3.43$\pm$0.20 & 0.20$\pm$0.04 & 2 & -3.75$\pm$0.40 & 2 \\
Mg & -4.40$\pm$0.04 &  -4.11$\pm$0.22 & 0.07$\pm$0.02 & 8 & -4.45$\pm$0.12 & 5 \\
Al & -5.55$\pm$0.03 & -6.18$\pm$0.24 & -- & 8 & -5.83$\pm$0.30 & 1\\
Si & -4.49$\pm$0.04 &  -4.12$\pm$0.28 & 0.21$\pm$0.05 & 8 & -4.55$\pm$0.14 & 5 \\
P & -6.59$\pm$0.03 &  -6.03$\pm$0.30 & 2.01$\pm$0.83 & 1 & -- & -- \\
S & -4.88$\pm$0.03 & -5.29$\pm$0.25 & 0.37$\pm$0.03 & 8 & -4.91$\pm$0.30 & 3\\
Cl & -6.50$\pm$0.30 &  -4.94$\pm$0.21 & 0.34$\pm$0.11 & 3 & -- & -- \\
Ca & -5.66$\pm$0.04 &  -5.89$\pm$0.30 & 0.12$\pm$0.03 & 1 & -- & -- \\
Ti & -7.05$\pm$0.05 &  -5.52$\pm$0.30 & 0.45$\pm$0.14 & 2 & -6.81$\pm$0.30 & 1 \\
Cr & -6.36$\pm$0.04 &  -5.05$\pm$0.27 & 0.64$\pm$0.20 & 8  & -6.25$\pm$0.30 & 2 \\
Fe & -4.50$\pm$0.04 &  -3.83$\pm$0.16 & 0.21$\pm$0.07 & 8 & -4.55$\pm$0.26 & 5 \\
Ni & -5.78$\pm$0.04 &  -5.28$\pm$0.19 & -- & 8 & -- & -- \\
Ba & -9.82$\pm$0.09 &  -7.79$\pm$0.40 & -- & 1 & -- & -- \\
Ce & -10.42$\pm$0.04 &  -6.64$\pm$0.20 & -- & 1 & -- & -- \\
Pr & -11.28$\pm$0.04 &  -8.16$\pm$0.21 & -- & 8 & -- & -- \\
Nd & -10.58$\pm$0.04 &  -7.43$\pm$0.25 & -- & 8 & -- & -- \\
\hline
\hline
\end{tabular}
}
\label{ab_fvar_tab}
\end{table}

We performed an abundance analysis of both HR 2949 and HR 2948 using the spectrum synthesis program {\sc zeeman} \citep{landstreet1988, wade2001}.  {\sc zeeman} performs polarized radiative transfer under the assumption of LTE.  Atomic data was taken from the Vienna Atomic Line Database VALD2 \citep{piskunov1995, ryabchikova1997, kupka1999, kupka2000}, using an `extract stellar' request.  Model atmospheres computed with ATLAS9 \citep{kurucz1993} were used, assuming solar abundances.  Surface chemical abundances, as well as \vsini~and \teff, were determined by directly fitting synthetic spectra to observed spectra, using the $\chi^2$ minimization algorithm described by \cite{folsom2012}.  

The observed spectra for HR 2949 and HR 2948 used here were created by averaging all available ESPaDOnS data for each star.  The averaged spectra were normalized by fitting a low order polynomial through continuum points of each spectral order, then dividing the order by that polynomial.  This averaging process has the advantage of minimizing the impact of surface chemical spots, providing results closer to the mean global surface chemical abundances.  

For HR 2949 the fitting process was performed on 9 independent spectral windows, covering most of the optical range (391.5--394, 440--460, 460--480, 490--510, 510--530, 530--550, 550--570, 570--600, and 600--650 nm), excluding H lines, the strongest He lines, and regions contaminated by telluric features. The results were then averaged to produce the final best-fit values, with the standard deviation taken as the uncertainty. For elements with constraints from less than 4 spectral windows, the uncertainties were increased to include the full window to window scatter, potential normalization errors, and the impact of any potential blending lines. We used a fixed $\log{g} = 4.1$, based on the Balmer line analysis. 

While a major strength of {\sc zeeman} is that it can include the effect of magnetic fields and perform polarized radiative transfer (integrated across a stellar disk), doing so increases the run time by 3 orders of magnitude (see \citealt{folsom2012}). Due to computational limitations we can either perform an extremely accurate analysis of a few elements and spectral lines, or a more approximate analysis of many more elements and spectral lines.  Since we are interested in diagnosing the full range of peculiarities in HR 2949, as well as determining \teff~based on lines with a wide range of ionization states and excitation potentials, breadth is more important for this analysis than extreme precision. Therefore the analysis of HR 2949 was performed using a non-magnetic model, with broadening due to Zeeman splitting approximated by 2 \kms~of microturbulence. In order to check the validity of this approximation, we conducted a second analysis in the 490-510 nm spectral window for He, N, Si, S, Cl, Cr, Fe, Ni, and Nd, assuming \bd~$= 2.7$ kG in the geometry derived in Sect 7.2 for phase 0 (and using the same fitting procedure as above). The abundances determined using magnetic vs. non-magnetic models typically agree to better than 0.02 dex, and always better than 0.05 dex. Thus, the impact of the magnetic field is much smaller than our other uncertainties.

We find HR 2949 has a best fit \teff~of $18.3 \pm 0.8$ kK, and a \vsini~of $63 \pm 3$ \kms. This \teff~is in agreement with our photometric and ion strength ratio measurements in Section 4.  We find strong photospheric chemical peculiarities in HR 2949, reported in Table \ref{ab_fvar_tab}, and plotted relative to the solar abundances of \cite{asplund2009} in Fig. \ref{abundances}.  We find strong overabundances of Fe-peak elements, by 1--2 dex above solar, and very strong overabundances of the rare earths Ce, Pr and Nd, by 3--4 dex above solar. Conversely, we find an underabundance of He, as well as Al and S.  We also find strong overabundances of Ba, Cl, and possible overabundances of Ne and P.  This pattern of chemical abundances is characteristic of magnetic Ap/Bp stars (e.g. \citealt{folsom2007, bailey2014}), thus we conclude HR 2949 is chemically a Bp star.  

We performed a similar abundance analysis on HR 2948, for comparison.  Somewhat larger spectral windows were used (440--480, 490--530, 530--570, 570--610, and 610--650 nm), due to the higher \vsini~producing weaker, more blended lines.  A $\log{g} = 4.24$ was used, based on the Balmer line analysis.  We find a \teff~of $13.6 \pm 1.2$ kK and a \vsini~of $117 \pm 2$ \kms.  We attempted to fit microturbulence as well, however the value is poorly constrained, since the large \vsini~leads to a lack of unblended weak lines (on the linear part of the curve of growth). Thus we only constrain microturbulence to be $\leq 1.2$ \kms. The chemical abundances we find for HR 2948 are entirely consistent with solar (see Table \ref{ab_fvar_tab} and Fig. \ref{abundances}).  Therefore we conclude that HR 2948 is a chemically normal B star.  

\begin{figure}
\centering
\includegraphics[width=8.5cm]{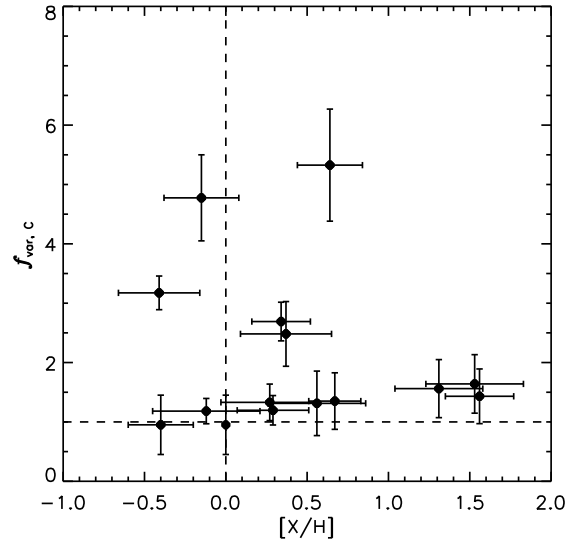} 
\caption{The corrected variability index $f_{\rm var, C}$ as a function of relative solar abundance. Vertical and horizontal dashed lines indicate, respectively, solar abundance, and the mean level of variability for a line of the strength used to measure $f_{\rm var}$.}
\label{ab_fvar}
\end{figure}

A full comparison of variability and abundance cannot be performed, as strong, unblended lines do not exist for many of the elements examined in the abundance analysis, particularly the rare-earth elements. For those elements which can be so analyzed, Fig. \ref{ab_fvar} compares the abundances to the quantity $f_{\rm var, C}$ from equation \ref{fvar} and Fig. \ref{varind}. There is no clear relationship between the variability of a given element and its abundance. The spectral lines of elements with solar, sub-solar and super-solar abundances are in some cases highly variable, and in other cases not.

\section{Magnetic field}

\begin{figure}
\centering
\includegraphics[width=8cm]{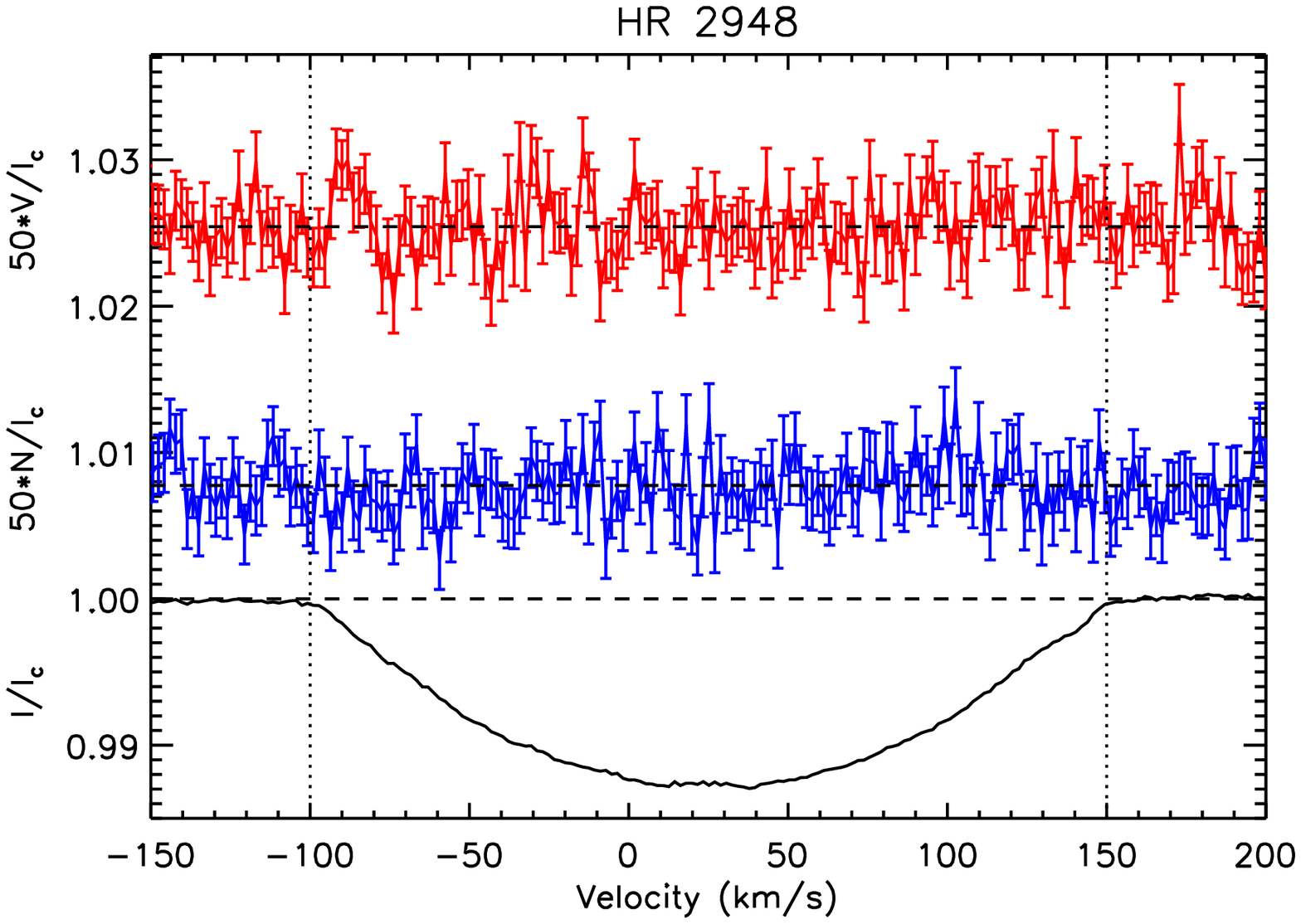} 
\includegraphics[width=8cm]{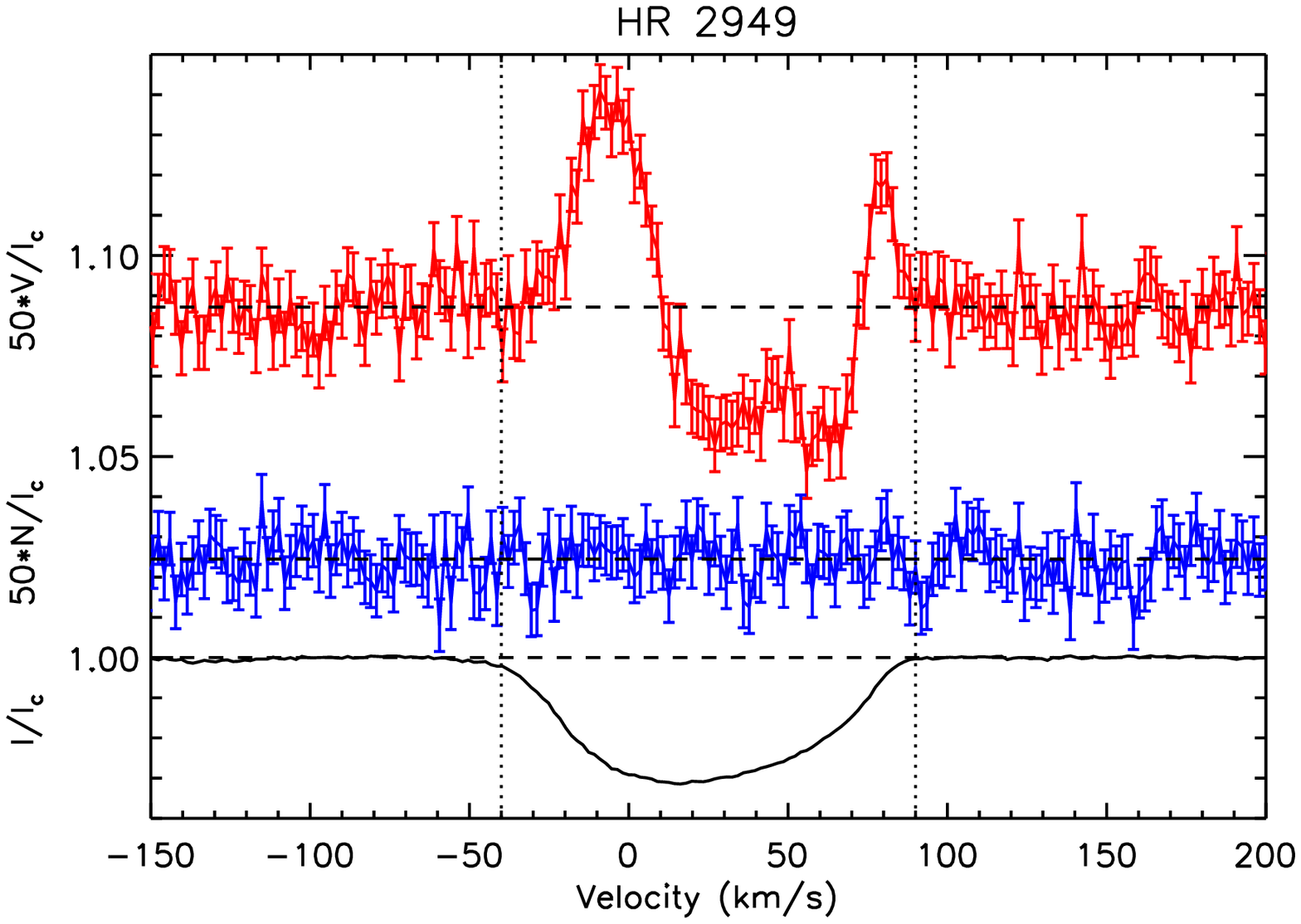} 
\caption{Representative LSD profiles for HR 2948 (top) and HR 2949 (bottom). The HR 2949 LSD profile was made using a line mask including all metallic lines and the 2010-01-01 ESPaDOnS observation. Stokes $I$ (bottom) is in black, Stokes $V$ (top) in red, and the diagnostic null $N$ (middle) in blue. Integration ranges used to measure \bz~are indicated with dotted vertical lines. Note the strong Stokes $V$ signature in HR 2949, in contrast to that of HR 2948, which is consistent with noise.}
\label{lsd_stars}
\end{figure}

\begin{figure}
\centering
\includegraphics[width=9cm]{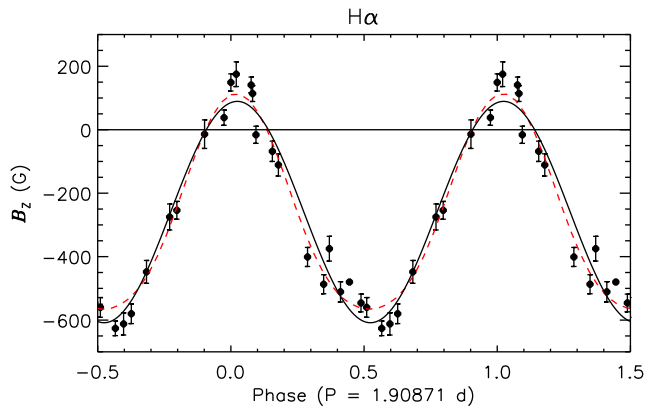}
\caption{Single-line longitudinal magnetic field measurements using \halp~(solid black circles). The solid (black) curve is the best-fit sinusoid; the dashed (red) curve is the best-fit offset dipole model with $a = 0.33$.}
\label{bell_ha_met}
\end{figure}

In order to maximize the signal-to-noise ratio, the ESPaDOnS observations were analyzed using Least Squares Deconvolution (LSD) \citep{d1997} as implemented in the iLSD package developed by \cite{koch2010}. 

As in the abundance analysis, line lists were downloaded from VALD2 (\citealt{piskunov1995, ryabchikova1997, kupka2000}). For HR 2949, the line list was constructed for a star with \teff$= 18$ kK, \lgg$=4.0$, and a line-strength threshold of 10\% beneath the continuum, yielding 462 lines. The line list for HR 2948 was constructed for a star with \teff$= 14$ kK and \lgg$=4.0$, and a 10\% threshold, yielding 561 lines. 

The line lists were cleaned by removing all H and He lines as well as all lines blended in their wings, along with all lines strongly contaminated with telluric lines. While in most cases the effect on Stokes $V$ of these blends is negligible, cleaning the lists in this fashion ensures that the mean Stokes $I$ profiles are not affected by lines with strong departures from the mean photospheric line profile. The remaining lines were then empirically adjusted such that their depths reflect the observed line intensities. 

For HR 2949, LSD profiles were created using a mask including all metallic lines, as well as single-element line masks for C, N, O, Ne, S, Mg, Si, and Fe. A He line mask was also tested. No single-element masks were created for HR 2948, although He was excluded in order to avoid distortion of the Stokes $I$ profile. The number of lines in each mask is given with the longitudinal field measurements in Table \ref{bz_1}. Representative LSD profiles created using metallic masks are shown for both stars in Fig. \ref{lsd_stars}. 

To determine whether or not a magnetic field is detected, a statistical test is applied \citep{d1997}. A detection is considered `definite' if the signal within the Stokes $V$ line profile departs from the noise, as diagnosed by both Stokes $V$ outside the line profile, and by the diagnostic null $N$, with the False Alarm Probability ${\rm FAP} < 0.001\%$. If ${\rm FAP} > 0.1\%$, the observation is formally considered a non-detection. Values between these limits are considered marginal. Integration ranges, determined by the extent of the absorption lines, are indicated by dotted vertical lines in Fig. \ref{lsd_stars}. For HR 2949, all measurements are definite detections, with ${\rm FAP} \ll 0.001$. For HR 2948, all individual observations are non-detections, as is the co-added observation. 

The longitudinal magnetic field \bz~was measured from the LSD profiles by taking the first-order moment of the Stokes $V$ profile (e.g. \citealt{mat1989}):

\begin{equation}\label{blong}
\langle B_z  \rangle  = -2.14\times 10^{11}\frac{\int \! vV(v)\mathrm{d}v}{\lambda_0 g_0 c\int \left[I_{\rm{c}}-I(v)\right] \mathrm{d}v},
\end{equation}

\noindent where $v$ is the Doppler velocity in \kms, and $\lambda_0$ and $g_0$ are the normalization values of the wavelength and Land\'e factor used to scale the Stokes $V$ profile. For all LSD profiles the same parameters were used to extract the LSD profiles and to measure \bz: $\lambda_0 = 500$ nm and $g_0$ = 1.2. The integration ranges are shown as dotted lines in Fig. \ref{lsd_stars}. The same measurement can be applied to the diagnostic $N$ profile, yielding an equivalent `null longitudinal magnetic field' \nz, with which \bz~can be compared. 

LSD profile measurements of \bz~for HR 2949 are given in the Appendix in Table \ref{bz_1}. For HR 2948, the single measurement gives \bz$= -19 \pm 35$ G and \nz$= -69 \pm 35$ G, consistent with a non-detection. 

Because Zeeman signatures are visible in many of HR 2949's spectral lines, single-line measurements of \bz~were also performed for a selection of lines: Ca {\sc ii} 393.4 nm and Ti {\sc ii} 457.2 nm (since there are too few lines of these species to perform LSD); Si {\sc iii} 455.2 nm and Fe {\sc ii} 516.9 nm (for comparison of single-line results to LSD results), and H$\alpha$ (since the relatively uniform distribution of this element across the photosphere should result in Stokes $V$ profiles that are most representative of the real global magnetic field geometry). These measurements, together with the rest wavelengths $\lambda_0$ and Land\'e factors $g$, are given in the Appendix in Table \ref{bz_2}. 

The H$\alpha$ \bz~measurements are shown phased with the rotation period in Fig. \ref{bell_ha_met}. The best-fit sinusoid is overplotted as a thick black line. The overall shape of the \bz~curve is somewhat non-sinusoidal, being sharply peaked near the positive pole and relatively broad near the negative pole. However, only two points are incompatible with a sinusoidal fit at the 3$\sigma$ level. 


\subsection{Magnetometry using different elements}

\begin{figure}
\centering
\begin{tabular}{cc}
\includegraphics[width=4cm]{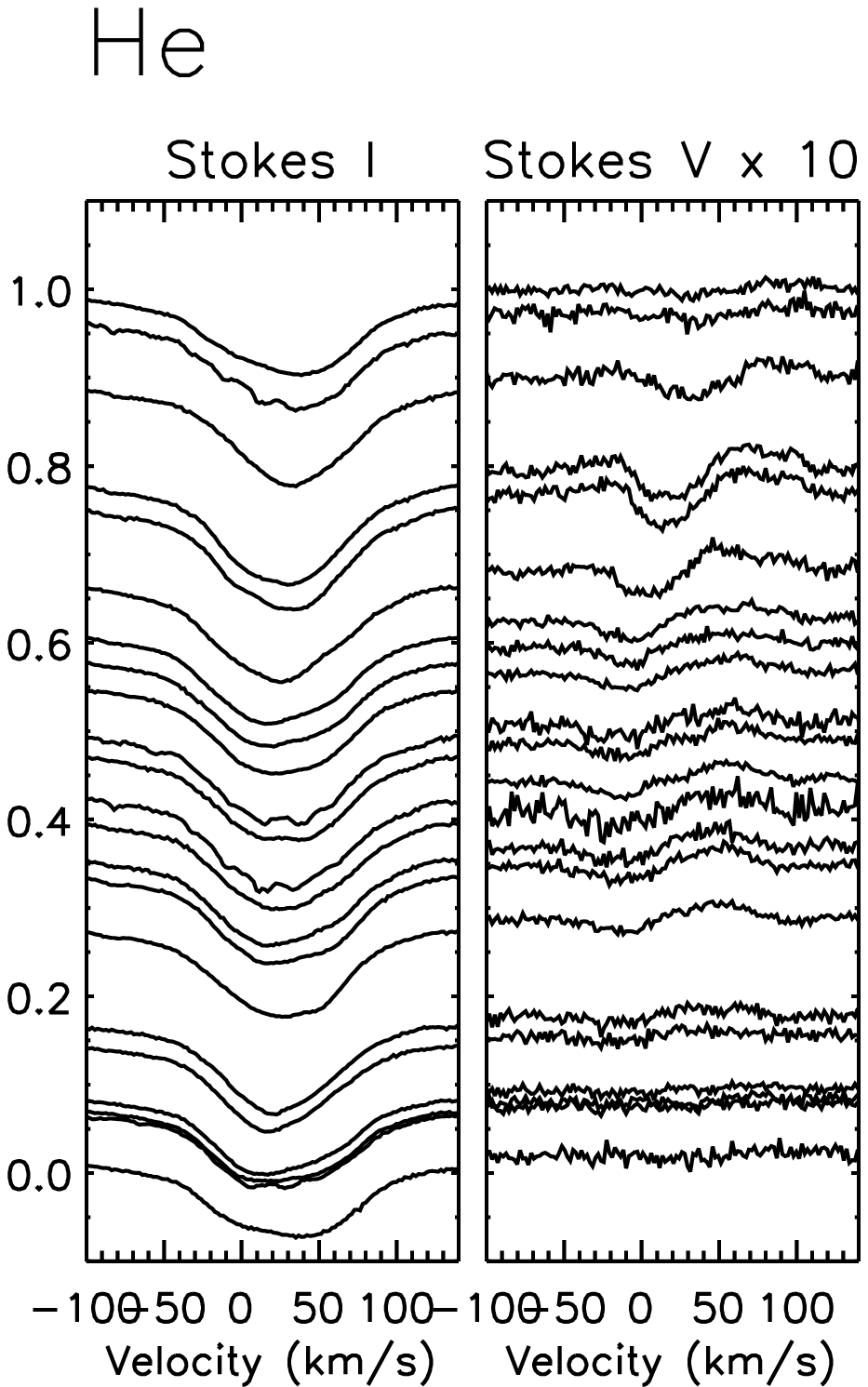} &
\includegraphics[width=4cm]{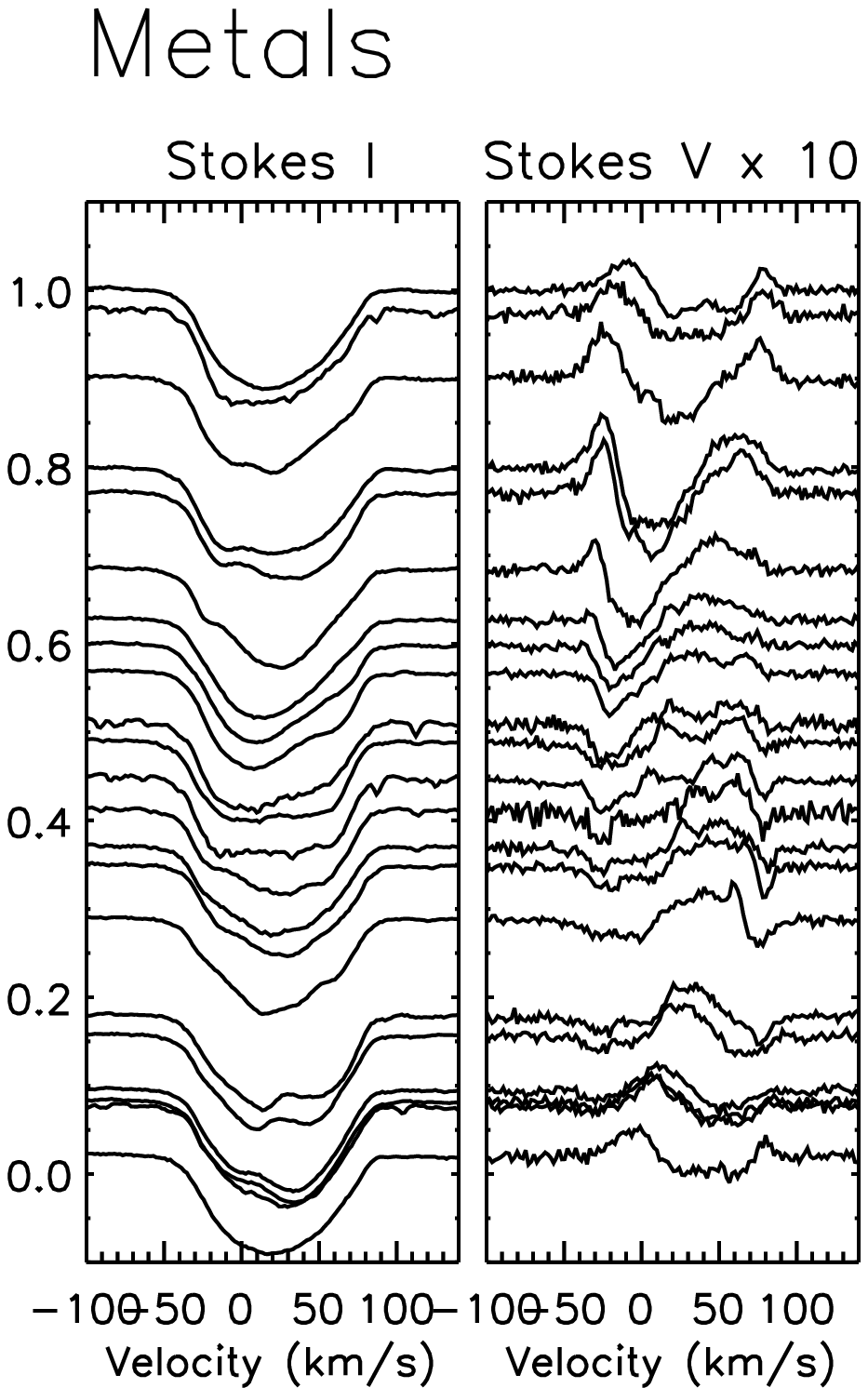} \\
\includegraphics[width=4cm]{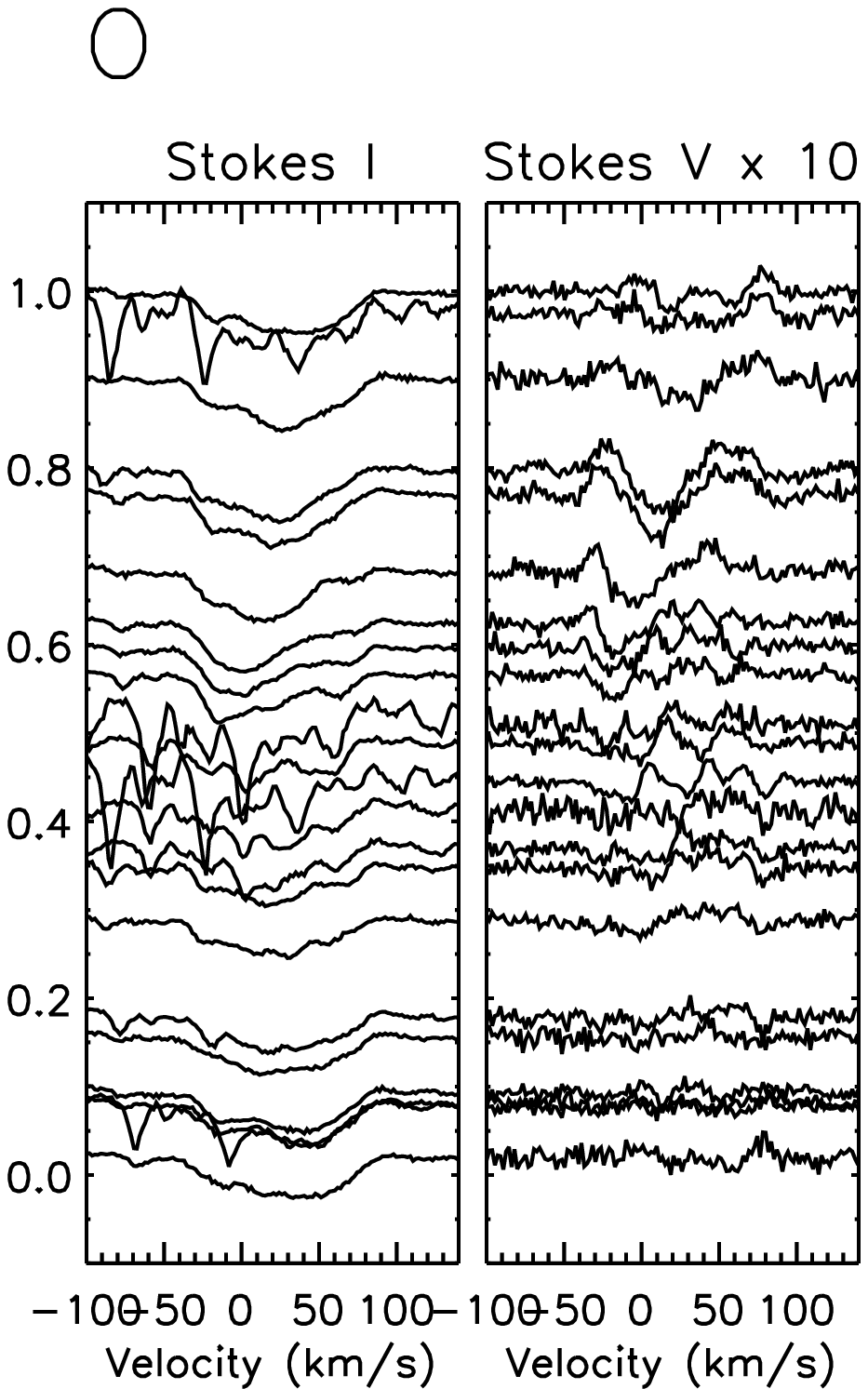} &
\includegraphics[width=4cm]{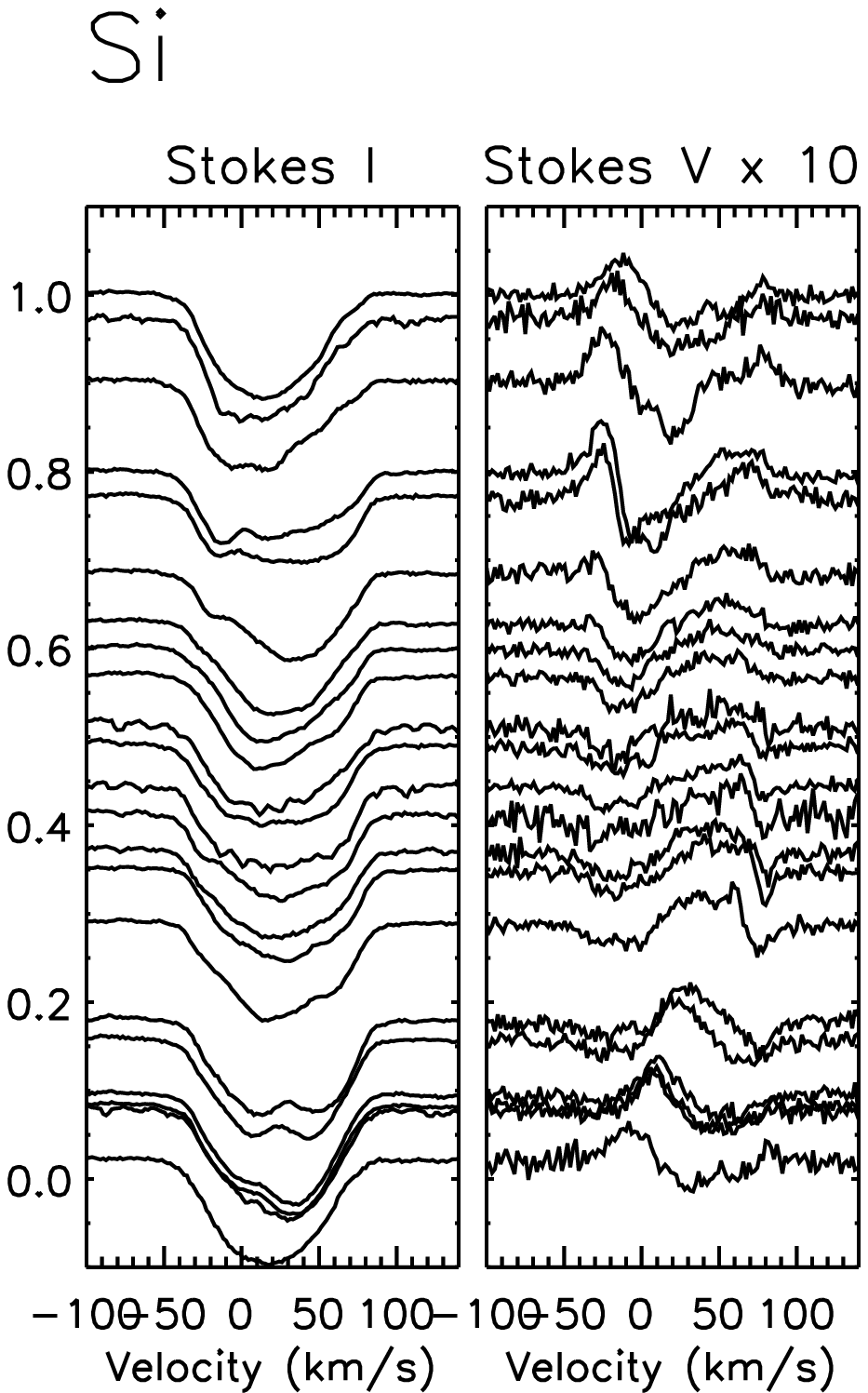} \\
\includegraphics[width=4cm]{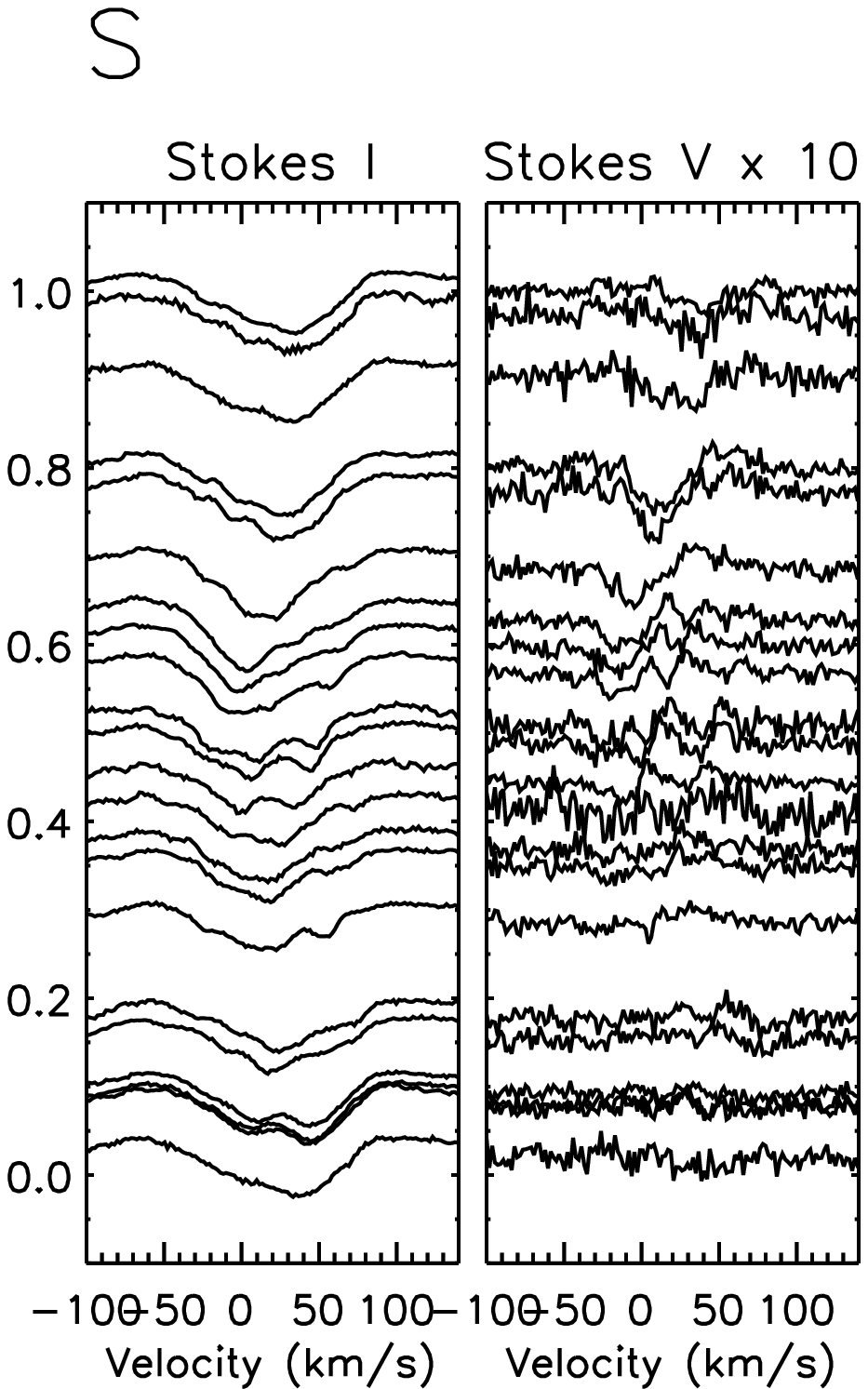} &
\includegraphics[width=4cm]{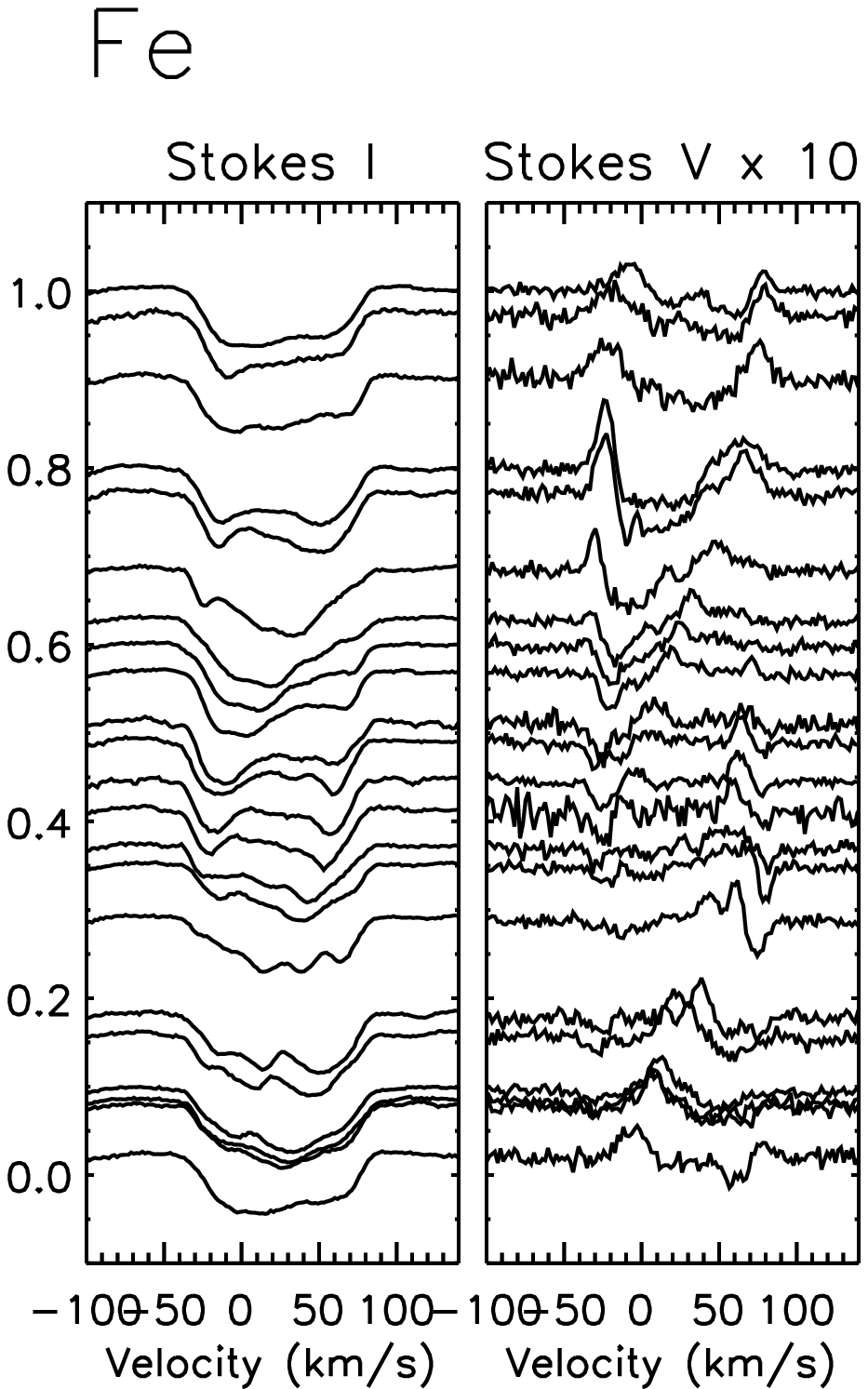} \\
\end{tabular}
\caption{Phased LSD profiles, with the continuum level set to the rotational phase. The Stokes $V$ scaling factor is relative to Stokes $I$.}
\label{lsdprofs-1}
\end{figure}

\begin{figure}
\centering
\includegraphics[width=8cm]{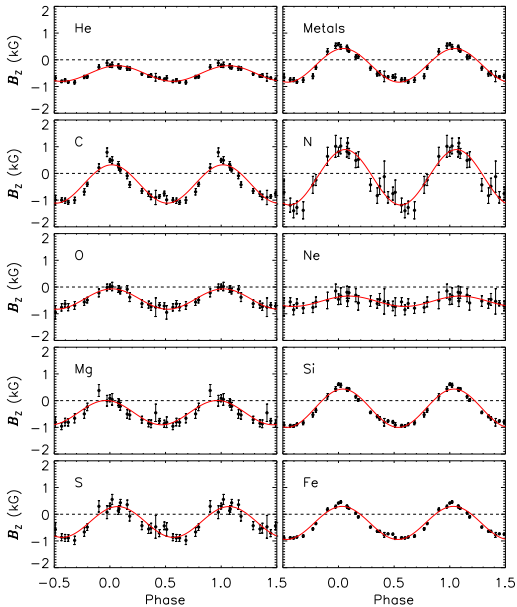}
\caption{LSD profile longitudinal magnetic field measurements for single-element masks.}
\label{bell_1}
\end{figure}

\begin{figure}
\centering
\includegraphics[width=8cm]{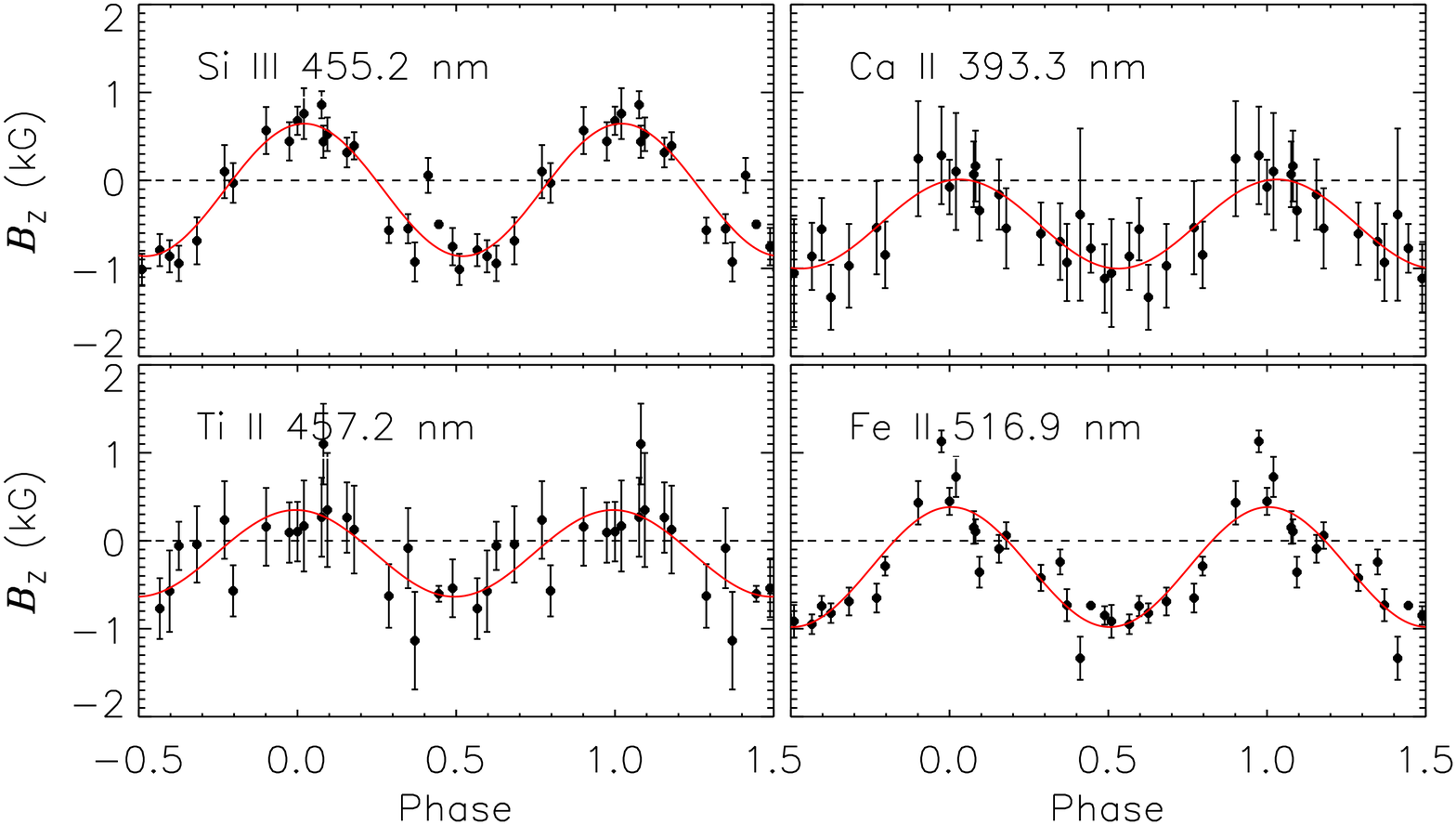}
\caption{Single line longitudinal magnetic field measurements.}
\label{bell_3}
\end{figure}

The LSD Stokes $I$ and $V$ profiles of HR 2949 created from He, metal, O, Si, S, and Fe masks are shown in Fig. \ref{lsdprofs-1}. The LSD profiles are stacked vertically, with the vertical offset proportional to the rotational phase. The LSD profiles created using all metallic elements are most similar to those created with Si and Fe masks, suggesting the former is dominated by these lines. 

Comparing the Si and Fe LSD profiles, Fe lines are consistently more complex than those of Si lines. This is likely a reflection of the greater complexity of the photospheric Fe abundance distribution, as is apparent in both the Stokes $I$ profiles and the dynamic spectra in Figs. \ref{dyn_he} and \ref{dyn_3}. 

The strongest difference between the Fe and Si lines (right column of Fig. \ref{lsdprofs-1}), and the He, O, and S lines (left column of Fig. \ref{lsdprofs-1}), occurs near the positive magnetic pole. Here, the Stokes $V$ signatures of the latter group disappear almost entirely, while those of the former group show cross-over signatures followed by a reversed polarity. 

O and S are, however, quite different from He: while He shows a simple S-shaped Stokes $V$ profile near the negative magnetic pole, O and S both display more complex Stokes $V$ profiles. 

Fig. \ref{bell_1} shows mosaics of phased LSD \bz~measurements from Table \ref{bz_1}. Best-fit sinusoids are shown in red. Measurements of He, O, and Ne are apparently well-fit by a sinusoid; however, they do not reverse polarity, while most other elements (except Mg) do. Of the remaining elements, some, such as C, Si, and Fe, are clearly non-sinusoidal. 

Fig. \ref{bell_3} shows a mosaic of \bz~measurements performed on individual spectral lines (see also Table \ref{bz_2}). The Si {\sc iii} and Fe {\sc ii} measurements are similar to the LSD measurements for these species, albeit with greater scatter and larger error bars. While the best-fit sinusoid to the Ti {\sc ii} 457.2 nm measurements indicates polarity reversal, the error bars of most measurements near the positive pole are consistent with zero. Ca {\sc ii}, however, clearly belongs to the non-polarity reversing group. 

Comparing spectral and magnetic diagnostics, there appears to be an absorption excess of He associated with the longitude of the negative magnetic pole, and a weakening of absorption associated with the positive pole. This leads to an apparent weakening of the positive magnetic pole when viewed in He lines. \cite{yakunin2014} have reported similar behaviour for HD 184927, and have shown that by self-consistently mapping the abundance distributions and magnetic field using Zeeman Doppler Imaging (e.g. \citealt{pk2002}) the divergent \bz~measurements from various chemical species can be reproduced with a single magnetic model.

\subsection{Magnetic Field Geometry}

\begin{table}
\centering
\caption[]{Sinusoidal fitting parameters $B_0$ and $B_1$, and resultant dipolar magnetic field model parameters \bd~and $\beta$ for single-element \bz~measurements (see eqns. \ref{ibeta}--\ref{bd_min}).}
\begin{tabular}{lrrrr}
\hline
\hline
Element & $B_0$ & $B_1$ & $B_{\rm d}$ & $\beta$ \\[2 pt]
 & (G) & (G) & (kG) & ($^\circ$) \\[2 pt]
\hline
H  & -278$\pm$14 &  376$\pm$18 & 2.4$^{+0.3}_{-0.2}$ & 50$\pm$16 \\[2 pt]
He & -514$\pm$14 &  291$\pm$19 & 3.2$^{+1.3}_{-0.4}$ & 27$\pm$16 \\[2 pt]
C & -396$\pm$47 &  716$\pm$60 & 4.2$^{+1.3}_{-0.3}$ & 57$\pm$15 \\[2 pt]
N & -144$\pm$64 & 1037$\pm$82 & 5.1$^{+1.9}_{-1.0}$ & 80$\pm$5 \\[2 pt]
O & -450$\pm$20 &  -69$\pm$26 & 2.5$^{+1.3}_{-0.7}$ & 8$\pm$6 \\[2 pt]
Ne & -533$\pm$27 &  193$\pm$36 & 3.2$^{+1.5}_{-0.6}$ & 19$\pm$13 \\[2 pt]
Mg & -450$\pm$37 & -454$\pm$50 & 3.4$^{+0.8}_{-0.2}$ & 42$\pm$17 \\[2 pt]
Si & -271$\pm$25 &  726$\pm$32 & 3.9$^{+1.2}_{-0.4}$ & 66$\pm$11 \\[2 pt]
S & -289$\pm$42 &  575$\pm$55 & 3.3$^{+0.8}_{-0.3}$ & 59$\pm$14 \\[2 pt]
Ca & -495$\pm$57 &  506$\pm$73 & 3.8$^{+0.9}_{-0.3}$ & 42$\pm$17 \\[2 pt]
Fe & -209$\pm$29 &  632$\pm$38 & 3.3$^{+1.1}_{-0.4}$ & 68$\pm$10 \\[2 pt]
\hline
\hline
\end{tabular}
\label{b0b1_tab}
\end{table}

\cite{preston1967} showed that, for a dipolar magnetic field inclined at an angle $\beta$ from the rotational axis and an inclination $i$ of the rotational axis from the line of sight, $i$ and $\beta$ are related by 

\begin{equation}\label{ibeta}
\tan{\beta} = \left(\frac{1-r}{1+r}\right)\cot{i},
\end{equation}

\noindent where $r$ is the ratio

\begin{equation}\label{r}
r = \frac{|B_0| - B_1}{|B_0| + B_1},
\end{equation}

\noindent and $B_0$ and $B_1$ are defined by the sinusoidal fit to the phased \bz~variation:

\begin{equation}\label{ibeta}\label{sinfit}
B_{\rm Z} = B_0 + B_1\sin{2\pi(\phi - \phi_0)}.
\end{equation}

As the \bz~variation differs depending on the spectral lines with which it is measured, so too do $B_0$ and $B_1$. 
We base our primary analysis upon \halp~measurements, as these are expected to be the least affected by LPV. Fitting parameters for different single-element measurements are tabulated in Table \ref{b0b1_tab}. 

The best-fit sinusoid to the \halp~measurements has $B_0~=~-278\pm14$~G and $B_1~=~376~\pm~18$ G, giving $r~=~-0.15~\pm~0.05$. With the inclination determined above, we then have $\beta~=~50\pm~16^\circ$. 

The strength of the magnetic dipole can then be derived from the relation \citep{preston1967}

\begin{equation}\label{bd_min}
B_{\rm d} = B_{z}^{\rm max}\left(\frac{15 + u}{20(3 - u)}(\cos{\beta}\cos{i} + \sin{\beta}\sin{i})\right)^{-1},
\end{equation}

\noindent where $u$ is the limb darkening coefficient, which we take to be 0.32, the mean of the values obtained from the tables calculated by \cite{vanhamme1993} and \cite{diazcordoves1995}. This yields $B_{\rm d} = 2.4^{+0.4}_{-0.3}$ kG.

If metal lines are used instead, equations \ref{ibeta} and \ref{bd_min} of course give very different results for $\beta$ and $B_{\rm d}$, which differ outside the formal uncertainties. These are summarized in Table \ref{b0b1_tab}. This demonstrates that, in the presence of surface abundance inhomogeneities, the use of particular sets of spectral lines can introduce significant systematic error into the determination of the magnetic field strengths and geometries of Bp stars.

\begin{figure}
\centering
\includegraphics[width=9cm]{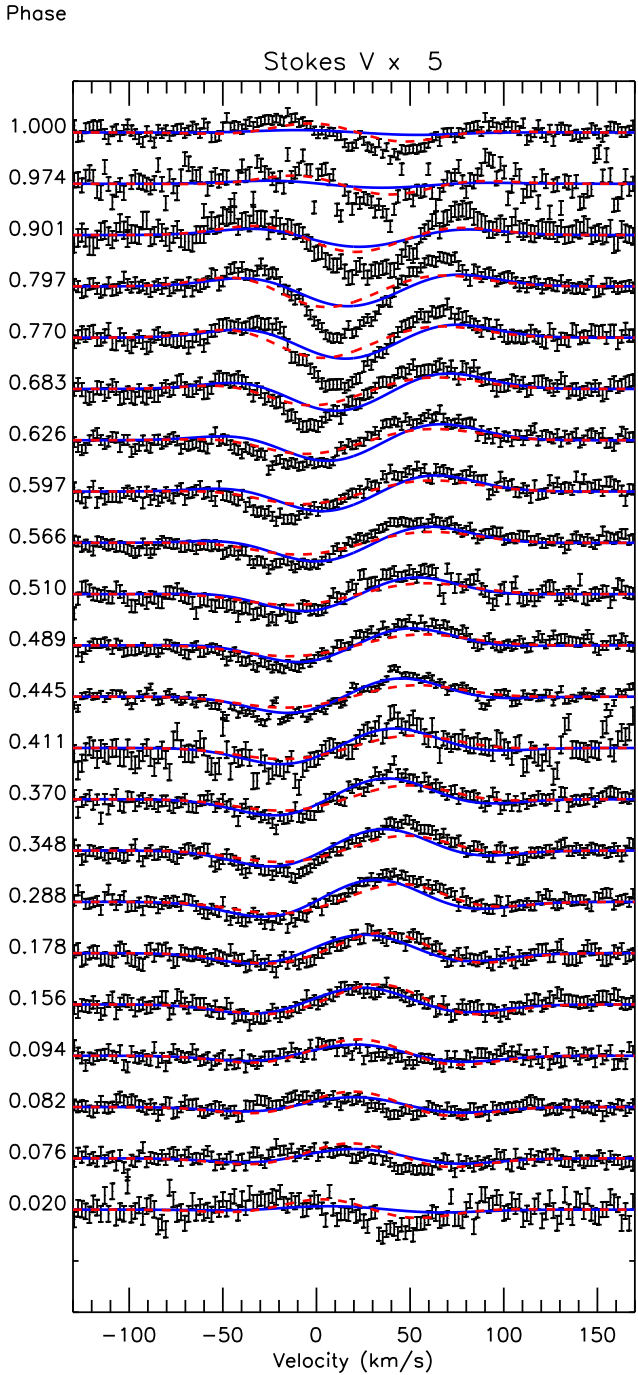}
\caption{Stokes $V$ profiles from H$\alpha$ (black) arranged in order of phase. Solid blue lines are synthetic Stokes $V$ spectra created for HR 2949's stellar parameters and a dipolar model matching the best fit model to the H$\alpha$ \bz~measurements. The model agrees well with observations near the negative magnetic pole (phase 0.5), but cannnot reproduce observations near the positive magnetic pole (phase 0.0). Dashed red lines are synthetic profiles created for an offset dipole model with $a=0.33$: while this model provides a somewhat better fit near the positive pole, the fit near the negative pole is substantially worse than the dipolar model.}
\label{halpha_stokesv}
\end{figure}

Fig. \ref{halpha_stokesv} shows a comparison between synthetic Stokes $V$ and observed \halp~Stokes $V$ profiles. The model synthetic profiles were calculated using a disk integration method similar to that described by \cite{petit2012a}. We examine two models: a dipolar model (solid blue lines) using the parameters above, and the best-fit offset dipole model (dashed red lines). The observed profiles (black points) exhibit a much simpler variation than is seen in He or metallic lines (compare to Fig. \ref{lsdprofs-1}). The dipolar profiles are able to reproduce the observed profiles quite well around the negative magnetic pole (rotational phases 0.35--0.5), however the fit is not as successful in the vicinity of the positive pole (rotational phases 0.8--1.2). 

To determine the parameters of the offset dipole model, we employed a goodness-of-fit test on the H$\alpha$ \bz~measurements. The \bz~curve is shown in Fig. \ref{bell_ha_met}. The model parameters are $\beta = 60^\circ, B_{\rm d} = 2.4$ kG, with the dipole offset by $a = 0.33 R_*$ along the magnetic axis. As is clear from Fig. \ref{halpha_stokesv}, even this large offset is unable to satisfactorily reproduce Stokes $V$: while the fit is somewhat improved near the positive pole, it is somewhat worse near the negative pole. In both cases, the fit is especially poor between phases 0.6 and 0.9.

While H lines are not as subject to differential flux redistribution as He and metallic lines, it is clear from both dynamic spectra and EW measurements that H is so affected, likely due to variations in the local H to He ratio (see Fig. \ref{dyn_he}). Furthermore, it is precisely at those phases that are most poorly fit that the line strength is at its strongest. From this we conclude that H line Stokes $V$ profiles are likely subject to warping by differential line strengths, and that Magnetic Doppler Imaging (MDI) will be necessary to reconstruct the magnetic field topology.


\section{Discussion}

\subsection{Binarity}

There is a discrepancy of $\sim$100 Myr in the apparent ages of HR 2948 and HR 2949 (see Table \ref{params}). Since binary pairs are expected to be coeval, this motivates a re-examination of the assumption that the stars are a binary pair, as, if this is not correct, some of the stellar parameters found above may need to be revised. Here we explore four different scenarios to attempt to explain this age discrepancy: 

\noindent 1) The Hipparcos parallax is unreliable, and hence the distance and luminosities found above are incorrect. 

\noindent 2) The isochrones used to determine the ages are inappropriate: pre-main sequence (PMS) isochrones should be used instead. 

\noindent 3) The stars are not gravitationally bound: their proximity on the sky is a chance coincidence, and they are not in fact at identical distances. 

\noindent 4) The stars are not coeval, but form a binary pair due to gravitational capture. 


Keeping \teff~the same and setting the minimum luminosity of HR 2949 at the ZAMS, the minimum distance to the star is 112$\pm$6 pc. If HR 2948's luminosity is recalculated at this distance, it's age is 89$^{+48}_{-31}$ Myr, i.e. the star is well off the ZAMS, and the age discrepancy is not resolved. Clearly, if instead the distance to the stars is greater than that determined from the Hipparcos parallax, the discrepancy is impossible to resolve: HR 2949 would have to be much cooler to be on the same isochrone as HR 2948. Therefore scenario 1) does not resolve the problem.

To explore scenario 2), we use the PMS isochrones presented by \cite{tognelli2011}. Only the 1 Myr isochrone is consistent with the luminosities and effective temperatures of both stars. This isochrone is shown by a dash-dotted line in Fig. \ref{hrd}. In this scenario, HR 2949 is essentially at the ZAMS, while HR 2948 has not yet reached it. 

However, HR 2948/9 is not located in or near to any star-forming regions, as might be expected if the system is indeed 1 Myr old. Furthermore, HR 2948 has no \halp~emission. Therefore it seems unlikely that the system is newly-formed.

To explore scenario 3), we note first that the radial velocities are identical within error bars, as expected for a weakly gravitationally bound system. The proper motions of the two stars differ by 8.8$\pm$2.6 mas yr$^{\rm -1}$ in right ascension and 4.8$\pm$2.6 mas yr$^{\rm -1}$ in declination (\citealt{hog2000}; see also Table \ref{params}), for a total relative proper motion of 10.0$\pm$3.5 mas yr$^{\rm -1}$. At the Hipparcos distance of 139~pc, this corresponds to a relative motion of 6.6$^{+3.8}_{-2.9}$ \kms. The stars are separated on the sky by 7.3'', corresponding to a physical separation of 1014$^{+175}_{-131}$ AU. With the masses above, and assuming a circular orbit, the orbital velocity is then $v_{\rm orb} \sim \sqrt{G(M_{\rm 1} + M_{\rm 2})/r} = 3.0 \pm 0.3$ \kms. In order for the orbital velocity to equal the relative motions, the semi-major axis of the system would have to be $a = 90^{+190}_{-55}$~AU, clearly inconsistent with the observed separation. The proper motions thus argue against the two stars being gravitationally bound. This analysis assumes the orbital plane is in the plane of the sky. If the inclination of the orbital axis from the line of sight is substantially greater than 0$^\circ$, the physical separation would of necessity be larger, implying a lower Keplerian velocity. At the same time, the deprojected relative velocity would be larger, thus increasing the discrepancy. However, the disagreement between the orbital velocity and the proper motions is relatively small, so we do not view this as conclusive evidence that the two stars are not associated. 

Regarding scenario 4), it is worth noting that the system is a wide binary, and not strongly gravitationally bound, making a gravitational capture scenario more plausible. 

\subsection{Variability}


The most likely explanation for the LPV seen in HR 2949's He and metallic lines are surface abundance variations. Chemical spots are commonly seen in Ap/Bp stars (e.g. \citealt{michaud1970, michaud1981}), indeed the presence of such spots is one of the most reliable indirect indicators of magnetism amongst early-type stars. Differences in the corrected variability index $f_{\rm var, C}$ are almost certainly related to this: elements with $f_{\rm var, C} > 1$ likely show more extreme surface abundance variations than elements with $f_{\rm var, C} = 1$. 

Comparing Figs. \ref{dyn_he}, \ref{bell_ha_met}, and \ref{bell_1}, the strongest He, O, and S line profiles are associated with the negative magnetic pole, while the positive magnetic pole is associated with absorption minima. The weakness of these lines near the positive pole may explain why their Zeeman signatures seem to disappear almost entirely near the positive pole (Figs. \ref{lsdprofs-1} and \ref{bell_1}). The intensity of the regions of the photosphere associated with the positive pole is reduced, decreasing the contribution of the magnetic pole to the polarized flux. 

A similar effect in metallic lines is likely the reason for the observed inconsistency in \bz~as measured using different elements. The strong asymmetry observed in the LSD Stokes $V$ profiles (see Fig. \ref{lsdprofs-1}) is associated with asymmetries in Stokes $I$. It is also possible that the apparent departures of \bz~from the simple sinusoid expected for a purely dipolar magnetic field are likewise a consequence of this effect, with greater relative weight given to the positive magnetic pole magnifying \bz~at these phases. Such an effect was first suggested by \cite{bl1977} for the magnetic Ap star $\alpha^2$ CVn, who noted that magnetometry conducted using H lines did not show the non-dipolar character of the \bz~measurements conducted using metallic lines.

Conversely, the departures of \halp~measurements from purely sinusoidal behaviour, along with the failure of a dipolar model to reproduce the Stokes $V$ profiles near the positive magnetic pole in this line, point towards a non-dipolar component to the magnetic field. 

Examination of the $H_{\rm p}$ magnitudes in Fig. \ref{hip} show that the star is brightest between phases 0.0 and 0.2, and dimmest between phases 0.4 and 0.5. This corresponds quite well to the line-strength extrema of He and Si. It thus seems likely that the photometric variability of HR 2949 is a consequence of the variable brightness of photospheric abundance patches, typical of magnetic chemically peculiar stars (e.g. \citealt{wraight2012}). Efforts to model the photometric  variations of some Bp stars by taking into account the variable optical brightness caused by flux redistribution due to chemical spots have successfully accounted for a majority or all of the light curve variability in many cases, e.g. for the He-strong Bp star HD 37776 \citep{2007AA...470.1089K}, the Ap star $\epsilon$ UMa \citep{2010AA...524A..66S}, the Si star HR 7224 \citep{2009AA...499..567K}, and the Ap star CU Vir \citep{2012AA...537A..14K}.

While the strongest H variability is seen in the core of the lines, and is almost certainly also a consequence of photospheric abundance variations, the origin of the LPV observed in the wings of the H Balmer lines (see Figs. \ref{dyn_he} and \ref{tvs}) is not so obvious. 

Early-type Bp stars often show emission in the wings of their Balmer lines originating in their corotating `centrifugal magnetospheres' (e.g. \citealt{town2005b, petit2013}). The magnetospheric parameters of HR 2949 are derived in the following subsection. Here, we note that the Balmer line variability is extremely weak; shows no signs of the blue-to-red variation associated with magnetospheric plasma (e.g., \citealt{grun2012b}, \citealt{rivi2013}); and shows no sign of a Balmer decrement \citep{ws1988}, as would be expected for a transluscent circumstellar plasma (indeed, \halp~is {\em less} variable than the higher-numbered Balmer lines, precisely the opposite of expectations for circumstellar emission\underline{}). Therefore we conclude that the Balmer wing LPV is not a consequence of a magnetosphere. 

Another possibility is that variation in the wings is related to the chemical spots. This could arise due to different degrees of line blanketing at different points of the surface \citep{khanshulyak2007}, producing small differences in atmospheric structure in different regions of the photosphere. In this case, the Balmer wing strength should correlate with Fe, and anticorrelate with Si \citep{khanshulyak2007}, approximately matching observations. 

A third possibility related to chemical spots is that the LPV is a pressure effect originating in local abundance enhancements or depletions. Such an effect is suspected in the magnetic Bp star HD 184927 (B2V, \citealt{yakunin2014}). The variations in HR 2949's wings are much weaker than those in HD 184927, however HD 184927 is a He-strong star, with large surface abundance variations. 

The final possibility is that the LPV is a consequence of variable magnetic pressure originating due to induced electric currents in the photosphere. This phenomenon has been suggested as the reason for Balmer line LPV in the He-weak Si star HR 7224 (B9p, \bd$< 0.4$ kG \citealt{lehmann2007}). In this case chemical spots were specifically excluded as the possible source of the LPV. Model atmospheres including the Lorentz force were shown to provide the best explanation for Balmer line LPV in the A0p star $\theta$ Aur \citep{shulyak2007} and the B6p star 56 Ari \citep{shulyak2010}, both of which host moderately strong photospheric magnetic fields (\bd$\sim 1$ kG). HR 2949's Balmer line wing LPV is similar in character to that seen in HR 7224, $\theta$ Aur, and 56 Ari, suggesting that a similar mechanism may be at work.

\subsection{Magnetosphere}

\begin{table}
\centering
\caption[Magnetospheric Parameters]{Magnetospheric parameters calculated assuming mass-loss rates from \cite{vink2001} and \cite{krticka2014}.}
\begin{tabular}{lrr}
\hline
\hline
\rk~($R_*$) & \multicolumn{2}{c}{3.5$^{+1.7}_{-1.2}$}  \\[2 pt]
$W$ & \multicolumn{2}{c}{0.14$^{+0.10}_{-0.07}$}  \\[2 pt]
\hline
Mass-loss prescription & Vink & Krticka \\[2 pt]
\\
$\log$\mdot~($M_\odot~{\rm yr}^{-1}$) & -9.7$^{+0.4}_{-0.3}$ & -10.5$^{+0.3}_{-0.4}$ \\[2 pt]
\vinf~(\kms) & 1070$^{+60}_{-80}$ & 1100$^{+800}_{-600}$ \\[2 pt]
$\log\eta_*$ & $4.7^{+0.6}_{-0.7}$ & $5.5^{+1.3}_{-0.7}$ \\[2 pt]
\ra~($R_*$) & 15$^{+7}_{-5}$ & 23$^{+24}_{-9}$ \\[2 pt]
$\log{(R_{\rm A}/R_{\rm K})}$ & 0.63$\pm0.35$ & 0.82$^{+0.49}_{-0.39}$ \\[2 pt]
$\tau_{\rm M}$ (Myr) & $3.0^{+3.0}_{-1.8}\times 10^4$ & $1.8^{+2.9}_{-0.9}\times 10^5$ \\[2 pt]
$\tau_{\rm J}$ (Myr) & 11$\pm1$ & 32$\pm13$ \\[2 pt]
$t_{\rm S}^{\rm max}$ (Myr) & 22$\pm 3$ & 63$\pm 35$ \\[2 pt]
\hline
\hline
\end{tabular}
\label{mag_prop}
\end{table}

When contained within a magnetic field, the spherical symmetry of a stellar wind is strongly modified if the energy density of the magnetic field exceeds the kinetic energy density of the wind \citep{bm1997, ud2002}. Confined plasma is forced to flow along closed magnetic loops. At the tops of the loops, the plasma collides, releasing high-energy X-rays in wind shocks (e.g. \citealt{bm1997,ud2014, naze2014}). The plasma may also become detectable as rotationally-modulated emission in wind-sensitive spectral lines, particularly in the ultraviolet (e.g. \citealt{shorebrown1990, smithgroote2001, schnerr2008, henrichs2013}), but also in optical lines such as \halp~(e.g. \citealt{walborn1974, pederson1979, leone2010, grun2012, rivi2013, petit2013}). 

As discussed above, we do not detect any sign of \halp~emission in HR 2949's spectra. 

There have been no published X-ray observations of HR 2949. 

There is a single high-dispersion IUE spectrum, acquired with the large aperture, a 10$\times$20'' oval which almost certainly included HR 2948. The flux is much too high to be associated with a single star of $V\sim 4.5$. However, it is compatible with the combined magnitude of the binary system. Using the ephemeris in equation \ref{ephemeris}, the IUE observation was acquired at a rotational phase of 0.27$\pm$0.12. Thus, the observation was either acquired during cross-over, at which point the magnetic equator is visible, or close to the negative magnetic pole. Approximating the magnetosphere as a torus in the plane of the magnetic equator, in the former case it should be detectable as enhanced absorption in the cores of wind-sensitive resonance lines, as the plasma will be occulting the disk. In the latter case it should appear as emission peaking at high velocity, as the plasma is projected on either side of the disk \citep{smithgroote2001}.

\begin{figure}
\centering
\includegraphics[width=8cm]{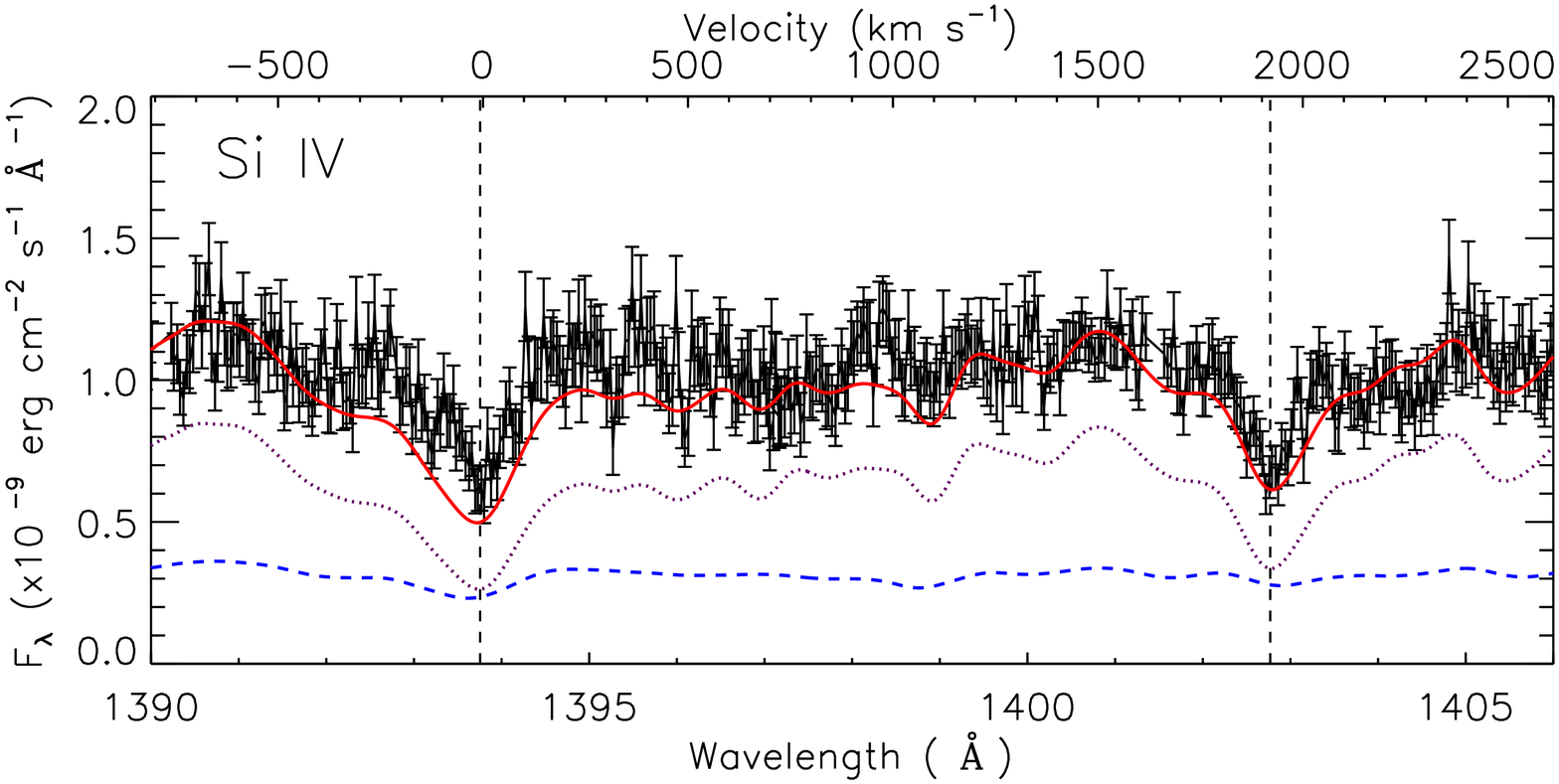}
\includegraphics[width=8cm]{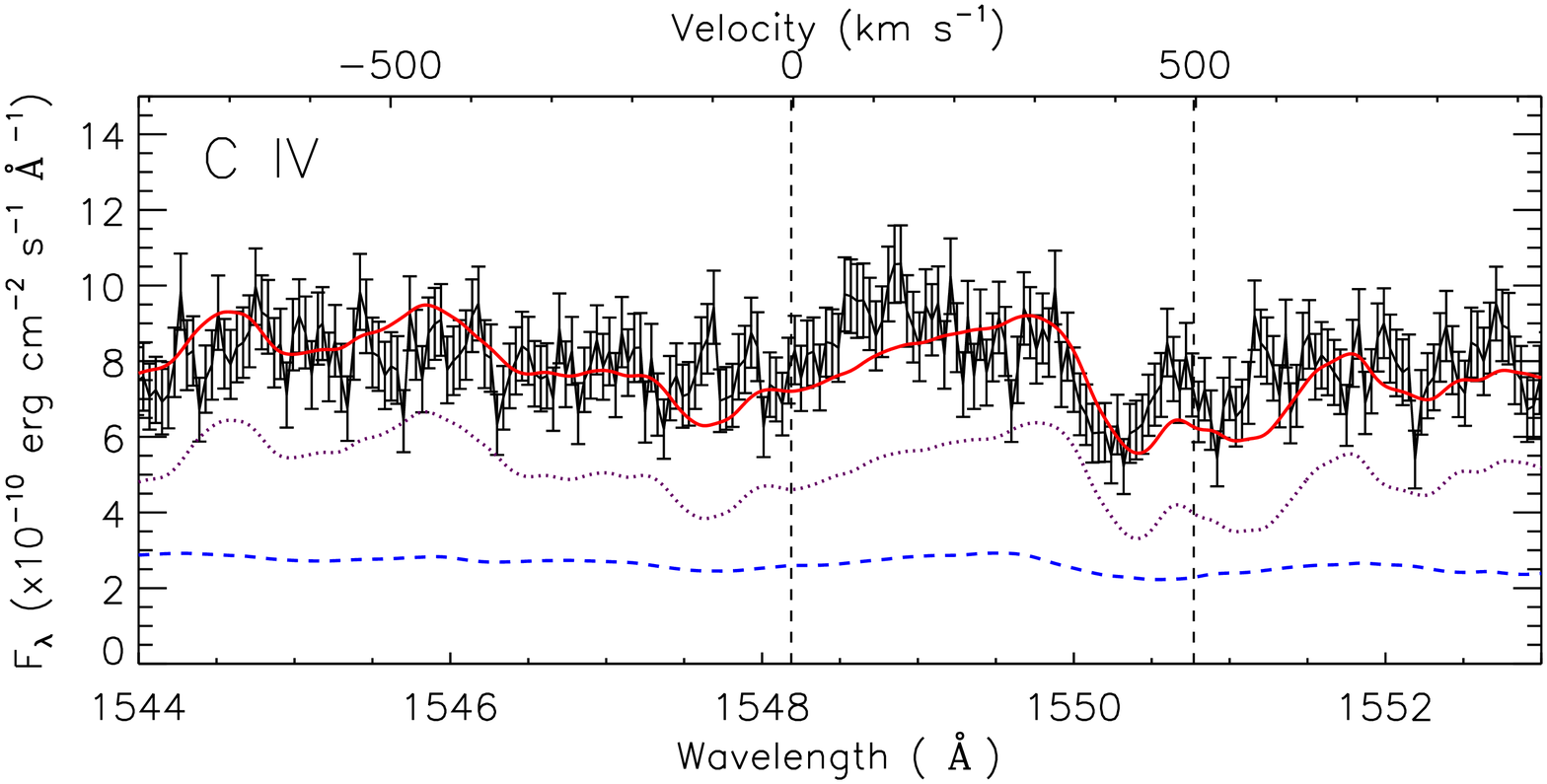}
\includegraphics[width=8cm]{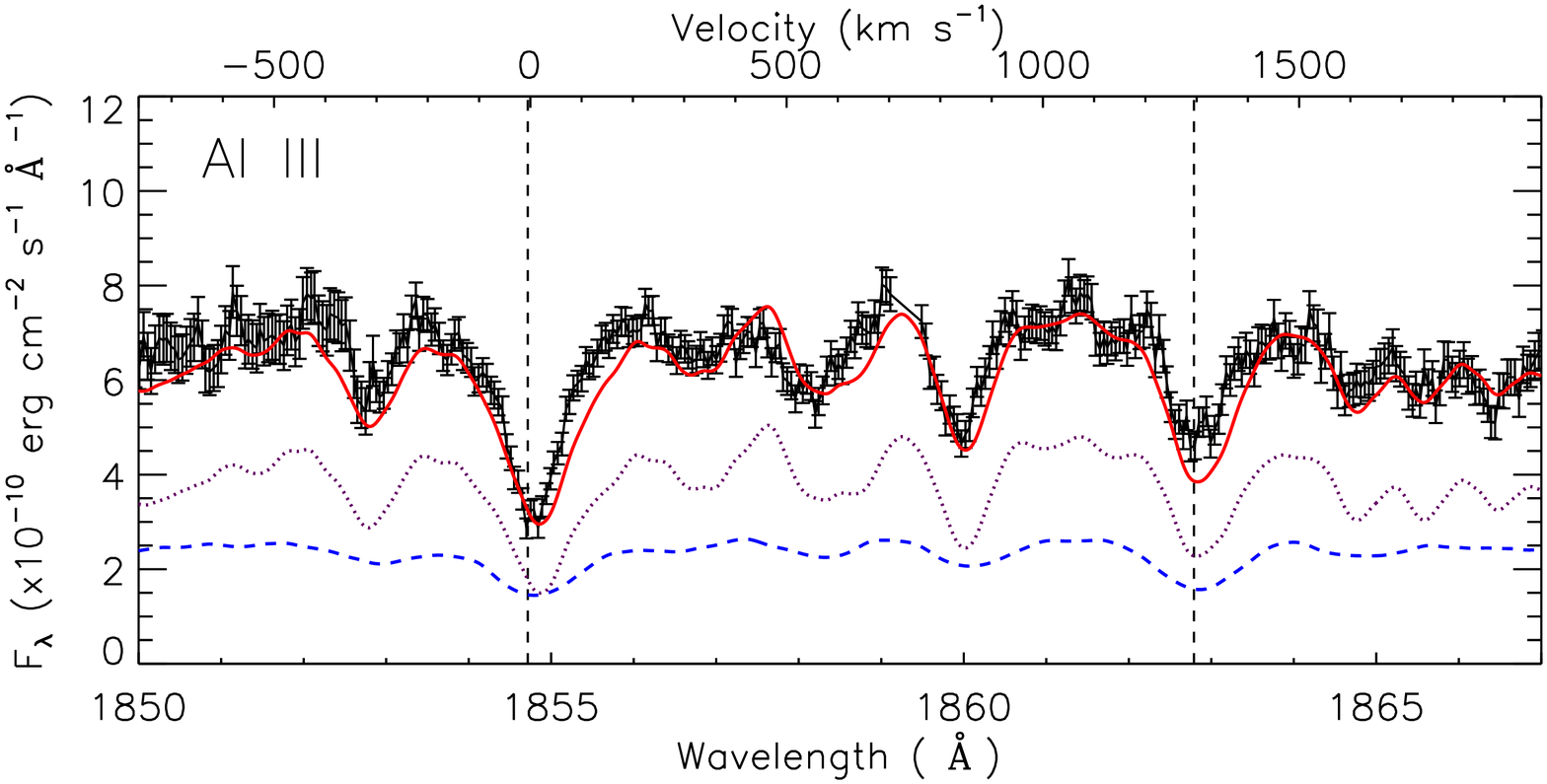}
\caption{The single IUE high-dispersion observation (black), with photospheric model spectra for HR 2948 (dashed blue), HR 2949 (dotted magenta), and the combined system (solid red). The three panels show wind-sensitive UV doublets; short vertical lines indicate the doublets' rest wavelengths.}
\label{iue}
\end{figure}

Fig. \ref{iue} shows the Si {\sc iv}, C {\sc iv}, and Al {\sc iii} doublets in the IUE observation, as compared to a photospheric model created for HR 2948+2949 by combining {\sc zeeman}/{\sc atlas9} models generated using the stellar parameters in Table \ref{params}, line lists from VALD, and the abundances in Table \ref{ab_fvar_tab}. This version of {\sc zeeman} included modifications introduced by \cite{2011AA...528A.132L}, who added extra continuous opacity sources for the UV. The models were rotationally broadened using the \vsini~of each star, and convolved to the FWHM of the IUE data (0.1 \AA). Fluxes were weighted by stellar radii (4.2 \rsun~and 3.7 \rsun~for HR 2948 and HR 2949, respectively), and dereddened assuming the same small reddening ($E(B-V) = 0.04$) used for the photometric analysis in Section 4. 

These three doublets are sensitive to the stellar wind, and are commonly seen in emission in magnetic early-type stars (e.g., \citealt{barker1982, smithgroote2001, oskinova2011, henrichs2013}). In this case, there is no evidence of emission in any of these lines. 

There is thus no direct evidence for the existence of a magnetosphere. However, its expected properties can be investigated based upon HR 2949's magnetic and stellar parameters. The magnetospheric parameters derived below are summarized in Table \ref{mag_prop}.

\subsubsection{Theoretical Characteristics}

The theory of magnetic wind confinement is treated in detail by \cite{bm1997}, \cite{ud2002}, and \cite{ud2008}. The specific case of the rigidly rotating or `centrifugal magnetosphere' found in magnetic mid B-type stars such as HR 2949, is given by \cite{town2005c}. Summaries of magnetic wind confinement theory are provided by \cite{petit2013} and \cite{yakunin2014}, whose basic outline we follow in determining the following magnetospheric parameters: the magnetic wind confinement parameter $\eta_*$ (which gives the ratio of magnetic to wind kinetic energy density); the Alfv\'en radius \ra~(which gives the maximum radius of closed magnetic loops); the rotation parameter $W$ (the ratio of the equatorial to the critical rotational velocities); the Kepler radius \rk~(the radius at which centrifugal force balances gravity); and the spindown timescale $\tau_{\rm J}$.




To determine $\eta_*$ we must first estimate the wind momentum \mdot\vinf, which we evaluate from the stellar parameters determined above (Table \ref{params}), using both the recipe of \cite{vink1999, vink2000, vink2001}, and the mass-loss rates of \cite{krticka2014}, who tailored the calculations specifically for chemically peculiar stars. 

The \citeauthor{vink2001} recipe yields $\log{(\dot{M})} = -9.7^{+0.4}_{-0.3}$ and \vinf$=1070^{+60}_{-80}$~\kms. 

\citeauthor{krticka2014}'s mass-loss rates are highly sensitive to abundances. The elements to which the wind is most sensitive (C, N, O, Si, S, and Fe) are either consistent with solar abundances (C and O), slightly underabundant (S), or only slightly overabundant (N, Si, and Fe). Moreover, while in general increased abundances increase \mdot\vinf, at low \teff~increased Fe abundances {\em decrease} the wind momentum. To make the most use of \citeauthor{krticka2014}'s tables, we therefore interpolated their calculated values of \mdot~and \vinf~according to the abundances in Section 6, then took the mean of the values calculated via the various tables. These are $\log{(\dot{M})}=-10.5^{+0.3}_{-0.4}$~\msun~yr$^{-1}$ and \vinf$=1070^{+814}_{-596}$ \kms. The uncertainty in \mdot~is largely a function of the uncertainties in \teff~and $R_*$, as there is little scatter between results for different elemental tables. The uncertainty in \vinf~reflects the enhancement of this parameter by the Si overabundance. 

Using \citeauthor{vink2001} mass-loss yields $\log{\eta_*}=4.7^{+0.6}_{-0.7}\gg1$, while using \citeauthor{krticka2014} mass-loss gives $\log{\eta_*}=5.5^{+1.2}_{-1.0}$. In either case, $\eta_*\gg1$, therefore the wind is magnetically confined. The Alfv\'en radius is then $R_{\rm A}=15^{+7}_{-5} R_*$ or $R_{\rm A}=23^{+24}_{-9} R_*$, respectively, i.e. in either case $R_{\rm A}>10 R_*$, and HR~2949 is predicted to possess a magnetosphere of significant spatial extent. 






Taking the equatorial rotational velocity, radius, and mass found in Section 4 yields \rk$=3.5^{+1.7}_{-1.2} R_*$ and $W=0.14^{+0.10}_{-0.07}$, respectively. Since \rk$<$\ra, HR 2949 is expected to possess a centrifugal magnetosphere. 

The ratio \rark~provides a dimensionless measure of the magnetosphere's volume. In this case, \rark~spans a wide range of values, from 0.28 to 1.31, with \rark~likely below 1 using \citeauthor{vink2001} mass loss rates and likely above 1 using \citeauthor{krticka2014} mass-loss rates. \cite{petit2013}, who base their analysis upon \citeauthor{vink2001} mass-loss rates, give a value near unity. Our lower value is due to the downward revision of $B_{\rm d}$, and the upward revisions of \rs, $\log{L}$, and \teff, as compared to the values reported by \cite{petit2013}: the weaker magnetic field, and stronger wind, lead to a reduced magnetic confinement volume.

The magnetic wind confinement theory developed by \cite{ud2002} deals with the simplest magnetic topology, i.e. a dipole aligned with the rotational axis. One might therefore wonder if the substantial obliquity of HR 2949's magnetic field with respect to the rotational axis may be the reason for the lack of an observable magnetosphere. However, the semi-analytic Rigidly Rotating Magnetosphere (RRM) formalism developed by \cite{town2005c} is able to predict the circumstellar plasma distributions and associated observational diagnostics for arbitrary values of $\beta$, and has been successful in qualitatively explaining the rotationally modulated H$\alpha$ emission seen in $\delta$ Ori C (B3 Vp, \citealt{leone2010}) and HD 176582 (B5 IV, \citealt{bohl2011}), both of which have $\beta$ close to 90$^\circ$. The obliquity of HR 2949's magnetic field is therefore an unlikely explanation for the lack of an observed magnetosphere.

A major consequence of the coupling between stellar winds and magnetic fields is rapid angular momentum loss, as the corotating plasma acts as an extended moment arm \citep{wd1967, ud2009, petit2013}. From the internal structure models of \cite{claret2004}, the moment of inertia $f\sim0.06$ for a 6 \msun~star at an age of $\sim$11 Myr, with very little variation during its previous evolution. Taking $f$, the mass-loss timescale $\tau_{\rm M}$, and \ra~to be constant, this then yields a spindown timescale $\tau_{\rm J} = 11 \pm 1$ Myr or $32 \pm 13$ Myr, depending on whether the \citeauthor{vink1999} or \citeauthor{krticka2014} mass-loss rates are used. 






We can use $\tau_{\rm J}$ and $W$ to infer the star's spindown age $t_{\rm S}$ \citep{ud2009, petit2013}. However, we note that the assumption that the initial rotational parameter $W_0 = 1.0$, made by \cite{petit2013} in their estimation of the maximum spindown age, is implausible. While magnetic torques are unable to spin down massive magnetic YSOs \citep{rosen2012}, gravitational torques during star formation enforce an upper limit of $W_0 = 0.5$ \citep{lin2011}. Intermediate mass, non-emission line B-type stars such as HR 2949 typically rotate even more slowly than this on the ZAMS, with their probability distributions peaking around $W_0 = 0.2$ \citep{huang2010}. 

\begin{figure}
\centering
\includegraphics[width=8cm]{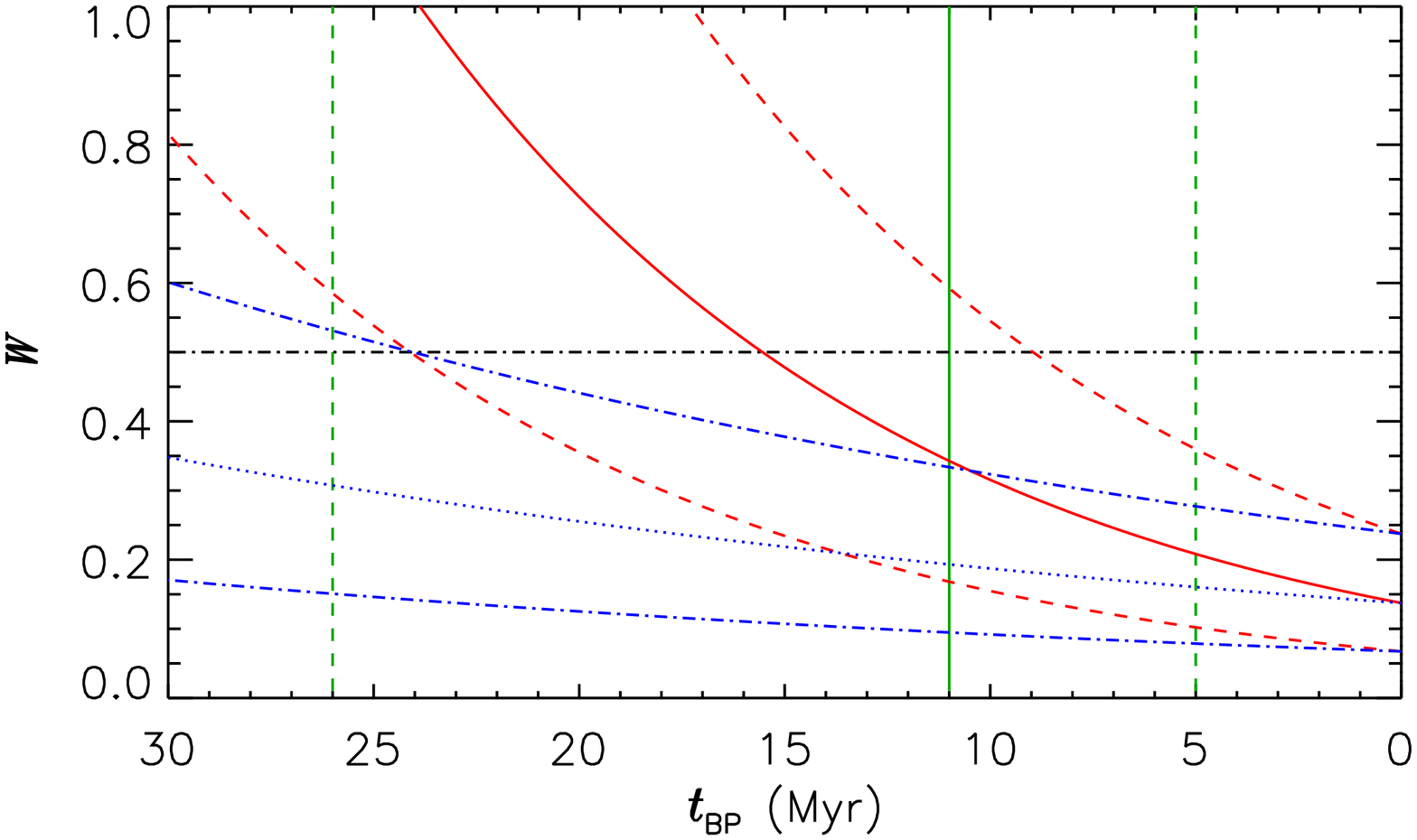}
\caption{The rotation parameter $W$ as a function of the time before present, $t_{\rm BP}$, for magnetospheric parameters calculated with the mass-loss rates of \citeauthor{vink2001} (solid red line) and \citeauthor{krticka2014} (dotted blue line). Vertical solid (green) lines indicate the stellar age, with the uncertainty indicated with dashed lines. Horizontal dash-dotted (blacl) lines $W_0 = 0.5$, the theoretical maximum \citep{lin2011}. $W(t_{\rm BP})$ is calculated such that $W(0) = W_*$. Dashed lines indicate the uncertainty in $W$. In the case of the \citeauthor{vink2001} mass-loss rates, $W_0 = 1$ can be accomodated within the uncertainty in the stellar age. Using \citeauthor{krticka2014} mass-loss rates, the maximum value of $W_0$ compatible with the stellar age is 0.55.}
\label{spindown}
\end{figure}

Rather than try to estimate the maximum spindown time, we instead use the stellar age from Table \ref{params} and the magnetospheric parameters from Table \ref{mag_prop} to calculate $W_0(t_{\rm BP})$:

\begin{equation}\label{wtbp}
W_0 = e^{(t_{\rm BP}/\tau_J + {\rm ln}W)},
\end{equation}

\noindent where $t_{\rm BP}$ is the time before present in Myr. The results are shown in Fig. \ref{spindown}, where equation \ref{wtbp} is solved using $\tau_{\rm J}$ as calculated using both \citeauthor{vink2001} and \citeauthor{krticka2014} mass-loss rates. 

Using \citeauthor{vink2001}'s mass-loss rates, initially critical rotation is compatible with the extreme upper bound on HR 2949's age. If the age is closer to the nominal value of 11 Myr, however, equation \ref{wtbp} predicts $W_0$ closer to the theoretical upper limit of $0.5$. Using \citeauthor{krticka2014}'s mass-loss rates, $W_0 < 0.5$ at almost all plausible ages. 

The calculated spindown time, empirical and theoretical bounds on $W_0$, and the stellar age, are mutually consistent, and could be interpreted as indirect evidence for a magnetosphere. However, the rotation parameter, $W = 0.14^{+0.10}_{-0.07}$, is also consistent with the probability peak observed by \cite{huang2010}, and hence with no spindown having occurred. Therefore, we conclude that while the stellar rotation is consistent with a magnetic spindown scenario, on its own this does not constitute evidence for the presence of a magnetosphere.

\section{Conclusions}

We have analyzed high-resolution spectropolarimetric observations of the optical double system HR 2948+2949. The principal results of this paper are that:
\\\\
\noindent 1) HR 2949 is a member of the class of He-weak, magnetic Bp stars;
\\\\
\noindent 2) Virtually all spectral lines, including H, are variable, and this variability furthermore distorts the magnetic diagnostics;
\\\\
\noindent 3) There is no sign of a magnetospheric signature in the commonly used optical or UV diagnostic lines.
\\\\
Stellar parameters and rotational properties were derived from spectral line profile modeling, photometry, and evolutionary models. A probable misidentification of photometric measurements in the literature leads to different stellar parameters from those previously reported, with HR 2949 being hotter, more luminous, and larger than previously believed.

Placement of HR 2948 on the Hertzsprung-Russell diagram indicates it to be significantly older than HD 2949. We have considered different scenarios to explain this discrepancy, and conclude that the most likely are that either the binary system is not primordial, but formed via gravitational capture, or the stars are not physically associated.

No magnetic field is detected in the single LSD measurement of HR 2948, with a longitudinal magnetic field error bar from the LSD profile of $\sim$35~G. 

Zeeman signatures are detected in all of the 22 observations of HR 2949. We find a rotational period of 1.90871 d, and a dipolar model with $i = 43^{+23\circ}_{-12}$, $\beta = 50\pm16^\circ$, and $B_{\rm d} = 2.4 \pm 0.3$ kG. This magnetic model is subject to revision, as the \bz~curve itself is non-sinusoidal  when measured in both metallic lines and H lines. While a simple offset dipole model produces a better fit to Stokes $V$ and \bz~than a centred dipole, the fit is still unsatisfactory. Without performing Magnetic Doppler Imaging (MDI), departures from sinusoidal behaviour due to flux redistribution cannot be ruled out. The moderate inclination of the rotational axis, moderate \vsini, the large spectropolarimetric data-set, and the detectability of Zeeman splitting in numerous individual spectral lines, make HR 2949 a prime candidate for Doppler and Magnetic Doppler Imaging. MDI has been performed for only a small sample of Bp stars to date, and we still have a poor understanding of the prevalence of departures from dipolar magnetic fields in these stars, and whether these correlate in any way to e.g. the stellar age. 

Our abundance analysis indicates that HR 2949 is a He-weak Bp star, with significant overabundances of Fe-peak and rare earth elements, while HR 2948 is a chemically normal solar-abundance star. Precise abundances have been determined for only a very few Bp stars, making this analysis a valuable contribution. Line-profile variability in He and metallic lines is almost certainly associated with photospheric chemical spots. Weak variability in the Balmer wings may be associated with variable displacement of H by other elements; variations in the strength of line blanketing; or a Lorentz force contribution to the photospheric pressure. Further modeling is required to distinguish between these possibilities. 

While there may be indirect evidence for the presence of a magnetosphere in the stellar rotation, as the maximum spindown age due to angular momentum loss via the magnetized wind is in reasonable agreement with the age inferred from evolutionary tracks, there is no {\em direct} evidence for a magnetosphere, with H Balmer lines and UV resonance lines in absorption. Previously, this absence was something of a mystery, as the star's magnetospheric parameters were quite similar to others with clear magnetospheric signatures. With our reanalysis of the stellar parameters, resulting in a substantially smaller Alfv\'en radius, HR 2949's characteristics are now within the range of other non-emission line magnetic B stars. Determining the conditions under which magnetically confined winds becomes visible requires studies, such as the present one, that determine the stellar and magnetic properties of stars without emission, in addition to examination of stars with emission. Only then can the two groups be clearly separated in parameter space, an essential condition for future clarification of the physics missing from current models.




\section*{Acknowledgements}
This work has made use of the VALD database, operated at Uppsala University, the Institute of Astronomy RAS in Moscow, and the University of Vienna. This work is based on observations obtained at the Canada-France-Hawaii Telescope (CFHT) which is operated by the National Research Council of Canada, the Institut National des Sciences de l'Univers of the Centre National de la Recherche Scientifique of France, and the University of Hawaii, and on observations made with ESO telescopes at the La Silla Observatory under programs 073.C-0337 and 076.C-0164. 

G.A.W. acknowledges Discovery Grant support from the Natural Sciences and Engineering Research Council. CPF acknowledges support from the French ANR grant "Toupies: Towards understanding the spin evolution of stars". RHDT acknowledges support from NASA award NNX12AC72G. MES thanks Paula R. T. Coelho for her assistance in providing model true continuum SEDs.

\bibliography{bib_dat.bib}{}

\newpage

\appendix

\section{Other variable spectral lines}

\begin{figure*}
\centering
\begin{tabular}{cc}
\includegraphics[width=8cm]{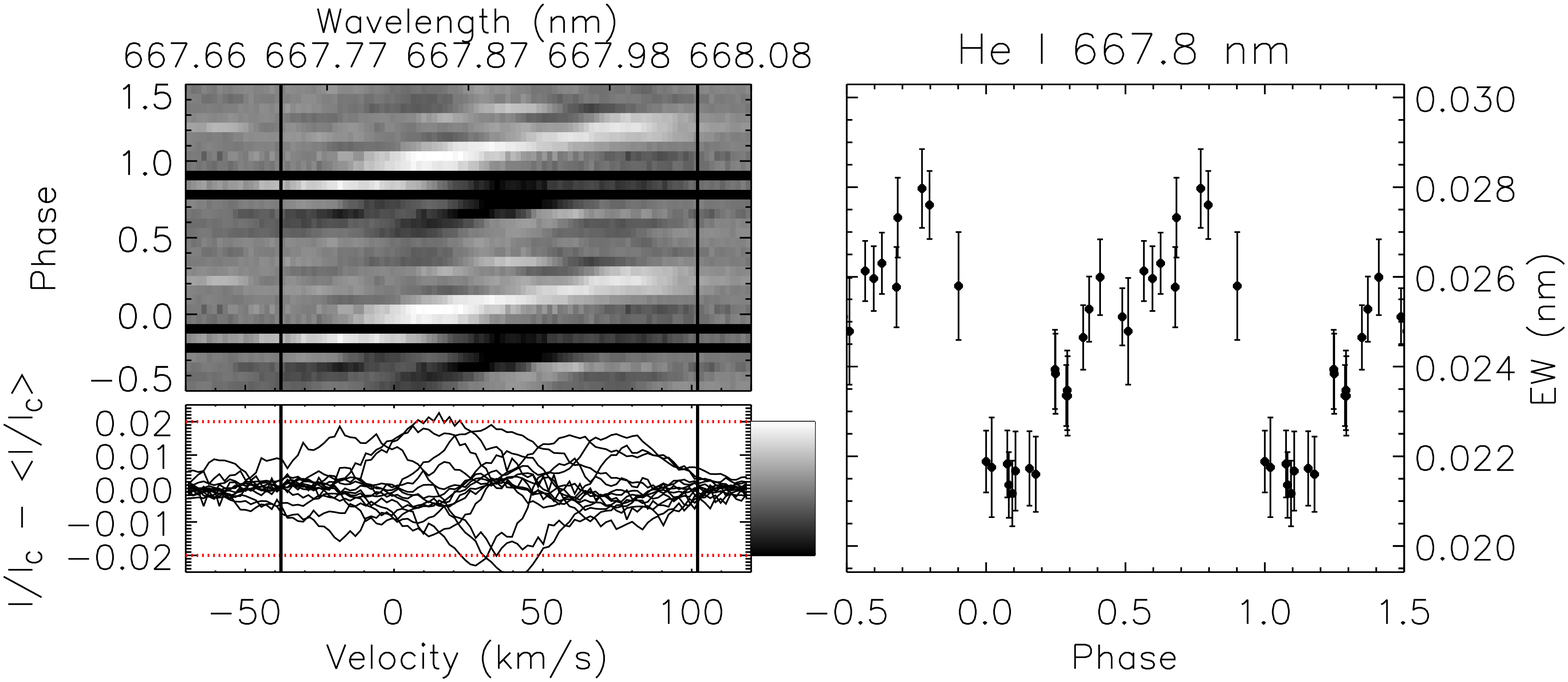} &
\includegraphics[width=8cm]{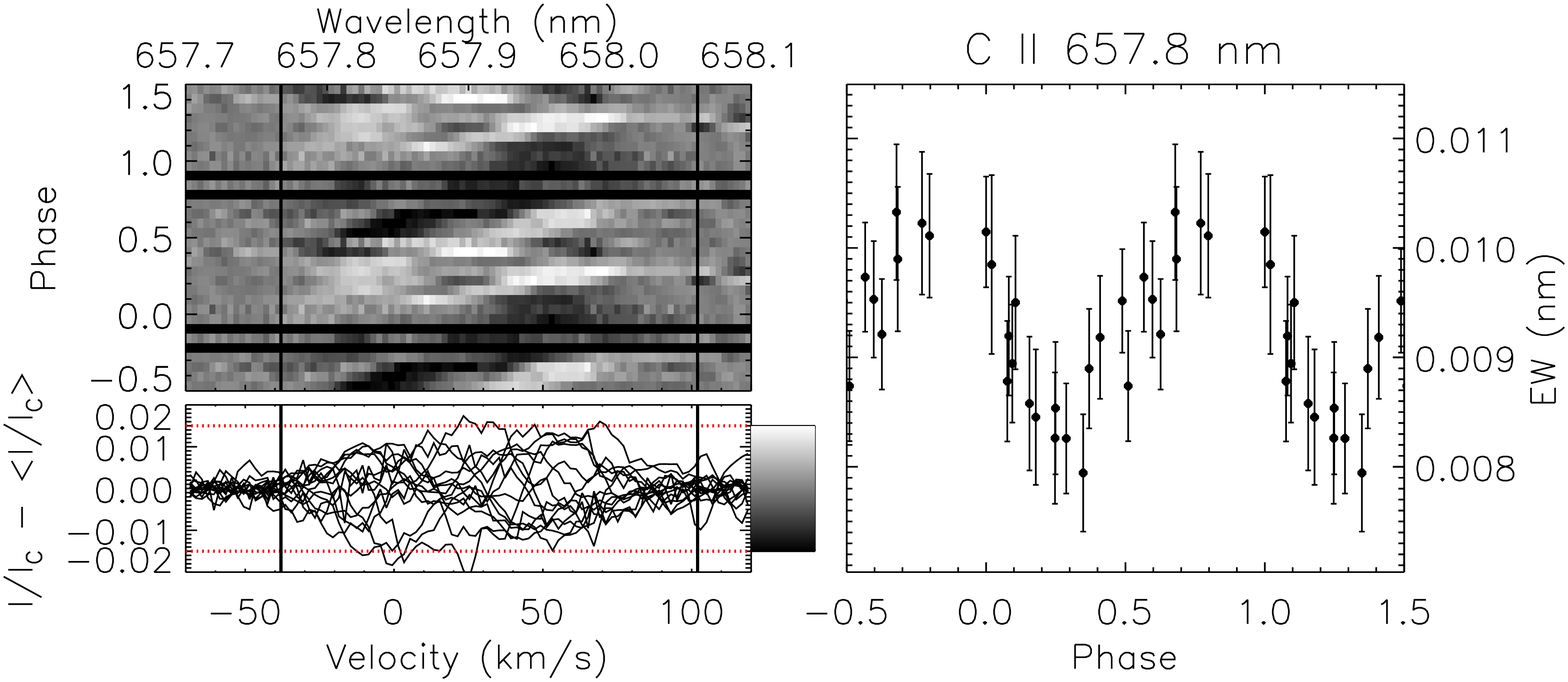} \\
\includegraphics[width=8cm]{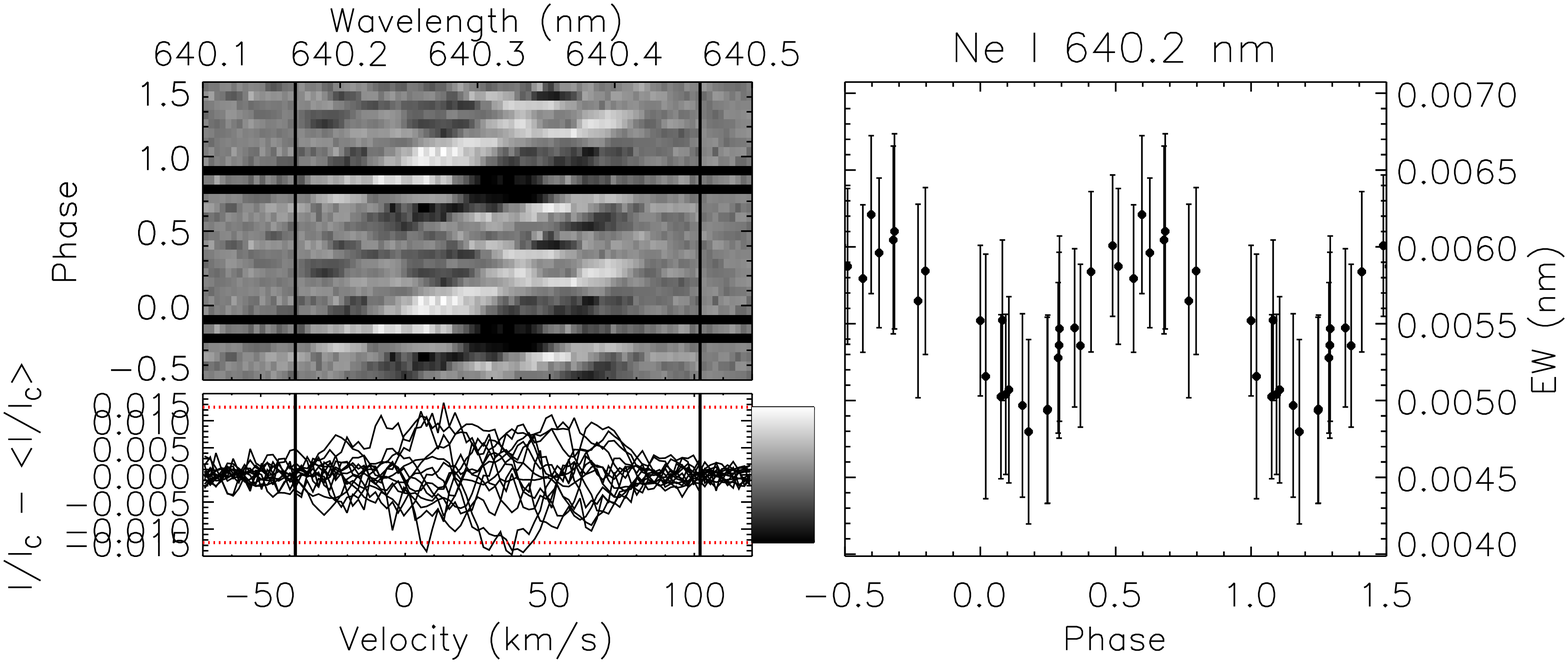} &
\includegraphics[width=8cm]{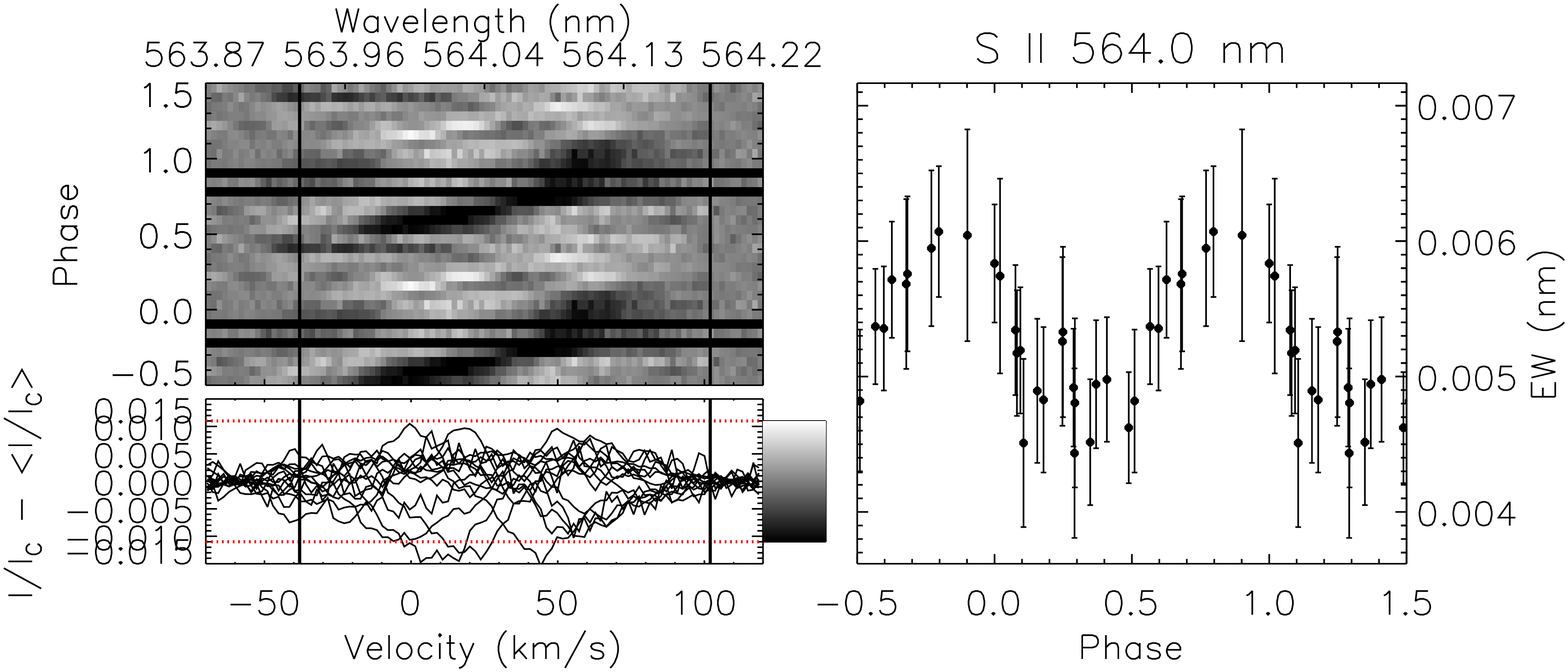} \\
\end{tabular}
\caption{As Fig. \ref{dyn_he}, for lines correlating with He.}
\label{dyn_2}
\end{figure*}

\begin{figure*}
\centering
\begin{tabular}{cc}
\includegraphics[width=8cm]{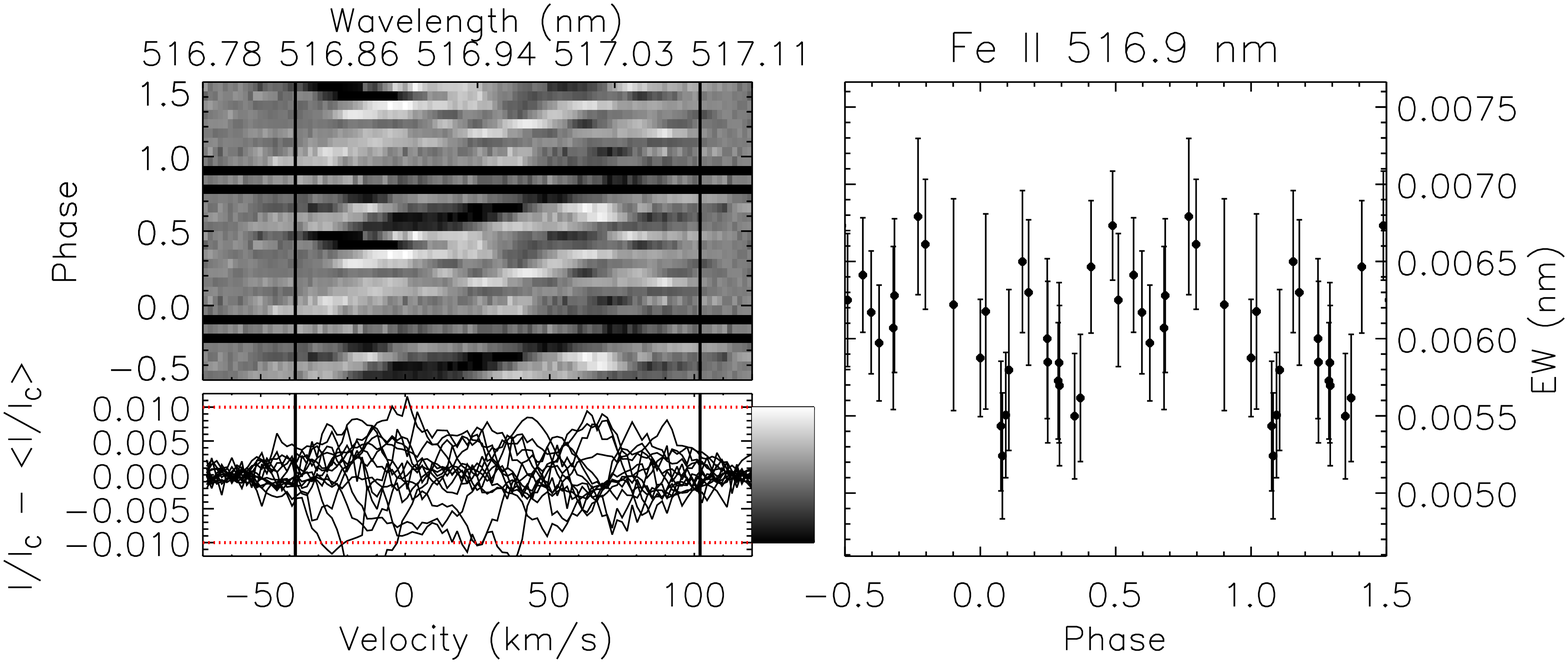} &
\includegraphics[width=8cm]{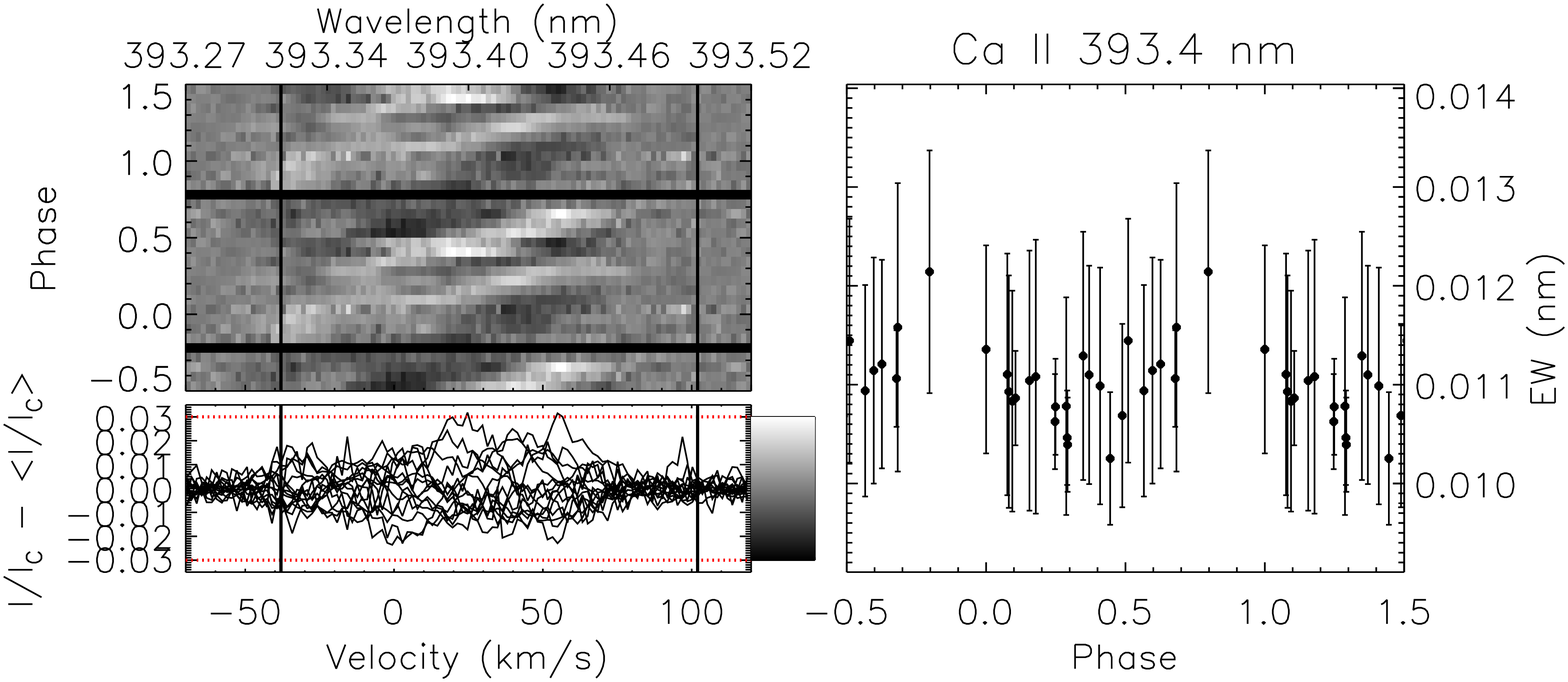} \\
\includegraphics[width=8cm]{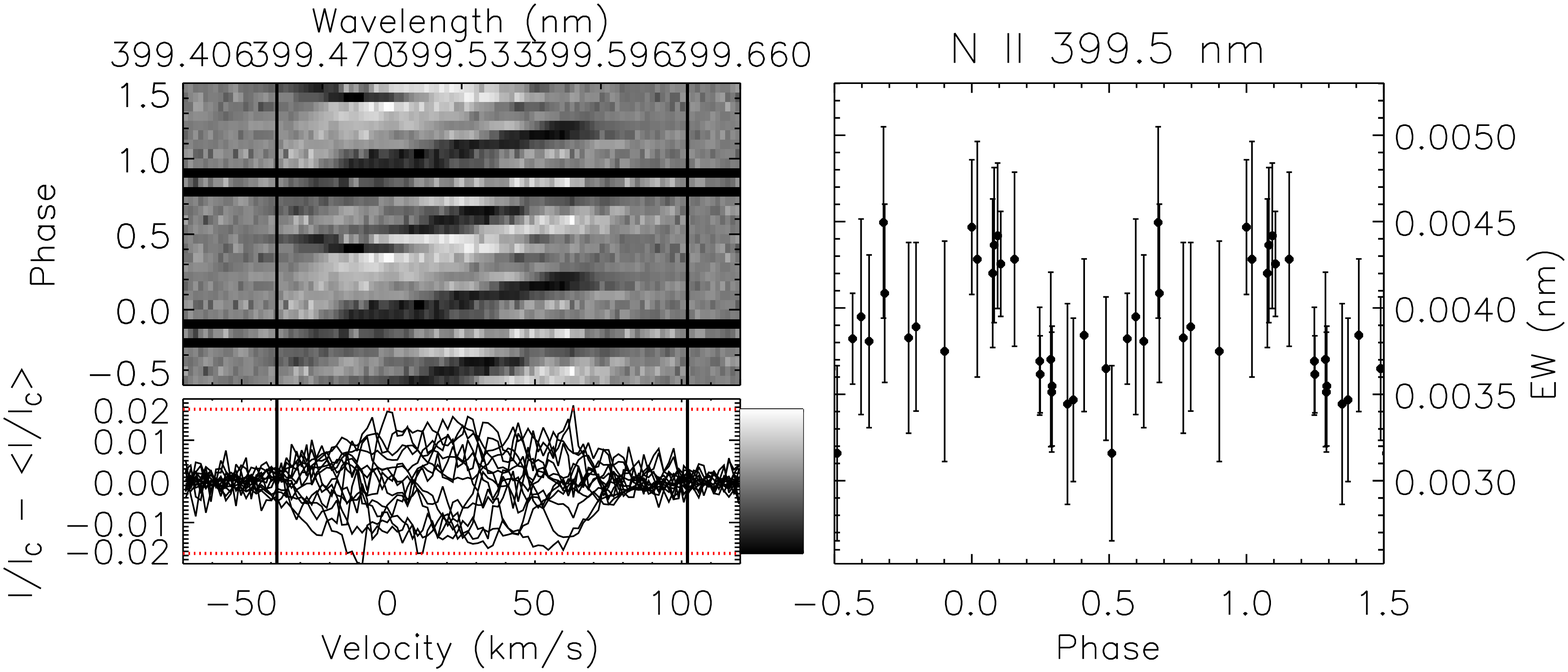} &
\includegraphics[width=8cm]{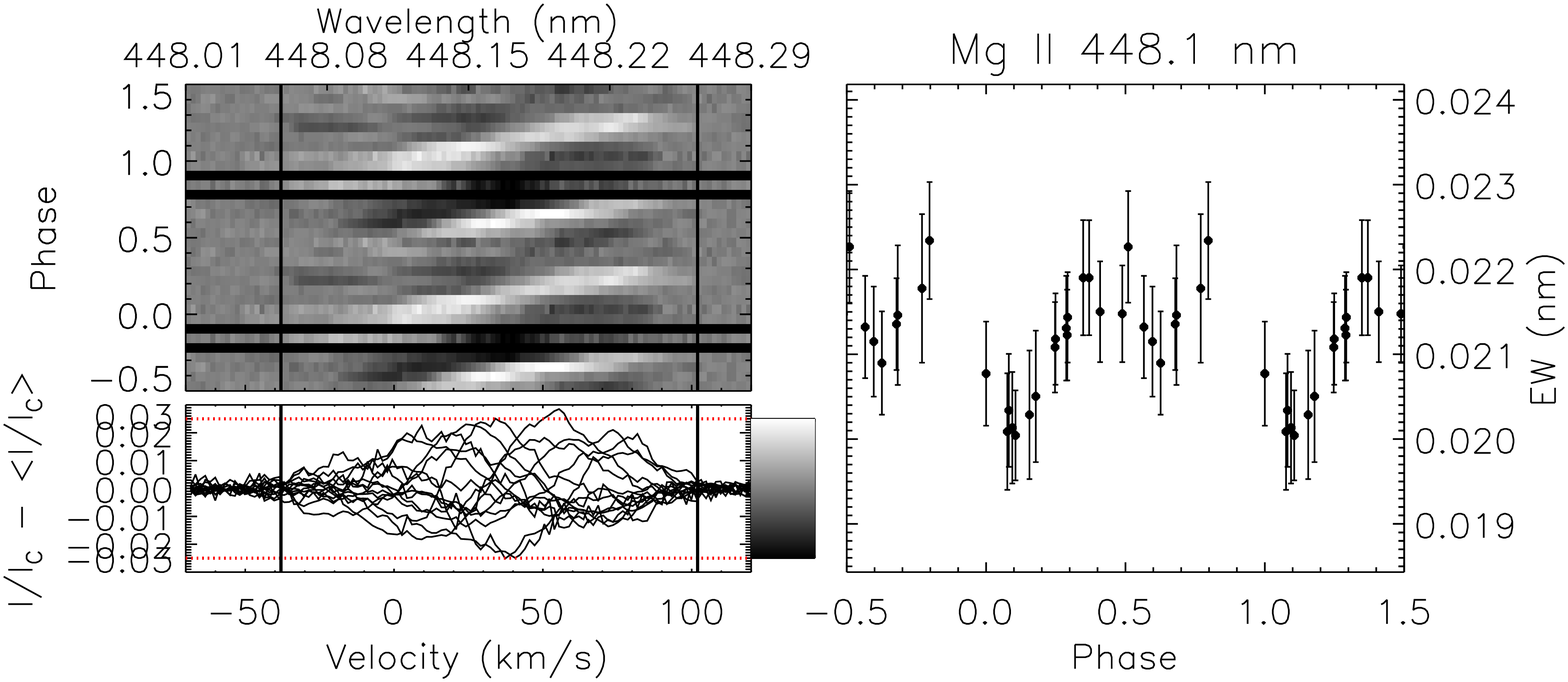} \\
\includegraphics[width=8cm]{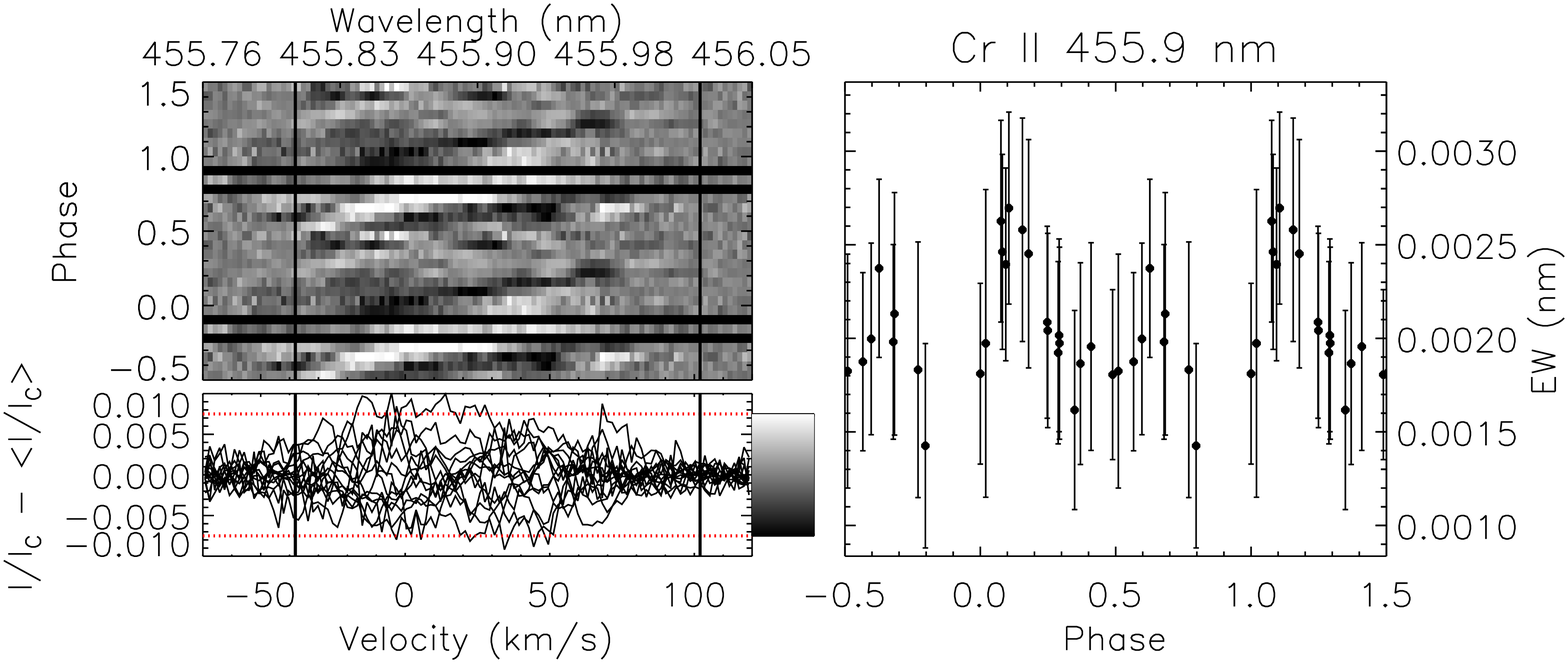} &
\includegraphics[width=8cm]{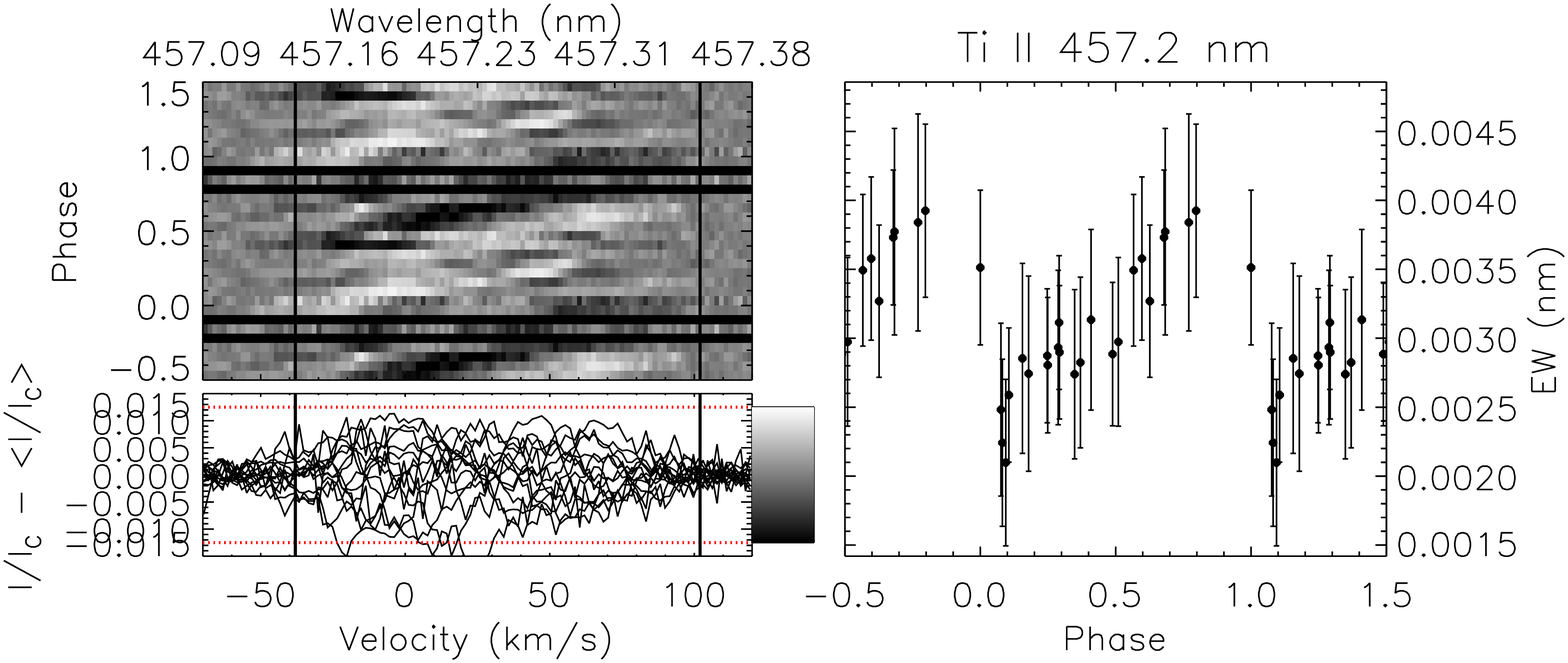} \\
\includegraphics[width=8cm]{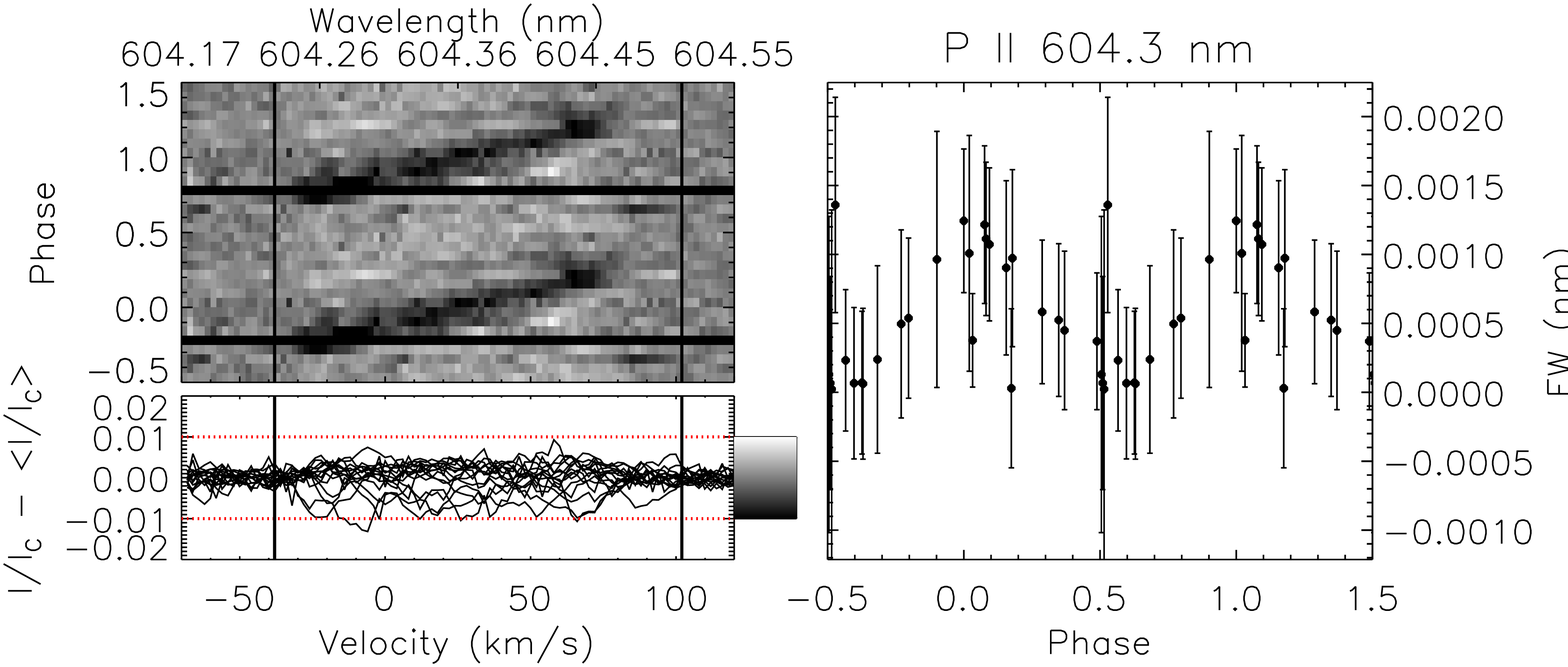} & 
\includegraphics[width=8cm]{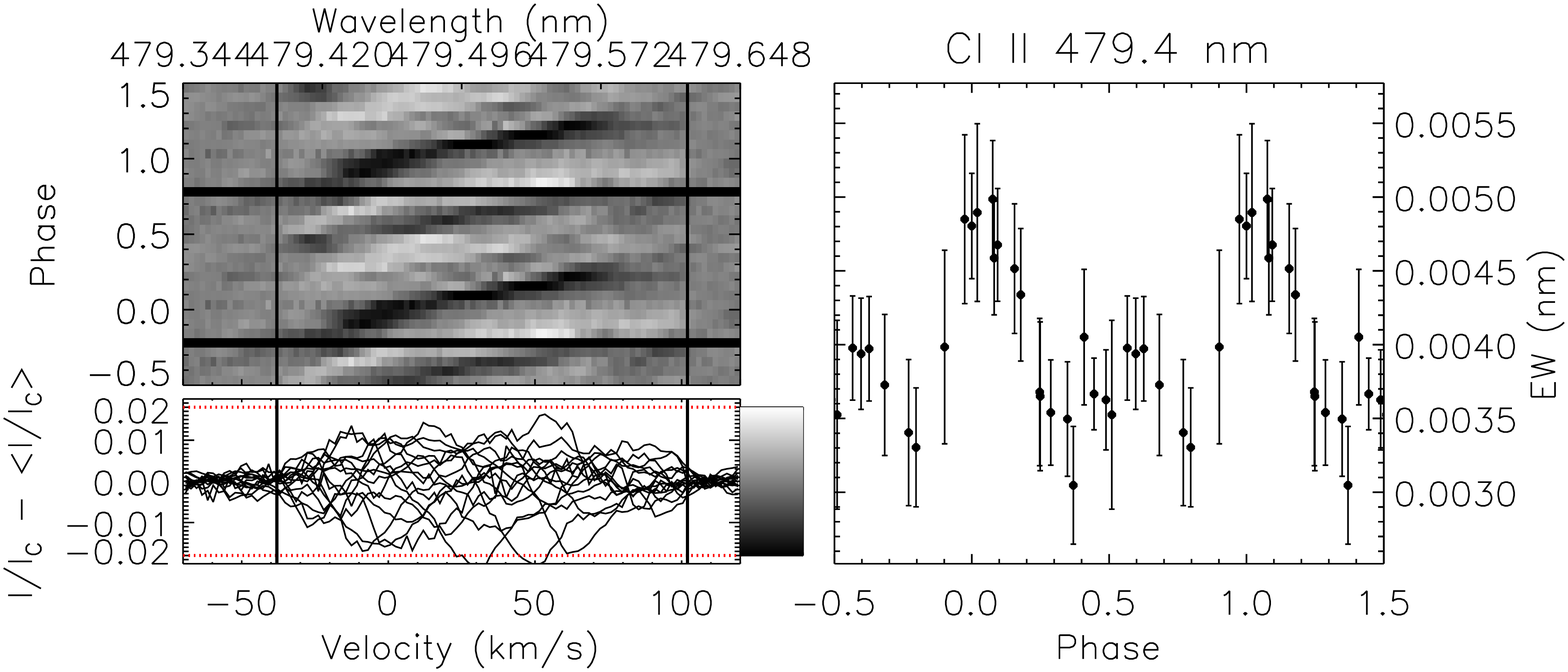} \\
\end{tabular}
\caption{As Fig. \ref{dyn_he}, for lines exhibiting complex variations that neither correlate nor anti-correlate with He.}
\label{dyn_3}
\end{figure*}

\section{Tabulated Longitudinal Magnetic Field Measurements}

\begin{table*}
\centering
\caption[Longitudinal magnetic field measurements]{Longitudinal magnetic field measurements \bz~and diagnostic null measurements \nz~from LSD profiles. Measurements are given in order of acquistion; sets of measurements are ordered by decreasing number of lines included in the mask, given beneath the mask name.}
\resizebox{18 cm}{!}{
\begin{tabular}{rrrrrrrrrrr}
\hline
\hline 
& \multicolumn{2}{c}{Metals} &  \multicolumn{2}{c}{Fe} &  \multicolumn{2}{c}{He} &   \multicolumn{2}{c}{S} &  \multicolumn{2}{c}{Si} \\
 & \multicolumn{2}{c}{190} & \multicolumn{2}{c}{ 50} & \multicolumn{2}{c}{ 47} & \multicolumn{2}{c}{ 40} & \multicolumn{2}{c}{ 23} \\
Phase & \bz~(G) & \nz~(G) & \bz~(G) & \nz~(G) & \bz~(G) & \nz~(G) & \bz~(G) & \nz~(G) & \bz~(G) & \nz~(G) \\
\hline
0.020 &  446$\pm$  45 &  -11$\pm$  45 &  547$\pm$  89 &   19$\pm$  89 & -190$\pm$ 102 &   -6$\pm$ 102 &  554$\pm$ 188 &  -16$\pm$ 188 &  610$\pm$  89 &   78$\pm$  89\\
0.076 &  314$\pm$  30 &   29$\pm$  30 &  514$\pm$  56 &    1$\pm$  56 & -219$\pm$  67 &   -8$\pm$  67 &  130$\pm$ 136 &   15$\pm$ 136 &  420$\pm$  55 &   53$\pm$  55\\
0.566 & -853$\pm$  29 &   14$\pm$  29 & -769$\pm$  58 &    5$\pm$  58 & -809$\pm$  47 &  -23$\pm$  47 & -862$\pm$ 129 &  -10$\pm$ 129 & -925$\pm$  58 &  -57$\pm$  58\\
1.000 &  439$\pm$  26 &   -4$\pm$  26 &  672$\pm$  51 &  -42$\pm$  51 & -219$\pm$  57 &    9$\pm$  57 &  307$\pm$ 110 &   -9$\pm$ 110 &  660$\pm$  53 &  -23$\pm$  53\\
0.082 &  243$\pm$  28 &  -31$\pm$  28 &  375$\pm$  53 &  -18$\pm$  53 & -275$\pm$  59 &    5$\pm$  59 &  198$\pm$ 130 & -115$\pm$ 130 &  351$\pm$  51 &    6$\pm$  51\\
0.597 & -855$\pm$  31 &   -5$\pm$  31 & -736$\pm$  64 &    8$\pm$  64 & -760$\pm$  58 &   10$\pm$  58 & -888$\pm$ 132 &    7$\pm$ 132 & -919$\pm$  62 &   45$\pm$  62\\
0.094 &  227$\pm$  28 &   38$\pm$  28 &  323$\pm$  55 &   29$\pm$  55 & -283$\pm$  62 &   21$\pm$  62 &  430$\pm$ 125 &  162$\pm$ 125 &  421$\pm$  51 &  -18$\pm$  51\\
0.626 & -848$\pm$  28 &  -25$\pm$  28 & -840$\pm$  55 &  -93$\pm$  55 & -832$\pm$  45 &    8$\pm$  45 & -891$\pm$ 123 &   69$\pm$ 123 & -861$\pm$  57 &  -19$\pm$  57\\
0.178 &  -95$\pm$  34 &   49$\pm$  34 &   82$\pm$  71 &   14$\pm$  71 & -334$\pm$  64 &   19$\pm$  64 &  -74$\pm$ 179 & -133$\pm$ 179 &   17$\pm$  62 &   98$\pm$  62\\
0.683 & -777$\pm$  39 &   23$\pm$  39 & -764$\pm$  74 &  103$\pm$  74 & -847$\pm$  58 &  -29$\pm$  58 & -978$\pm$ 157 & -140$\pm$ 157 & -706$\pm$  81 &  -24$\pm$  81\\
0.156 &    9$\pm$  33 &    3$\pm$  33 &   87$\pm$  66 &  -76$\pm$  66 & -316$\pm$  74 &   45$\pm$  74 &  358$\pm$ 172 &  -41$\pm$ 172 &   66$\pm$  62 &  -24$\pm$  62\\
0.770 & -580$\pm$  41 &  -38$\pm$  41 & -605$\pm$  76 &  -81$\pm$  76 & -643$\pm$  61 &   32$\pm$  61 & -665$\pm$ 164 &   56$\pm$ 164 & -429$\pm$  92 &  -12$\pm$  92\\
0.288 & -489$\pm$  28 &   42$\pm$  28 & -384$\pm$  66 &  -33$\pm$  66 & -526$\pm$  47 &   28$\pm$  47 & -428$\pm$ 150 &  -69$\pm$ 150 & -516$\pm$  52 &  -16$\pm$  52\\
0.797 & -369$\pm$  32 &   48$\pm$  32 & -226$\pm$  60 &   17$\pm$  60 & -634$\pm$  53 &  -34$\pm$  53 & -481$\pm$ 124 &   88$\pm$ 124 & -227$\pm$  76 &    5$\pm$  76\\
0.348 & -630$\pm$  31 &   31$\pm$  31 & -579$\pm$  75 &   59$\pm$  75 & -605$\pm$  56 &   17$\pm$  56 & -554$\pm$ 166 &  -42$\pm$ 166 & -632$\pm$  59 &   18$\pm$  59\\
0.901 &  179$\pm$  52 &   40$\pm$  52 &  407$\pm$  97 &  145$\pm$  97 & -276$\pm$  86 &   47$\pm$  86 &  296$\pm$ 213 &   27$\pm$ 213 &  188$\pm$ 113 & -196$\pm$ 113\\
0.489 & -712$\pm$  31 &   -5$\pm$  31 & -645$\pm$  63 &   66$\pm$  63 & -737$\pm$  52 &  -21$\pm$  52 & -435$\pm$ 145 &   93$\pm$ 145 & -810$\pm$  60 &   11$\pm$  60\\
0.370 & -636$\pm$  37 &  -26$\pm$  37 & -682$\pm$  87 &   80$\pm$  87 & -586$\pm$  64 &  -78$\pm$  64 & -520$\pm$ 184 & -240$\pm$ 184 & -665$\pm$  70 &  -64$\pm$  70\\
0.510 & -866$\pm$  54 &   24$\pm$  54 & -787$\pm$ 112 &  -58$\pm$ 112 & -685$\pm$  87 &   30$\pm$  87 & -568$\pm$ 248 &  145$\pm$ 248 & -936$\pm$ 101 &  -31$\pm$ 101\\
0.411 & -739$\pm$  87 &   91$\pm$  87 & -636$\pm$ 186 &  163$\pm$ 186 & -654$\pm$ 144 &   10$\pm$ 144 & -471$\pm$ 428 &  -13$\pm$ 428 & -971$\pm$ 168 &   26$\pm$ 168\\
0.974 &  400$\pm$  49 &   56$\pm$  49 &  764$\pm$  97 &   91$\pm$  97 & -119$\pm$ 101 &   11$\pm$ 101 &   76$\pm$ 241 &  143$\pm$ 241 &  533$\pm$ 102 &   18$\pm$ 102\\
0.445 & -672$\pm$  22 &   20$\pm$  22 & -667$\pm$  51 &   70$\pm$  51 & -683$\pm$  38 &   13$\pm$  38 & -709$\pm$ 131 &   31$\pm$ 131 & -759$\pm$  46 &   21$\pm$  46\\
\hline
\hline
 &      \multicolumn{2}{c}{O} &   \multicolumn{2}{c}{C} &   \multicolumn{2}{c}{N} &  \multicolumn{2}{c}{Mg} &  \multicolumn{2}{c}{Ne} \\
 & \multicolumn{2}{c}{ 21} & \multicolumn{2}{c}{ 13} & \multicolumn{2}{c}{ 10} & \multicolumn{2}{c}{  6} & \multicolumn{2}{c}{  3} \\
Phase & \bz~(G) & \nz~(G) & \bz~(G) & \nz~(G) & \bz~(G) & \nz~(G) & \bz~(G) & \nz~(G) & \bz~(G) & \nz~(G) \\
\hline
0.020 &    19$\pm$  158 &    75$\pm$  158 &   492$\pm$  135 &   -90$\pm$  135 &  1026$\pm$  293 &  -125$\pm$  293 &    47$\pm$  217 &   -21$\pm$  217 &  -338$\pm$  308 &   232$\pm$  308\\
0.076 &   -76$\pm$  123 &    31$\pm$  123 &   344$\pm$  102 &    -7$\pm$  102 &   815$\pm$  201 &     2$\pm$  201 &   -59$\pm$  134 &  -110$\pm$  134 &  -196$\pm$  229 &  -225$\pm$  229\\
0.566 &  -761$\pm$  136 &    65$\pm$  136 & -1000$\pm$   85 &   -59$\pm$   85 & -1188$\pm$  263 &   239$\pm$  263 &  -958$\pm$  129 &   -54$\pm$  129 &  -556$\pm$  155 &  -170$\pm$  155\\
1.000 &    14$\pm$   97 &  -117$\pm$   97 &   491$\pm$   72 &   123$\pm$   72 &   807$\pm$  171 &   169$\pm$  171 &    89$\pm$  133 &  -128$\pm$  133 &  -460$\pm$  199 &  -289$\pm$  199\\
0.082 &   -76$\pm$  116 &   -96$\pm$  116 &   211$\pm$   84 &   -21$\pm$   84 &  1136$\pm$  196 &   121$\pm$  196 &  -113$\pm$  141 &   -55$\pm$  141 &  -521$\pm$  210 &  -131$\pm$  210\\
0.597 &  -631$\pm$  125 &  -119$\pm$  125 & -1024$\pm$  101 &   -41$\pm$  101 & -1398$\pm$  264 &    84$\pm$  264 &  -802$\pm$  142 &  -167$\pm$  142 &  -852$\pm$  188 &   139$\pm$  188\\
0.094 &  -151$\pm$  107 &   -22$\pm$  107 &   139$\pm$   81 &    -5$\pm$   81 &   747$\pm$  189 &    75$\pm$  189 &  -211$\pm$  146 &    25$\pm$  146 &  -214$\pm$  218 &    77$\pm$  218\\
0.626 &  -755$\pm$  133 &   155$\pm$  133 & -1085$\pm$   77 &   -11$\pm$   77 & -1275$\pm$  239 &   -68$\pm$  239 &  -812$\pm$  138 &    87$\pm$  138 &  -615$\pm$  175 &  -116$\pm$  175\\
0.178 &  -382$\pm$  163 &   216$\pm$  163 &  -423$\pm$  105 &    69$\pm$  105 &   521$\pm$  289 &    33$\pm$  289 &  -537$\pm$  158 &   -19$\pm$  158 &  -596$\pm$  268 &   215$\pm$  268\\
0.683 &  -690$\pm$  145 &   -38$\pm$  145 & -1001$\pm$  110 &   -31$\pm$  110 & -1382$\pm$  320 &   281$\pm$  320 &  -662$\pm$  170 &   -27$\pm$  170 &  -791$\pm$  194 &    27$\pm$  194\\
0.156 &   -83$\pm$  135 &    56$\pm$  135 &   -91$\pm$  109 &    17$\pm$  109 &   487$\pm$  272 &   -34$\pm$  272 &  -501$\pm$  160 &   160$\pm$  160 &  -306$\pm$  245 &  -401$\pm$  245\\
0.770 &  -564$\pm$  124 &   -82$\pm$  124 &  -689$\pm$  116 &    28$\pm$  116 &  -429$\pm$  318 &   -80$\pm$  318 &  -496$\pm$  168 &   -53$\pm$  168 &  -767$\pm$  190 &   -97$\pm$  190\\
0.288 &  -622$\pm$  137 &    -9$\pm$  137 &  -658$\pm$   87 &   -17$\pm$   87 &  -427$\pm$  237 &    13$\pm$  237 &  -712$\pm$  128 &   147$\pm$  128 &  -525$\pm$  198 &   129$\pm$  198\\
0.797 &  -489$\pm$  107 &    -1$\pm$  107 &  -407$\pm$   77 &    75$\pm$   77 &  -248$\pm$  207 &   -35$\pm$  207 &  -236$\pm$  140 &    41$\pm$  140 &  -558$\pm$  186 &    18$\pm$  186\\
0.348 &  -657$\pm$  131 &    14$\pm$  131 &  -763$\pm$   97 &   -11$\pm$   97 &  -832$\pm$  290 &  -203$\pm$  290 &  -848$\pm$  132 &    86$\pm$  132 &  -619$\pm$  176 &  -151$\pm$  176\\
0.901 &  -219$\pm$  150 &   164$\pm$  150 &   213$\pm$  146 &   256$\pm$  146 &   638$\pm$  340 &   103$\pm$  340 &   379$\pm$  224 &   111$\pm$  224 &  -503$\pm$  310 &    19$\pm$  310\\
0.489 &  -708$\pm$  117 &    20$\pm$  117 &  -753$\pm$   89 &     6$\pm$   89 &  -756$\pm$  244 &  -187$\pm$  244 &  -865$\pm$  134 &  -152$\pm$  134 &  -709$\pm$  180 &    96$\pm$  180\\
0.370 &  -739$\pm$  136 &  -109$\pm$  136 &  -880$\pm$  103 &  -105$\pm$  103 &  -484$\pm$  306 &   143$\pm$  306 &  -869$\pm$  144 &    18$\pm$  144 &  -464$\pm$  222 &   184$\pm$  222\\
0.510 &  -949$\pm$  252 &    -1$\pm$  252 &  -989$\pm$  156 &    -9$\pm$  156 &  -707$\pm$  459 &    10$\pm$  459 &  -810$\pm$  199 &   -75$\pm$  199 &  -737$\pm$  335 &   -83$\pm$  335\\
0.411 &  -756$\pm$  349 &   345$\pm$  349 &  -901$\pm$  267 &  -300$\pm$  267 &  -122$\pm$  585 &   846$\pm$  585 &  -456$\pm$  326 &  -215$\pm$  326 &  -539$\pm$  473 &  -328$\pm$  473\\
0.974 &   -16$\pm$  127 &   107$\pm$  127 &   791$\pm$  162 &    44$\pm$  162 &  1014$\pm$  411 &   734$\pm$  411 &    -4$\pm$  201 &   123$\pm$  201 &  -149$\pm$  260 &   220$\pm$  260\\
0.445 &  -677$\pm$   93 &   -55$\pm$   93 &  -861$\pm$   70 &   -15$\pm$   70 &  -895$\pm$  214 &   -70$\pm$  214 &  -756$\pm$   90 &    79$\pm$   90 &  -608$\pm$   99 &   -10$\pm$   99\\
\hline
\hline
\end{tabular}
}
\label{bz_1}
\end{table*}

\begin{table*}
\centering
\caption[Longitudinal magnetic field measurements]{Longitudinal magnetic field measurements \bz~and diagnostic null measurements \nz~from individual lines, arranged in order of increasing rest wavelength.}
\resizebox{18 cm}{!}{
\begin{tabular}{rrrrrrrrrrr}
\hline
\hline 
 & \multicolumn{2}{c}{    Ca II} & \multicolumn{2}{c}{   Si III} & \multicolumn{2}{c}{    Ti II} & \multicolumn{2}{c}{    Fe II} & \multicolumn{2}{c}{H$\alpha$} \\
$\lambda_0$ & \multicolumn{2}{c}{393.3663} & \multicolumn{2}{c}{455.2622} & \multicolumn{2}{c}{457.1968} & \multicolumn{2}{c}{516.9000} & \multicolumn{2}{c}{656.2797} \\
$g$ & \multicolumn{2}{c}{1.170} & \multicolumn{2}{c}{1.250} & \multicolumn{2}{c}{1.572} & \multicolumn{2}{c}{1.330} & \multicolumn{2}{c}{1.000} \\
\hline
Phase & \bz~(G) & \nz~(G) & \bz~(G) & \nz~(G) & \bz~(G) & \nz~(G) & \bz~(G) & \nz~(G) & \bz~(G) & \nz~(G) \\
0.020 &    101$\pm$   665 &    129$\pm$   665 &    759$\pm$   289 &    113$\pm$   289 &    168$\pm$   517 &    386$\pm$   517 &    726$\pm$   228 &    -21$\pm$   228 &    265$\pm$    46 &     74$\pm$    46\\
0.076 &     69$\pm$   373 &   -186$\pm$   373 &    860$\pm$   154 &     46$\pm$   154 &    268$\pm$   449 &    319$\pm$   449 &    151$\pm$   185 &    202$\pm$   185 &    183$\pm$    29 &    -51$\pm$    29\\
0.566 &   -864$\pm$   382 &   -411$\pm$   382 &   -793$\pm$   182 &    -22$\pm$   182 &   -771$\pm$   345 &   -181$\pm$   345 &   -949$\pm$   112 &   -150$\pm$   112 &   -733$\pm$    26 &    -70$\pm$    26\\
1.000 &    -75$\pm$   310 &    133$\pm$   310 &    678$\pm$   160 &     19$\pm$   160 &    104$\pm$   339 &   -333$\pm$   339 &    448$\pm$   154 &   -138$\pm$   154 &    197$\pm$    31 &      1$\pm$    31\\
0.082 &    163$\pm$   402 &    552$\pm$   402 &    440$\pm$   183 &    105$\pm$   183 &   1100$\pm$   459 &  -1367$\pm$   459 &    106$\pm$   137 &   -190$\pm$   137 &    116$\pm$    29 &    -49$\pm$    29\\
0.597 &   -555$\pm$   354 &    -77$\pm$   354 &   -863$\pm$   183 &   -198$\pm$   183 &   -573$\pm$   463 &    191$\pm$   463 &   -742$\pm$   115 &     82$\pm$   115 &   -708$\pm$    41 &     38$\pm$    41\\
0.094 &   -343$\pm$   341 &    122$\pm$   341 &    524$\pm$   191 &   -260$\pm$   191 &    350$\pm$   648 &    239$\pm$   648 &   -357$\pm$   175 &   -172$\pm$   175 &      0$\pm$    31 &      7$\pm$    31\\
0.626 &  -1329$\pm$   370 &   -230$\pm$   370 &   -944$\pm$   201 &    216$\pm$   201 &    -58$\pm$   274 &   -576$\pm$   274 &   -823$\pm$   111 &     -8$\pm$   111 &   -661$\pm$    36 &    -43$\pm$    36\\
0.178 &   -545$\pm$   456 &    386$\pm$   456 &    393$\pm$   155 &      5$\pm$   155 &    127$\pm$   499 &     -6$\pm$   499 &     60$\pm$   151 &    164$\pm$   151 &   -165$\pm$    41 &     26$\pm$    41\\
0.683 &   -971$\pm$   476 &    292$\pm$   476 &   -687$\pm$   267 &     66$\pm$   267 &    -41$\pm$   434 &   -272$\pm$   434 &   -690$\pm$   158 &    -38$\pm$   158 &   -500$\pm$    41 &    -24$\pm$    41\\
0.156 &   -161$\pm$   399 &    -35$\pm$   399 &    317$\pm$   169 &   -298$\pm$   169 &    264$\pm$   400 &   -445$\pm$   400 &    -91$\pm$   158 &     33$\pm$   158 &    -88$\pm$    36 &     34$\pm$    36\\
0.770 &   -538$\pm$   530 &    478$\pm$   530 &    100$\pm$   302 &    185$\pm$   302 &    236$\pm$   441 &   -340$\pm$   441 &   -651$\pm$   164 &     21$\pm$   164 &   -258$\pm$    47 &     -5$\pm$    47\\
0.288 &   -606$\pm$   355 &    480$\pm$   355 &   -567$\pm$   144 &     49$\pm$   144 &   -628$\pm$   361 &    332$\pm$   361 &   -422$\pm$   148 &    -52$\pm$   148 &   -513$\pm$    35 &     17$\pm$    35\\
0.797 &   -847$\pm$   378 &   -227$\pm$   378 &    -29$\pm$   226 &   -178$\pm$   226 &   -570$\pm$   291 &    225$\pm$   291 &   -288$\pm$   107 &    -46$\pm$   107 &   -221$\pm$    31 &     -2$\pm$    31\\
0.348 &   -696$\pm$   435 &     -8$\pm$   435 &   -549$\pm$   166 &    107$\pm$   166 &    -84$\pm$   455 &    118$\pm$   455 &   -241$\pm$   142 &    148$\pm$   142 &   -596$\pm$    35 &     75$\pm$    35\\
0.901 &    247$\pm$   654 &   -393$\pm$   654 &    567$\pm$   268 &   -200$\pm$   268 &    159$\pm$   442 &    -44$\pm$   442 &    432$\pm$   247 &    298$\pm$   247 &     55$\pm$    52 &    131$\pm$    52\\
0.489 &  -1116$\pm$   390 &   -287$\pm$   390 &   -753$\pm$   212 &   -159$\pm$   212 &   -540$\pm$   329 &    400$\pm$   329 &   -849$\pm$   105 &   -346$\pm$   105 &   -616$\pm$    32 &     39$\pm$    32\\
0.370 &   -933$\pm$   439 &    427$\pm$   439 &   -927$\pm$   222 &   -414$\pm$   222 &  -1136$\pm$   555 &    527$\pm$   555 &   -732$\pm$   182 &    152$\pm$   182 &   -465$\pm$    43 &      8$\pm$    43\\
0.510 &  -1054$\pm$   613 &    332$\pm$   613 &  -1012$\pm$   178 &     11$\pm$   178 &  -2105$\pm$   573 &  -1449$\pm$   573 &   -916$\pm$   187 &   -763$\pm$   187 &   -620$\pm$    32 &     76$\pm$    32\\
0.411 &   -389$\pm$   979 &    827$\pm$   979 &     57$\pm$   199 &    -25$\pm$   199 &  -3686$\pm$  1551 &   1512$\pm$  1551 &  -1336$\pm$   245 &   -489$\pm$   245 &   -582$\pm$    36 &    -33$\pm$    36\\
0.974 &    284$\pm$   554 &   -601$\pm$   554 &    444$\pm$   218 &   -158$\pm$   218 &     94$\pm$   343 &    350$\pm$   343 &   1130$\pm$   124 &   -318$\pm$   124 &     34$\pm$    27 &     21$\pm$    27\\
0.445 &   -773$\pm$   277 &    -27$\pm$   277 &   -500$\pm$    38 &    -32$\pm$    38 &   -603$\pm$    90 &   -515$\pm$    90 &   -737$\pm$    24 &     98$\pm$    24 &   -503$\pm$     6 &    -22$\pm$     6\\
\hline
\hline
\end{tabular}
}
\label{bz_2}
\end{table*}

\end{document}